\pdfoutput=1
\documentclass[twocolumn, linenumbers]{aastex631}

\usepackage{graphicx} 
\usepackage{amsmath}
\usepackage{rotating}    
\usepackage{physics}
\usepackage{bbding}
\usepackage{lineno}
\AtBeginDocument{%
  \DeclareRobustCommand{\sout}{\bgroup\ULdepth=-.5ex\relax\ULset}%
}
\usepackage[pass]{geometry}
\usepackage{multirow}
\usepackage[flushleft]{threeparttable}

\newcommand\blfootnote[1]{%
  \begingroup
  \renewcommand\thefootnote{}\footnote{#1}%
  \addtocounter{footnote}{-1}%
  \endgroup
}
\nolinenumbers
\begin{document}

\title{The Galactic Black Hole mass distribution \footnote{Released on }}

\author[0000-0003-4543-0912]{Lívia S. Rocha$^{a}$}
\affiliation{Instituto de Física, Universidade de São Paulo (USP)\\  R. do Matão, 1371, Cidade Universitária \\ São Paulo, SP 05508-090, Brazil}
\author[0009-0000-7480-5395]{Nathalia Pires$^{b}$}
\affiliation{Instituto de Astronomia, Geofísica e Ciências Atmosféricas (IAG), Universidade de São Paulo (USP)\\
R. do Matão, 1226, Cidade Universitária\\
São Paulo, SP 05587-050, Brazil}
\author[0000-0003-3109-9042]{Lucas M. de Sá}
\affiliation{Instituto de Astronomia, Geofísica e Ciências Atmosféricas (IAG), Universidade de São Paulo (USP)\\
R. do Matão, 1226, Cidade Universitária\\
São Paulo, SP 05587-050, Brazil}
\affiliation{Universit\"at Heidelberg, Zentrum f\"ur Astronomie, Institut f\"ur Theoretische Astrophysik \\
Albert Ueberle Str. 2, Heidelberg, 69120, Germany}

\author[0009-0002-9944-9889]{Bianca B. Martins}
\affiliation{Instituto de Astronomia, Geofísica e Ciências Atmosféricas (IAG), Universidade de São Paulo (USP)\\
R. do Matão, 1226, Cidade Universitária\\
São Paulo, SP 05587-050, Brazil}
\author[0009-0000-3268-9058]{Lucas G. Barão}
\affiliation{Instituto de Astronomia, Geofísica e Ciências Atmosféricas (IAG), Universidade de São Paulo (USP)\\
R. do Matão, 1226, Cidade Universitária\\
São Paulo, SP 05587-050, Brazil}
\author[0009-0008-7350-7546]{Guilherme R. C. Sampaio}
\affiliation{Instituto de Astronomia, Geofísica e Ciências Atmosféricas (IAG), Universidade de São Paulo (USP)\\
R. do Matão, 1226, Cidade Universitária\\
São Paulo, SP 05587-050, Brazil}
\author[0009-0008-0984-7642]{Davi V. Rodrigues}
\affiliation{Instituto de Astronomia, Geofísica e Ciências Atmosféricas (IAG), Universidade de São Paulo (USP)\\
R. do Matão, 1226, Cidade Universitária\\
São Paulo, SP 05587-050, Brazil}
\author[0009-0007-7430-8146]{Rachel M.G. Takahashi}
\affiliation{Instituto de Astronomia, Geofísica e Ciências Atmosféricas (IAG), Universidade de São Paulo (USP)\\
R. do Matão, 1226, Cidade Universitária\\
São Paulo, SP 05587-050, Brazil}
\author[0000-0003-1771-1055]{M.G.B. de Avellar}
\affiliation{Instituto Nacional de Pesquisas espaciais (INPE)\\ Av. dos Astronautas, 1758\\ São José dos Campos, SP 12227-010, Brazil}
\author[0000-0003-4089-3440]{J. E. Horvath}
\affiliation{Instituto de Astronomia, Geofísica e Ciências Atmosféricas (IAG), Universidade de São Paulo (USP)\\
R. do Matão, 1226, Cidade Universitária\\
São Paulo, SP 05587-050, Brazil}
\author[0000-0002-5914-0556]{Antônio Bernardo}
\affiliation{Instituto de Astronomia, Geofísica e Ciências Atmosféricas (IAG), Universidade de São Paulo (USP)\\
R. do Matão, 1226, Cidade Universitária\\
São Paulo, SP 05587-050, Brazil}

\blfootnote{$^{a}$livia.silva.rocha@usp.br
\\ $^{b}$nathaliadutrapires$@$usp.br}
 
\begin{abstract}
The population of extragalactic black holes has been significantly expanded and better understood since the first detection of gravitational waves. Over the past decade, approximately 290 events have been confirmed to involve at least one black hole. In contrast, the galactic black hole population has grown at a much slower pace, with only 42 systems possessing mass estimates, while the vast majority remain as candidates. In this work, we present the first update of a Bayesian analysis of the mass distribution of galactic black holes since 2011, previously based on a sample of 15 low-mass X-ray binaries and 5 high-mass X-ray binaries. Our update includes 28 low-mass X-ray binaries, 6 high-mass X-ray binaries, 7 detached binaries and 1 isolated black hole, also present in the living catalog \textit{CatNoir} which we introduce. We apply a Bayesian framework using five parametric models -- Gaussian, two-Gaussian, log-normal, power-law, and decaying exponential -- and compare their performance in fitting the observed data. Furthermore, we inspect the effect of removing model-dependent systems from the sample. Our results suggest that the power law better describes the Combined sample, while the Gaussian is preferred for the LMXB-only sample, closely followed by the two-Gaussian and log-normal. We derive the $1\%$-quantile of the preferred model's posterior predictive distribution to access information about the lower mass gap. For the Combined and LMXB Samples, we find $M_{1\%} > 4.06\,M_\odot$ and $M_{1\%} > 2.56\,M_\odot$ (at $90\%$ C.I.), respectively. The LMXB result allows for the existence of a sparsely populated Gap, instead of a completely empty one.

\end{abstract}


\section{Introduction}
\label{sec:intro}

Black Holes (BHs) as endpoints of stellar evolution have a long and distinguished history in Astrophysics, starting in the first half of the 20th century. Shortly after the pioneering works of \cite{Oppenheimer1939}, a shift in the perception from a pure mathematical solution to a physical object developed, in which the works of \citet{Zeldovich} and \cite{Wheeler} were decisive, as well as contributions from many theoreticians and observers across the second half of that century. Around 1970, BHs were accepted as real astronomical objects, an extreme form of stellar structure where the importance of dense matter description is overcome by gravity. 

However, by their very nature and absence of internal radiation mechanisms, BHs prove difficult to study in Nature. Their indirect effects, mainly through gravitational fields, are still the main signal of their existence. If, from one side, the fact that they can be fully defined by their mass, angular momentum and electric charge simplifies their descriptions, from the other it limits constraints on the formation processes. X-ray binaries (XBs), where high-energy emission is produced by matter accreted onto a BH from a stellar companion, constitute the large majority of known BHs in our Galaxy. In fact, the recent detection of an isolated object through gravitational microlensing by two independent groups \citep{2022ApJ...933L..23L, 2022ApJ...933...83S} offers the first ever non-binary evidence of a galactic BH.

The number of stellar BHs --- i.e., first generation stellar collapse products --- accumulated along the stellar evolution history of the Milky Way is expected to have a lower bound of $\sim 10^{7}$ objects. Nevertheless, this population is not well-sampled, and nowadays the number of BHs confirmed with confidence in our galaxy is still limited to the order of a few dozens. Even accounting for the existing uncertain candidates, the mismatch between the expected and observed sample is striking. Although this limits our general picture about the formation and evolution of this population, inferences based on the known sample are still possible, similarly to what is done for Neutron Stars (NSs) \citep{2023Univ...10....3R, 2018MNRAS.478.1377A} --- for which around 4000 objects are known \citep{2005AJ....129.1993M}, but with a ``hidden'' population surely much larger and with new types revealed from time to time.

In contrast, the sample of known extragalactic BHs increases at a much greater pace, thanks to the efforts of the LIGO-Virgo-KAGRA Collaboration (LVK) and its gravitational-wave (GW) detectors. While prior to the start of operations there was an expectation of a much larger number of NS-NS mergers detections compared to BH-BH (or BBH) mergers, the observational runs revealed a different scenario. The fourth Gravitational Wave Transient Catalog (GWTC-4) \citep{Abac_2025}, comprising all detections from the first four observing runs (from 2015 to November 2025), has a total of 219 GW detections, with the vast majority of them being associated with BBH mergers, only 2 firmly associated with NS-NS merges and at least 7 where the least massive component has a mass within the \textit{Lower Mass Gap} (LMG) - a long-standing problem that emerged after the first analysis of transient black holes \citep{1998ApJ...499..367B} -, thus considered as unidentified objects. 

The LMG arose from an apparent absence of observed compact objects with masses between the most massive NS and the least massive BH, in the range between $\sim 2-5~M_\odot$. Whether this feature is a consequence of observational bias or a real feature from the mass distribution of compact objects remains under discussion, but what we now know for sure is that the gap is likely only partially ``depleted'', instead of ``desert'' \citep{2022ApJ...941..130D,RayHidingOutLowEnd2025}, given recent detections of mass gap objects, which means that although the formation of very light BH's might be rare, they do exist.

In this work, we update the galactic BHs catalog presented by \cite{galaxiesBHs2023deSa} by incorporating newly reported systems and revised orbital parameter estimates. In addition, we provide the first Bayesian analysis of the sample since the works by \citet{Ozel2010} and \citet{Farr2011}. Furthermore, we also provide a living catalog of these systems, named \texttt{CatNoir}\footnote{Available at \href{https://catnoir.iag.usp.br}{https://catnoir.iag.usp.br}. The content of Tables \ref{tab:BHs_ecc} and \ref{tab:list_BHs} are the main source of information to \texttt{CatNoir 1.0.0}, released along with this work.} (\texttt{version 1.0.0}), which will be continuously updated, and provides to the community a reliable source of information for galactic stellar-mass BH's. The current sample contains 42 objects with orbital parameter inferences, of which 28 are in Low-Mass X-ray Binaries (LMXBs), 6 in High-Mass X-ray Binaries (HMXBs), 7 in detached binaries, and 1 is an isolated BH. Additionally, we discuss the presence of a LMG in face of the updated galactic sample of BHs.

The structure of this work is as follows. In Section \ref{sec:sample} we discuss the mass determination methods as well as present one-by-one of the revised and newly discovered BHs since the work of \cite{galaxiesBHs2023deSa}. In Section \ref{sec:analysis} we discuss the Bayesian method and each parametrization used to model the mass distribution of BHs, whose results are presented in Section \ref{sec:results}, together with the discussion on the model selection. In Section \ref{sec:lower_mass_gap} we discuss the problem of the LMG in face of our results presented in Section \ref{sec:results}. Section \ref{sec:formation_channels} provides a discussion on the formation scenarios that might form the different classes of BHs we analyze here. Finally, we summarize our results and make our conclusions in Section \ref{sec:conclusions}.

\section{The BH Mass Sample} \label{sec:sample}
The majority of Galactic BHs discovered so far are in X-ray binaries, in which optical and X-ray observations can provide spectroscopical measurements of the companion star, providing us its orbital period and velocity semi-amplitude. Through Kepler's third law, these quantities define the BH's mass function:

\begin{equation}\label{eq:mass_function}
    f(M_{BH}) = \frac{P_b~K_{cp}^3}{2 \pi G} (1 - e^2)^{3/2} = \frac{M_{BH} ~ \sin^3 i} {(1 + q)^2},
\end{equation}

\noindent where $P_b$ is the orbital period, $K_{cp}$ the semi-amplitude of the companion's velocity curve, $e$ the eccentricity, and $G$ is the gravitational constant. Once the mass function ($f_{BH}$), the orbital inclination angle ($i$) and the mass-ratio ($q = M_{cp} / M_{BH}$) between the companion star mass ($M_{cp}$) and the BH mass ($M_{BH}$) are determined or constrained, a mass estimate for the BH can be obtained. 

While non-interacting BH \citep[often called ``dormant BH'', e.g.,][]{t2019,ElBadry2022,ElBadry2023,panuzzo2024discovery} do not provide direct observational evidence of their existence in the form of X-rays, their presence in a binary system can still be inferred through variations in the radial velocity and spectroscopic properties of their companions. Isolated BHs, on the other hand, do not emit radiation themselves and lack stellar companions, making it difficult to infer their properties through traditional methods. Accretion of matter from the Interstellar Medium can possibly produce X-rays or radio emission in dense environments \citep{Agol&Kamionkowski2002}, but no such detection has been confirmed yet. Currently, the only method to detect an isolated non-interacting BH is by gravitational microlensing.

The observation of background stars can be disrupted by the motion of a star or a compact object as it crosses the line of sight, and end up amplifying their brightness or causing shifts in their apparent position. Based on General Relativity, astrometric microlensing depends on the relative proper motion between the background source of brightness and the object (lens), as well as their relative parallax. These parameters define a characteristic time scale over which the source moves an angular distance equal to the Einstein ring radius, directly related to the lens mass. From light curves of background sources, it is possible to determine the parallax and the Einstein ring radius and, therefore, the mass of the object. 

The only BH confirmed through microlensing so far is OGLE-2011-BLG-0462 (also named MOA-2011-BLG-191), located in the direction of the galactic bulge and detected independently by the Optical Gravitational Lensing Experiment (OGLE) and Microlensing Observations in Astrophysics (MOA) surveys \citep{2022ApJ...933L..23L, 2022ApJ...933...83S}. Although an initial conflict was found in mass estimates, subsequent follow-up observations confirmed that the lensing object is consistent with an isolated stellar-mass black hole with a mass of $M_{BH} = 7.88(8) ~M_\odot$ \citep{mroz2022}, with later estimates still favoring a mass above the LMG \citep{LamReanalysisIsolatedBlackHole2023,sahu2025}. The object Gaia18ajz is also known as a candidate BH detected through microlensing \citep{2025A&A...694A..94H}, but since the conflict in its mass remains, we do not include it in our catalog. 

In this section we update the sample of BH masses collected by \citet{galaxiesBHs2023deSa}, including X-ray binaries, dormant BHs and the microlensing event. We included 9 new objects, revised 6 masses and excluded LB-1 and SS-433 from our sample, as discussed below. Furthermore, the BH candidates NGC 3201 \#12560, with $M\sin(i)\gtrsim7.68(50) M_{\odot}$ \citep{giesers2019}, and NGC 3201 \#21859, with $M\sin(i)\gtrsim4.36(41)M_{\odot}$ \citep{giesers2018}, were kept out of the sample since only minimum mass thresholds were evaluated for them.

In the following, we discuss all new systems included in the sample (Section \ref{subsec:new_bhs}), as well as the revised ones (Section \ref{subsec:revised_bhs}) since the work of \citet{galaxiesBHs2023deSa}. The remaining systems were already extensively presented in the mentioned paper, and it is not in the scope of this work to present them again. As discussed in \citet{galaxiesBHs2023deSa}, the nominal values and uncertainties of Keplerian parameters and BH masses have sometimes been reported in the literature with inconsistent or unclear definitions. We standardize our sample by explicitly treating each Keplerian parameter as either a Gaussian ${\cal N}(\mu, \sigma)$, with mean $\mu$ and standard deviation $\sigma$, an asymmetric Gaussian \citep[defined in][]{galaxiesBHs2023deSa} ${\cal AN}(m, \sigma_1, \sigma_2)$, with mode $m$ and upper (lower) standard deviation $\sigma_1$ ($\sigma_2$), uniform distribution ${\cal U}(a,b)$ defined between $a-b$, or an isotropic distribution ${\cal I}(a,b)$. Throughout this text, the values following a Gaussian or an Asymmetric Gaussian distribution will be represented using the notations $\mu(\sigma)$ and $\mu_{-\sigma_{2}}^{+\sigma_1}$, respectively. Table \ref{tab:BHs_ecc} shows the eccentricities for those systems in which they are non-negligible. Table \ref{tab:list_BHs} lists all the available Keplerian parameters and the BH masses originally reported in the literature ($M_{lit}$) for the 42 systems in our sample, followed by the individual masses we infered in this work. In Appendix \ref{apx:schematic_visu} we provide a schematic visualization of each BH in our sample and their companions, when applicable, through Figures \ref{fig:individual_BHs1} and \ref{fig:individual_BHs2}. 


\begin{table}
    \caption{List of 10 BHs for which a non-negligible eccentricity was found. The notation on the right column means that the eccentricities follow a Gaussian distribution.}
    \centering
    \begin{tabular}{l | c}
    Name & eccentricity \\
    \hline\hline
     Cyg X-1& 0.018(3) \\
LMC X-1 & 0.0256(66) \\
M33 X-7 & 0.0185(77) \\
2MASS J05215658+4359220 cp. &0.0048(26) \\
Gaia BH1 & 0.451(5) \\
Gaia BH2&	0.5176(9)\\
Gaia BH3&	0.7291(48)\\
VFTS 243&	0.017(12)\\
HD 130298&	0.457(7)\\
G3425&	0.05(1)\\
    \end{tabular}\label{tab:BHs_ecc}
\end{table}

\subsection{New objects}\label{subsec:new_bhs}
Below we discuss each of the systems included in the catalog since the work of \citet{galaxiesBHs2023deSa} (please refer to it or to \textit{CatNoir} for information about systems that are not discussed in this work).

\subsubsection{3A 1524-617}

The work of \citet{yanes2024evidence} exhibits the main characteristics of the binary 3A 1524-617 (KY TrA) and its dark companion from a study of optical spectroscopic and photometric data. They were able to determine an orbital period of $P_b =0.26(1) ~ d$ and a semi-amplitude velocity of $K_{cp} = 501(52)\; \mathrm{km}\; \mathrm{s}^{-1}$ (applying the FWHM-$K_2$ correlation \citet{casares2015fwhm}). Using the empirical correlation between inclination and $H\alpha$ line profile \citep{casares2022correlation}, they also found $i = 57(13)^\circ$. Finally, through Monte Carlo simulation, and adopting $q=\mathcal{U}(0.01,0.31)$, the BH mass was inferred to be $M_{BH} = 5.8^{+3.0}_{-2.4}\;\mathrm{M}_\odot$.
\subsubsection{Gaia BH2}
\citet{ElBadry2023} found a new BH candidate from \textit{Gaia} DR3, by studying the red giant astrometry companion of the compact object and with follow-up spectroscopy measurements. All the important results are shown in Table 2 of the mentioned work, and here we adopt those based on \textit{Gaia} and radial velocity together: $P_b =  1276.7(6) ~ d$, $K_{cp} = 25.23(4) \; \mathrm{km}\; \mathrm{s}^{-1}$, $i = 34.87(34)^{\circ}$, $M_{cp} = 1.07(19)\;\mathrm{M}_\odot$ and $M_{BH} = 8.94(34)\;\mathrm{M}_\odot$. Like Gaia BH1, this is one of the few objects with a measured eccentricity: $e = 0.5176(9)$. The mass ratio informed in Table \ref{tab:BHs_ecc} is estimated from $q = M_{cp}/M_{BH}$.

\subsubsection{Gaia BH3}
Gaia BH3 is a close ($\sim 590$ pc) and substantially massive (compared with the known Galactic BHs sample) compact object with a mass of $M_{BH} =  32.70(82)\;\mathrm{M}_\odot$, as reported by \citet{panuzzo2024discovery}. The system has an orbital period $P_b = 4253(98)~ d$ and eccentricity $e = 0.7291(48)$. Spectroscopic analysis of the companion star indicates a mass of $M_{cp} = 0.76(5)\;\mathrm{M}_\odot$.

\subsubsection{MAXI J0637-430}

\citet{soria20222}, combining the sets of published works of X-ray and optical results about the BH candidate MAXI J0637-430, were able to deduce the main properties of the system. Here, we adopt the given results $P_b = 0.092^{+0.034}_{-0.025}$ d, $q = \mathcal{U}(0.04,0.06)$ and $M_{BH} =  5.1(1.6)\;\mathrm{M}_\odot$.

\subsubsection{Swift J1727.8-1613}

With an optical spectroscopy campaign with the GTC telescope, \citet{sanchez2024} reported the dynamical parameters of the X-ray transient Swift J1727.8-1613, discovered in September 2023. They were able to determine $P_b=10.8038(10)$ h, $K_{cp}=390(4) \; \mathrm{km}\; \mathrm{s}^{-1}$ and $f(M_{BH})=2.77(9)\ M_{\odot}$. The authors constrain the inclination angle to $i < 74^\circ$, resulting in a minimum compact object mass of $M_{BH}>3.12(10)\ M_{\odot}$, assuming a limiting case where $q\sim0$. No lower limit is defined for the inclination angle, so we adopt $\mathcal{U}(5, 74)$. Since their analysis was not able to accurately determine the value of the mass ratio, we assume $q=\mathcal{U}(0.001,1)$.

\subsubsection{HD 130298}

\citet{m2022} reported the parameters of the system HD 130298, previously identified by \citet{peri2011}, with an orbital period of $14.62959(85)$ d. They evaluated an evolutionary mass of $M_{cp}=28.0^{+5.2}_{-4.1}\ M_{\odot}$ and, due to spectral lines that are not visible in the disentangled and composite spectra, the authors suggest that the primary object is expected to be a quiet stellar-mass BH. Constraints on the orbital inclination angle, $i=54(16)^{\circ}$, leads to a mass of $M_{BH}=8.8^{+3.5}_{-1.5}\ M_{\odot}$.

\subsubsection{G 3425}

\citet{wang2024} identified a BH candidate in the LMG range using spectroscopy from the Large Aperture Multi-Object Spectroscopic Telescope and astrometry data from Gaia. They derived a period $P_b=879.11^{+3.22}_{-2.57}$ d , $K_{cp}=22.91^{+0.13}_{-0.14} \; \mathrm{km}\; \mathrm{s}^{-1}$, $e = 0.05(1)$, $i=89^{\circ ~+15.48^\circ}_{~~~-10.08^\circ}$, $M_{BH}=3.58^{+0.80}_{-0.47}\ M_{\odot}$ and $M_{cp}=2.66^{+1.18}_{-0.68}\ M_{\odot}$, from which we estimated $q$.

\subsubsection{VFTS 243}

Located in the Large Magellanic Cloud, \citet{shenar2022} analyzed data from over 6 years from the Very Large Telescope FLAMES Tarantula Survey (VFTS) to study this non-interacting massive system. They could determine $P_b=10.4031(4)$ d, $e=0.017(12)$, $K_{cp}=81.4(1.3) \; \mathrm{km}\; \mathrm{s}^{-1}$ and $f=0.581(28)\ M_{\odot}$. The authors argue that the ellipsoidal variations detected in the system's light curve imply an orbital inclination of $i=\mathcal{U}(40,90)^\circ$, which leads to $M_{BH}=10.1(2.0)\ M_{\odot}$. Furthermore, since they infer $M_{cp} = 25.0(2.3)$, we define the mass ratio to be $q = 0.402(201)$.

\subsubsection{IC 10 X-1}

Located in the Local Group starburst galaxy IC 10, \citet{silverman2008} examined the system, which consists of a variable X-ray source and a Wolf-Rayet (WR) companion star. They estimated $P_b=34.93(4)$ h, $K_{cp}=370(20) \; \mathrm{km}\; \mathrm{s}^{-1}$ and $f(M)=7.64(1.26)\ M_{\odot}$. Since the system presents X-ray eclipses, they constrained the inclination angle to $i\gtrsim 78^{\circ}$, yielding a minimum BH mass $\geq 23.1(2.1)\ M_{\odot}$, by assuming a range of reasonable WR masses between $\mathcal{U}(17,35)\ M_{\odot}$. We adopt an upper inclination angle value of 90°, since \cite{clark2004} find that this system has deep X-ray eclipses. From WR mass estimates and the resultant BH mass, we infer the mass ratio to be $q = \mathcal{U}(0.6746,1.129)$.

\subsection{Revised BHs}\label{subsec:revised_bhs}
Below we present all the measurement updates of BH systems already present in \citet{galaxiesBHs2023deSa}, and justify the decision of removing LB-1 and SS-433 from the sample.

\subsubsection{GRS 1009-45}

For this source, we will adopt the same parameters as the ones in \citet{galaxiesBHs2023deSa}, however, we will adopt an uncertainty of 10\% on the literature's BH mass of 4.4 $M_{\odot}$, resulting in $M_{BH}=4.40(44)$. We did this because \citet{filippenko1999} present two different mass determination, depending on the inclination angle estimate. Then, for compatibility with the other orbital parameters, we selected $M_{BH}\approx4.4\ M_{\odot}$.

\subsubsection{GRO J0422+32}

\citet{cherepashchuk2024band} have used new photometric data in the I$_c$ band to update the parameters for GRO J0422+32. They found that their new measurements are inconsistent with the orbital period from \citet{webb2000tio}, finding a new value of $P_b = 0.2115220(5)$ d, consistent with older measurements by \citet{filippenko1995mass} and \citet{harlaftis1999keck}. Consequently, they assumed $f(M) = 1.21(6)$ from \citet{filippenko1995mass} and considered both $q' \approx9$ \citep{harlaftis1999keck} and $q' \approx 14$ \citep{petrov2017masses} ($q' = 1/q$), the latter being considered more reliable since it employs a more advanced model for the luminous companion. The $q' \approx 9$ case yields parameters close to the ones reported in \citet{galaxiesBHs2023deSa}, but the authors found that it yields masses for the donor star in excess of those expected for its spectral type (M2V), making the $q' \approx14$ case preferable.  We thus adopt their $P_\mathrm{orb}$ and $f(M)$, as well as the suggested interval for the inclination angle, $\mathcal{U}(33, 49)^\circ$, which yields $M_{cp} = 0.47(21) ~\mathrm{M}_\odot$ and $M_{BH} = 6.5(2.9) ~\mathrm{M}_\odot$. \citet{filippenko1995mass} finds $q=0.1093(86)$, which is the value adopted in this work.

Notably, this makes the BH in GRO J0422+32 a potential LMG candidate, if its mass can be determined with greater precision. The large uncertainty is connected to lack of available simultaneous X-ray observations to the author's optical data, which are key to further constraining the mass. 

\subsubsection{GRO J1655-40}

Using observations from the Small and Medium Aperture Research Telescope System (SMARTS), \cite{petretti2023} presented their analysis of the I-band photometry taken between 2006 and 2016. The authors used two different methods: a ``model-based'' approach -- using the Eclipsing Light Curve (ELC) code developed by \cite{oroszhauschildt2000}-- and a ``data-based'' approach -- using algorithms based on the data itself, such as PDM (developed by \cite{stellingwerf1978}). We assume here the results by the model-based approach because off the parameters determined in the light-curve fitting. From this method, they find $P_b=2.621928(4)$ d, $i=65.0(8)^{\circ}$, $q=\boldsymbol{0.263_{-0.019}^{+0.023}}$ and $K_{cp}=226.2(1.4) \; \mathrm{km}\; \mathrm{s}^{-1}$, with $1\sigma$-credible interval. However, for this system, we update only the values of $P_b$ and $K_{cp}$, since the authors highlighted that the I-band LC used by them was $\sim$ 0.2 mag fainter than the one used in \cite{greene2001} (where the previous results came from). Therefore, their results for $q$ and $i$ may be slightly skewed, despite their compatibility at the $3\sigma$-level. We inform in Table \ref{tab:BHs_ecc} the mass inferred in \citet{galaxiesBHs2023deSa}.

\subsubsection{XTE J1859+226}

\cite{yanesrizo2022} presented a refined mass measurement for XTE J1859+226 from the data obtained by 2 nights of observations using time-resolved spectroscopy from {\it Gran Telescopio Canarias} and {\it William Herschel Telescope} photometry. Cross-correlating the star's spectra against a late K-type spectral template they constrained the orbital period to $P_b=0.276(3)$ d. Using correlations between the orbital parameters and properties of the double-peaked H$\alpha$ emission-line they found: $i=66.6(4.3)^{\circ}$, $K_{cp}=562(40) \; \mathrm{km}\; \mathrm{s}^{-1}$ , $q=0.07(1)$, $M_{cp}=0.55(16)$ $M_{\odot}$ and $M_{BH}=7.8(1.9)$ $M_{\odot}$.

\subsubsection{GRS 1716-249}

\citet{casares2023orbital}, using photometric data and a FWHM-$K_2$ correlation to the disc $H\alpha$ emission line, estimated the semi-amplitude velocity $K_{cp} = 521(52) \; \mathrm{km}\; \mathrm{s}^{-1}$ and orbital period $P_b = 0.278(8)$ d to the transient GRS 1716-249, leading to a mass function $f(M) = 4.1(1.2)\;\mathrm{M}_\odot$. From the depth of the $H\alpha$ from the quiescent light curve they constrained $i = 61(15)$°. Via Monte Carlo simulation, the masses were estimated as $M_{cp} = 0.5(3)\;\mathrm{M}_\odot$ and $M_{BH} =  6.4^{+3.2}_{-2.0}\;\mathrm{M}_\odot$ at the $1\sigma$ level.

\subsubsection{M33 X-7}

\cite{ramachandran2022} used phase-resolved simultaneous Hubble Space Telescope (HST) and XMM-Newton observations for tracing the interaction of the stellar wind with the BH. They obtained $i=65.5$°, $M_{cp}=38^{+22}_{-10}\ M_{\odot}$, $M_{BH}=11.4^{+3.3}_{-1.7}\ M_{\odot}$. Since no uncertainties were given by the authors, we considered a standard deviation of 10\% of the reported value for $i=65.5(6.5)^{\circ}$, and $q=3.30(1.5)$ as a rough estimate based on the individual mass determinations.

\subsubsection{Cyg X-1}

\cite{new} revisited the case of Cyg X-1 and found a new distance using radio astrometry.
As a consequence, the mass of the black hole was revised upwards to $21.2(2.2) \, M_{\odot}$. While this new estimate pushes the black hole towards the higher interval of masses (see below), we have checked that the final effect on the mass distribution is not large and we kept the former value of the mass in the calculations. 

\subsubsection{OGLE-2011-BLG-0462}

\citet{sahu2025}, with three additional epochs of HST observations in relation to their previous study \citep{2022ApJ...933...83S} and photometry from 16 different telescopes, re-constrained the lens mass to $M_{BH} = 7.15(83)\ M_{\odot}$, which is still compatible with their previous determination of $7.1(1.3) \ M_{\odot}$. \citet{Lam_2022} found a BH mass in the range of $\mathcal{U}(1.6,4.4)\ M_{\odot}$. In this paper we consider the results from \citet{sahu2025}.

\subsubsection{LB-1}

LB-1 is a Galactic binary on a $P_b = 78.9(3)$ d orbit, with optical emission dominated by a B-type primary of approximately solar metallicity and velocity semi-amplitude $K_{cp} = 52.8(7) \; \mathrm{km}\; \mathrm{s}^{-1}$. The system displays a strong variable H$\alpha$ emission line. The original inference of the presence of a massive BH \citet{Liu2019}, included in \citet{galaxiesBHs2023deSa}, assumed this line originated in an accretion disk around the secondary, and under this assumption derived a velocity semi-amplitude $K_{BH} = 6.4 \; \mathrm{km}\; \mathrm{s}^{-1}$, which combined with the primary mass yields an estimated secondary mass of $68^{+11}_{-13}\,M_\odot$. Due to the lack of spectral features associated with the secondary, the authors concluded that the companion must be a $\sim 68\,M_\odot$ BH.

This result was immediately considered either problematic or intriguing, as there is no detected X-ray emission from the purported accretion disk, and the existence of such a massive BH in a solar-metallicity binary challenges our current understanding of supernovae and stellar evolution. Subsequent studies questioned both the mass estimate and the nature of the companion. \citet{AbdulMasih2020} showed that the variability in the H$\alpha$ emission line could be explained by the superposition of the B star’s H$\alpha$ absorption line and a static H$\alpha$ emission profile. This interpretation invalidates the $K_{BH}$ measurement from \citet{Liu2019} and removes the evidence for a massive BH, leaving the nature of the companion uncertain. \citet{ElBadry2020} similarly demonstrated that the observed variability in H$\alpha$ is consistent with a static emission component superimposed on the primary’s absorption line. They proposed that the absence of spectral features from a secondary still supports a BH companion, but with a mass between 6 and 20\,$M_\odot$ for the $87\%$ most likely range of inclinations.

\citet{Shenar2020} applied modern spectral disentangling techniques and showed that the optical flux previously attributed solely to the primary is actually split in an 11:9 ratio between the primary and secondary. They identified the secondary as a rapidly rotating ($v \sin i \approx 300 \; \mathrm{km}\; \mathrm{s}^{-1}$) B3\,V star with a disk, i.e., a Be star. The broadening of the secondary’s spectral lines due to rapid rotation likely caused its spectral features to be hidden within those of the primary. Finally, the authors concluded that LB-1 is a solar metallicity binary consisting of a $1.5(4)\,M_\odot$ stripped primary and a $7(2)\,M_\odot$ Be companion near critical rotation. We therefore have removed LB-1 from our sample.

\subsubsection{SS-433}

\citet{cherepashchuk2025} constrained the system mass to $M_{BH}=4.2(4)\ M_{\odot}$, which heavily differs from his previous result of $M_{BH}>7\ M_{\odot}$ \citep{cherepashchuk2019} reported in \citet{galaxiesBHs2023deSa}. The lack of dynamical confirmation of a BH in SS-433 has also kept the nature of its compact object in debate. Because the first estimate places the compact object in SS-433 inside the LMG, our analysis of the BH mass distribution could be highly sensitive to its inclusion. We therefore opt for a more conservative approach, and also remove SS-433 from our sample.


\begin{table*}[tbp]
\caption{\footnotesize Orbital and dynamical parameters for the updated sample of BHs, with 42 objects in total. As discussed in Section \ref{sec:sample}, we represent the parameters through distributions, which can be the normal (${\cal N}$), the asymmetric normal (${\cal AN}$) -- as described by \citet{galaxiesBHs2023deSa} --, or uniform (${\cal U}$). In the sixth column we present the BH masses as originally reported in the literature (see footnote for references). In the seventh column we report the result of a skewed Gaussian fit over the results of our mass recomputation (Section \ref{sec:estimated_masses}) as the median of the distribution accompanied by the $68\%$ credibility interval. The data compiled in this Table can also be found in \texttt{CatNoir v 1.0}.}
\label{tab:list_BHs}
\centering
\scriptsize
\begin{threeparttable}
 {\begin{tabular}{|l c c c c c c|}
   \hline
\textbf{Name} & \textbf{$P_{orb}$ (d)} & \textbf{$K_{cp}$ (km\,s$^{-1}$)}  & \textbf{\textit{i} (deg)} & \textbf{\textit{q}} & $M_\mathrm{lit}$ ($\mathrm{M}_\odot$) & $M_\mathrm{new}$ \\
\hline
4U 1543-475$^{1}$ & 1.116407(3) & $129.33^{+1.68}_{-1.79}$&  20.7(1.5) & $\mathcal{U}$(0.25-0.31) & $\mathcal{U}$(8.45,10.39) & $9.15_{-1.47}^{+2.50}$ \\ 
GRS 1915+105$^{2, 3}$ &33.5(1.5) & 140(15) & 70(2) & $0.08_{-0.02}^{+0.04}$ & 14(4)& $13.43_{-3.56}^{+5.72}$ \\ 
GS 1354-64$^{3, 4, 5}$ & 2.54451(8) & 279.0(4.7) & $\mathcal{U}$(5, 80) & $0.12^{+0.03}_{-0.04}$  & $> 7.6(7)$ & $8.46_{-0.46}^{+4.61}$\\ 
GRS 1124-684$^{6, 7}$ & 0.43260249(9) & 406.8(2.7) & $43.2^{+2.1}_{-2.7}$ & 0.079(7)  & $11^{+2.1}_{-1.4}$ & $10.92_{-1.00}^{+2.14}$ \\ 
XTE J1118+480$^8$ & 0.1699337(2) & 710.0(2.6) & 73.5(5.5) & 0.024(9) & $7.46^{+0.34}_{-0.69}$ & $7.34_{-0.33}^{+0.99}$\\ 
3A 0620-003$^8$ & 0.32301415(7) & 435.4(5) & 51.0(9) & 0.060(4) & $6.61^{+0.23}_{-0.17}$ & $6.63_{-0.26}^{+0.26}$\\
GS 2000+251$^{1,9}$ & 0.344086(2) & 519.5(5.1) & $\mathcal{U}$(54, 60) & $0.042(12)$ & $\mathcal{U}$(5.5,8.8) & $9.05_{-0.48}^{+0.90}$\\
MAXI J1659-152$^{10}$ & 0.1006(2) & 750(80) & $\mathcal{U}$(70, 80) & $\mathcal{U}(0.018, 0.067)$  & $\mathcal{U}$(3.3,7.5) & $7.96_{-2.09}^{+5.96}$ \\ 
MAXI J1305-704$^{3,11}$ & 0.394(4) & 554(8) & $72^{+5}_{-8}$ & $0.05(2)$& $9.0^{+1.5}_{-1.0}$ & $8.90_{-0.59}^{+1.92}$ \\
GS 2023+338$^{12,13}$ & 6.4714(1) & 208.5(7) & $67^{+3}_{-1}$ & $0.060^{+0.004}_{-0.005}$ & $9.0^{+0.2}_{-0.6}$ & $8.56_{-0.48}^{+0.29} $\\ 
XTE J1650-500$^{3,14}$ & 0.3205(7) & 435(30) & $\mathcal{U}$(50, 70) & $\mathcal{U}$(0.01, 0.50) & $\mathcal{U}$(4.0,7.3) & $5.24_{-1.13}^{+3.45}$ \\ 
GRO J0422+32$^{15,16,17}$ & 0.2115220(5) & 380.6(6.5) & $\mathcal{U}$(33, 49) & $\mathcal{U}(0.0714, 0.1111)$  & $6.5(2.9)$ & $4.13_{-0.70}^{+1.11}$\\ 
H 1705-250$^{18,19}$ & 0.5228(44) & 441(6) & $\mathcal{U}$(60, 80) & $0.014^{+0.019}_{-0.012}$ & $\mathcal{U}$(5.0,7.4) & $5.56_{-0.30}^{+1.34}$ \\ 
GRO J1655-40$^{3, 20}$ & 2.621928(4) & 226.2(1.4) & 65.0(8) & $0.263_{-0.019}^{+0.023}$ & $6.86_{-0.49}^{+0.52}$ & $6.84_{-0.49}^{+0.57}$ \\ 
XTE J1859+226$^{21}$ & 0.276(3) & 562(40) & $66.6(4.3)$ & $0.07(1)$ & $7.8(1.9)$ & $8.52_{-2.48}^{+6.20}$\\ 
MAXI J1803-298$^{3,22}$ & 0.29(3) & 515(55) & $\mathcal{U}$(65, 85) & $\mathcal{U}$(0.01, 0.20) & $\mathcal{U}(4.0,8.0)$ & $5.18_{-1.29}^{+2.76}$\\ 
MAXI J1820+070$^{23,24}$ & 0.68549(1) & 417.7(3.9) & $\mathcal{U}$(66, 81) & 0.072(12)  & $\mathcal{U}(5.96,8.06)$ & $6.59_{-0.28}^{+0.88}$\\ 
XTE J1550-564$^{3,25}$ & 1.5420333(24) & 363.14(5.97) & 74.7(3.8) & 0.033(3)  & 9.10(61) & $9.06_{-0.54}^{+0.81}$\\ 
GX 339-4$^{26}$ & 1.7587(5) & 219(3) & $\mathcal{U}$(37, 78) & 0.18(5)  & $\mathcal{U}$(2.3,9.5) & $3.49_{-0.08}^{+4.10}$\\ 
GRS 1009-45$^{3,27}$ & 0.2852060(14) & 475.4(5.9) & 78.0(7.8) & 0.137(15) & $4.40(44)$ & $4.35_{-0.21}^{+0.64}$\\ 
SAX J1819.3-2525$^{3,28,29}$ & 2.81678(56) & 211.0(3.1) & 72.3(4.1) & $0.447^{+0.048}_{-0.034}$  &$6.4(6)$ & $6.68_{-0.51}^{+0.82}$\\
3A 1524-617$^{30}$ & 0.26(1) & 501(52) & 57(13) & $\mathcal{U}$(0.01,0.31) &$5.8^{+3.0}_{-2.4}$ & $5.46_{-1.13}^{+5.90}$\\ 
Swift J1727.8-1613$^{31}$ & 0.45016(4) & 390(4) & $\mathcal{U}$(5, 74) & $\mathcal{U}$(0.001,1) & $> 3.12(10)$ & $3.71_{-0.71}^{+0.03}$ \\ 
GRS 1716-249$^{32}$ & 0.278(8) & 521(52) & 61(15)  &$\mathcal{U}$(0.02,0.25)& $6.4^{+3.2}_{-2.0}$  & $5.12_{-0.11}^{+0.11}$\\ 
MAXI J0637-430$^{33}$ & $0.092^{+0.033~*}_{-0.025}$ & -- & --& $\mathcal{U}$(0.04, 0.06) & $5.1\pm1.6^*$ & $5.1_{-1.6}^{+1.6}$\\ 
MAXI J1813-095$^{34}$ & -- & -- & --  &--& 7.4(1.5)$^*$ & $7.4_{-0.93}^{+0.90}$\\ 
MAXI J1535-571$^{35}$ & -- & -- & --   &--& 8.9(1.0)$^*$ & $8.9_{-1.0}^{+1.0}$\\ 
H 1743-322$^{36,37}$ & -- & -- & --   &--& $11.21^{+1.65~*}_{-1.96}$ & $11.13_{-2.00}^{+1.61}$ \\ 
Cyg X-1$^{3,38,39,40}$ & 5.599829(16) & 75.57(70) & 27.06(76) & $1.29^{+0.17}_{-0.14}$ &$14.81(98)$ & $14.06_{-1.87}^{+2.59}$ \\ 
LMC X-1$^{3, 41}$ & 3.90917(5) & 71.61(1.10) & 36.38(2.02) & $2.81^{+0.62}_{-0.41}$& $10.91(1.54)$ & $10.48_{-2.14}^{+4.65}$ \\ 
LMC X-3$^{3,42}$ & 1.7048089(11) & 241.1(6.2) & 69.24(72) & $0.512^{+0.099}_{-0.084}$& $6.98(56)$ & $6.93_{-0.86}^{+1.17}$\\ 
M33 X-7$^{3, 43}$ & 3.453014(20) & 108.9(5.7) & 65.50(6.55) & 3.3(1.5)& $11.4^{+3.3}_{-1.7}$ & $15.16_{-3.23}^{+5.23}$ \\ 
NGC 300 X-1$^{3,44,45}$ & 1.3663375(125) & 267.5(7.7) & $\mathcal{U}$(60, 75) & $1.42^{+0.70}_{-0.33}$ & $17(4)$ & $21.44_{-4.41}^{+16.25}$ \\ 
IC 10 X-1$^{46,47}$ & 1.45175(1) & 370(20) & $\mathcal{U}$(78, 90) & $\mathcal{U}$(0.6746, 1.129) & $ > 23.1(2.1)$ & $28.86_{-8.11}^{+14.61}$  \\ 
2MASS J0521+435$^{3,48}$ & 83.2(6) & 44.6(1) & $75.43^{+5.16}_{-2.35}$& $0.74^{+0.26}_{-0.17}$ & $3.3^{+2.8}_{-0.7}$ & $2.52_{-0.40}^{+0.92}$\\ 
HD 130298$^{49, 50}$ & 14.62959(85) & 71.78(68) & 54(16) & $3.180^{+0.720}_{-0.499}$ & $8.8^{+3.5}_{-1.5}$ & $11.57_{-3.84}^{+11.51}$\\ 
VFTS 243$^{51}$ & 10.4031(4) & 81.4(1.3) & $\mathcal{U}$(40, 90) & 0.404(201)& $10.1(2.0)$ & $8.82^{+8.88}_{-1.96}$ \\ 
Gaia BH1$^{3,51,52}$ & 185.59(5) & 66.7(6) & 126.6(4) & 0.097(5) & $9.62(18)$ & $9.42_{-0.30}^{+0.33}$ \\ 
Gaia BH2$^{53}$ & 1276.7(6) & 25.23(4) & 34.87(34) & 0.120(22) & $8.94(34)$ & $8.80_{-0.72}^{+0.94}$\\
Gaia BH3$^{54}$ & 4253(98) & -- & 110.580(95)  & -- & 32.70(82) & $32.7_{-0.8}^{+0.8}$\\ 
G3425$^{55}$ & $879.11^{+3.22}_{-2.57}$ & $22.91^{+0.13}_{-0.14}$ & $89.30^{+15.48}_{-10.08}$& $0.743^{+0.508}_{-0.388}$ & $3.58^{+0.80}_{-0.47}$ & $3.44_{-1.37}^{+2.45}$\\ 
OGLE-2011-BLG-0462$^{56}$ & -- & -- & --  &--& 7.15(83) & $7.15_{-0.83}^{+0.83}$ \\  \hline
\end{tabular}}
\begin{tablenotes}[flushleft]
\small
\item{\footnotesize $^*$ Quantities informed within $90\%$ credible intervals.}
\item {\footnotesize References: 
$^1$ \citet{orosz2003}. 
$^2$ \citet{greiner2001}.
$^3$ \citet{galaxiesBHs2023deSa}.
$^4$ \citet{casares2009}.
$^5$ \citet{Farr2011}.
$^6$ \citet{wu2015}.
$^7$ \citet{wu2016}.
$^8$ \citet{gonzalez2013}.
$^9$ \citet{ioannou2004}.
$^{10}$ \citet{torres2021}.
$^{11}$ \citet{matasanchez2021}.
$^{12}$ \citet{casares1994}.
$^{13}$ \citet{khargharia2010}.
$^{14}$ \citet{orosz2004}.
$^{15}$ \citet{cherepashchuk2024band}.
$^{16}$ \citet{filippenko1995mass}. 
$^{17}$\citet{petrov2017masses}.
$^{18}$ \citet{remillard1996}.
$^{19}$ \citet{harlaftis1997}.
$^{20}$ \citet{petretti2023}.
$^{21}$ \citet{yanesrizo2022}.
$^{22}$ \citet{matasanchez2022}.
$^{23}$ \citet{torres2019}.
$^{24}$ \citet{torres2020}.
$^{25}$ \citet{orosz2011a}.
$^{26}$ \citet{heida2017}.
$^{27}$ \citet{filippenko1999}.
$^{28}$ \citet{orosz2001}.
$^{29}$ \citet{macdonald2014}.
$^{30}$ \citet{yanes2024evidence}.
$^{31}$ \citet{sanchez2024}.
$^{32}$ \citet{casares2023orbital}.
$^{33}$ \citet{soria20222}.
$^{34}$ \citet{jana2021}.
$^{35}$ \citet{shang2019}.
$^{36}$ \citet{molla2017}.
$^{37}$ \citet{tsurnov2018}.
$^{38}$ \citet{brocksopp1999}.
$^{39}$ \citet{orosz2011b}.
$^{40}$ \citet{gies2003}.
$^{41}$ \citet{orosz2009}.
$^{42}$ \citet{orosz2014}.
$^{43}$ \citet{ramachandran2022}.
$^{44}$ \citet{binder2021}.
$^{45}$ \citet{crowther2010}.
$^{46}$ \citet{silverman2008}.
$^{47}$ \citet{prestwich2007orbital}.
$^{48}$ \citet{t2019}.
$^{49}$ \citet{m2022}.
$^{50}$ \citet{peri2011}
$^{51}$ \citet{shenar2022}.
$^{52}$ \citet{ElBadry2022}.
$^{53}$ \citet{ElBadry2023}.
$^{54}$ \citet{panuzzo2024discovery}.
$^{55}$ \citet{wang2024}.
$^{56}$ \citet{sahu2025}.
}
\end{tablenotes}
\end{threeparttable}
\end{table*}

\section{Statistical Analysis}\label{sec:analysis}
As in \citet{Ozel2010} and \citet{Farr2011}, we performed a Bayesian analysis of the mass distribution BHs over different parametric distributions to infer the underlying population parameters and identify the model that best describes the updated catalog listed in Table \ref{tab:list_BHs}. We consider two analysis cases, one named Combined sample, with all 42 BHs included, and the other where only the 28 LMXBs are considered. 

This section presents the statistical framework used in the analysis, while results of the Combined and LMXB sample's are discussed in Section \ref{sec:results}.

\subsection{Bayesian Inference}\label{sec:bayes}
Based on Bayes' theorem, Bayesian methods provide a probabilistic framework to estimate the posterior probability distribution of a set of parameters $\boldsymbol{\theta}$, according to:
\begin{equation}
    P(\boldsymbol{\theta} | d) = \frac{P(d | \boldsymbol{\theta}) ~ P(\boldsymbol{\theta})}{P(d)},
\end{equation}
where $P(\boldsymbol{\theta} | d)$ is called the \textit{posterior distribution}, $P(d | \boldsymbol{\theta})$ is the \textit{likelihood} describing the observational data, $P(\boldsymbol{\theta})$ is the \textit{prior} representing previous knowledge (or lack) about each parameter, and $P(d)$ is the \textit{evidence}, which acts as a normalization constant ensuring that $\int P(\boldsymbol{\theta} | d)\,d\boldsymbol{\theta} = 1$, and computed as:
\begin{equation}
    P(d) = \int P(d | \boldsymbol{\theta})\,P(\boldsymbol{\theta})\,d\boldsymbol{\theta}.
\end{equation}

Following a hierarchical approach, our likelihood is constructed by combining the individual mass distributions $P_i(d | M_{\rm BH})$, described in Section~\ref{sec:individual_like}, with a population model $P(M_{\rm BH} | \boldsymbol{\theta})$, which represents the mass distribution of the entire population of BHs. The hierarchical likelihood is given by:

\begin{equation}
    P(d | \boldsymbol{\theta}) = \prod_j \int P_j(d | M_{\rm BH})\,P(M_{\rm BH} | \boldsymbol{\theta})\,dM_{\rm BH}.
\end{equation} 

Each individual BH mass of the sample is treated as a latent variable and sampled following the treatment of \citet{Ozel2010}, as described in Section \ref{sec:individual_like}. For the population-level distribution, however, we follow the approach of \citet{Farr2011} and adopt the same parametrizations used in their work, discussed in Sections \ref{sec:parametrization_gaussian} - \ref{sec:parametrization_expo}, to evaluate whether the updated sample maintains the same statistical characteristics or reveals significant differences—particularly in relation to the LMG, and in face of the addition of non-X-ray-binary BHs. We furthermore account for the results from \citet{Farr2011}, in addition to the current status of the field, to set the priors on each model's parameter.

\subsubsection{Individual-level likelihoods}\label{sec:individual_like}
Following the methodology of \citet{Ozel2010}, we divide the sample into four categories (instead of three), depending on how well-constrained are the orbital parameters for each system, as we describe here.
\paragraph{Case 1}
In the cases where $i$ and $q$ are tightly constrained, the individual likelihood is described with a Gaussian
\begin{align}
    \nonumber
        P^1_j(d|M_{BH}) &= \frac{1}{\sqrt{2\pi}~\sigma_{M_{BH},j}} \\
        &\times \exp \left[ \frac{- (M-M_{BH_j,mean})^2}{2\sigma_{M_{BH},j}^2}\right],
\end{align}
where $M_{BH_j,mean}$ is the mean and $\sigma_{M_{BH},j}$ informed in the literature.
\paragraph{Case 2}
In the second case, we have only lower and upper constraints for both $i$ and $q$, which we assume to be independent and to have uniform and isotropic distributions. We treat $i$ in terms of an uniform distribution over $\cos i$. The final likelihood is thus given by
\begin{align}
\nonumber
        P_j(d|M)&=C_j \int_{q_{min}}^{q_{max}} dq \int_{(\cos i)_{min}}^{(\cos i)_{max}} d(\cos i)  \\
        &\times \exp \left[- \frac{\left[f_j - \frac{M \sin^3 i}{(1 + q)^2}\right]^2}{2 \sigma_{f,j}^2} \right].
\end{align}
The minimum and maximum values of $i$ and $q$ are the boundaries of $\mathcal{U}(a,b)$ and $\mathcal{I}(a,b)$ distributions given in the fourth and fifth columns of Table \ref{tab:list_BHs}.
\paragraph{Case 3}
The third case correspond to systems where the mass function and the orbital inclination are measured, but the mass ration is only constrained, and the individual likelihood is given by the equation below
\begin{align}
\nonumber
        P_j(d|M)&=C_j \int_{q_{min}}^{q_{max}} dq \int_{(\cos i)_{min}}^{(\cos i)_{max}} d(\cos i)  \\
        &\times \exp \left[- \frac{\left[f_j - \frac{M \sin^3 i}{(1 + q)^2}\right]^2}{2 \sigma_{f,j}^2} - \frac{(i - i_{j,mean})}{2\sigma_{i,j}^2} \right].
\end{align}

\paragraph{Case 4}
Finally, we have in the current sample a fourth case that was not covered by \citet{Ozel2010}, in which the mass function and the mass ratio are measured, but the inclination is only constrained. In this case, the distribution is
\begin{align}
\nonumber
        P_j(d|M)&=C_j \int_{q_{min}}^{q_{max}} dq \int_{(\cos i)_{min}}^{(\cos i)_{max}} d(\cos i)  \\
        &\times \exp \left[- \frac{\left[f_j - \frac{M \sin^3 i}{(1 + q)^2}\right]^2}{2 \sigma_{f,j}^2} - \frac{(q - q_{j,mean})}{2\sigma_{q,j}^2} \right].
\end{align}

The following subsections describe the population-level distributions considered. Unless stated otherwise, the same priors are used for both the Combined and LMXB samples.
\subsubsection{Gaussian}\label{sec:parametrization_gaussian}
As a first simple model, motivated by an apparent accumulation of masses near $\sim 10~M_\odot$, we adopt a single Gaussian distribution:
\begin{align}\label{eq:gaussian}
\nonumber
    P(M_{\mathrm{BH}} | \boldsymbol{\theta}) &= P(M_{\mathrm{BH}} | \mu, \sigma)  \\ &= \frac{1}{\sigma \sqrt{2 \pi}} \exp \left[ - \frac{(M_{\mathrm{BH}} - \mu)^2}{2\sigma^2} \right],
\end{align}
where the parameters to be inferred are the mean $\mu$ and standard deviation $\sigma$. We limit $\mu$ and $\sigma$ to be positive, and assume the priors $P(\mu) = {\cal N}(20, 10)$ and $P(\sigma) = {\cal N}(5, 2)$.

\subsubsection{Two-Gaussian}\label{sec:parametrization_twogaussian}
Given the inclusion of more massive objects in the current sample, as for example a BH with $\sim 32~M_\odot$, we test a bimodal model to investigate whether the presence of a secondary subpopulation of BHs is favored. The probability distribution is a weighted sum of two Gaussian components:
\begin{align}
    P(M_{\mathrm{BH}} | \boldsymbol{\theta}) &= P(M_{\mathrm{BH}} | r, \mu_1, \mu_2, \sigma_1, \sigma_2) \nonumber \\
    \nonumber
    &= \frac{r}{\sigma_1 \sqrt{2\pi}} \exp \left[ - \left( \frac{M_{\mathrm{BH}} - \mu_1}{\sqrt{2}\sigma_1} \right)^2 \right] \\
    &+ \frac{1 - r}{\sigma_2 \sqrt{2\pi}} \exp \left[ - \left( \frac{M_{\mathrm{BH}} - \mu_2}{\sqrt{2}\sigma_2} \right)^2 \right].
\end{align}
The parameters to be fitted include the mixing fraction $r$ of each component, the means $\mu_1$ and $\mu_2$, and standard deviations $\sigma_1$ and $\sigma_2$. We impose the constraint of $0< \mu_1 < \mu_2$, $\sigma_1, \sigma_2 > 0$ and $0 < r < 1$. For the Combined sample we assume the following priors: $P(r) = {\cal U}(0,1)$, $P(\mu_1) = {\cal N}(10, 5)$, $P(\mu_2) = {\cal N}(20,8)$, $P(\sigma_1) = {\cal N}(5,2)$ and $\sigma_2 = {\cal N}(10, 5)$. In the LMXB case we had to change some priors to achieve better convergence in the chains, and they are $P(r) = {\cal U}(0,1)$, $P(\mu_1) = {\cal N}(7, 2)$, $P(\mu_2) = {\cal N}(15,8)$, $P(\sigma_1) = {\cal N}(5,2)$ and $\sigma_2 = {\cal N}(5, 5)$.

\subsubsection{Log Normal}\label{sec:parametrization_lognormal}
Given the observed clustering of masses near $10~M_\odot$ and the presence of a long tail extending up to $30~M_\odot$, we also consider a log-normal distribution, which naturally incorporates asymmetry, with a rapid rise up to the peak and a slower falloff in a long tail. This parametrization is given by:
\begin{align}
    \nonumber
    P(M_{\mathrm{BH}} | \boldsymbol{\theta}) &= P(M_{\mathrm{BH}} | \mu, \sigma) \\
    &= \frac{1}{M_{\mathrm{BH}} \sigma \sqrt{2 \pi}} \exp \left[ - \frac{(\log M_{\mathrm{BH}} - \mu)^2}{2\sigma^2} \right].
\end{align}

The parameters $\mu$ and $\sigma$ correspond to the mean and standard deviation of the distribution in $\log M_{\mathrm{BH}}$. The mean and standard deviation of $M_{\mathrm{BH}}$ in linear space are

\begin{align}
    \langle M_{\mathrm{BH}} \rangle &= \exp \left( \mu + \frac{\sigma^2}{2} \right), \label{eq:lognormal_m} \\
    \sigma_{M_{\mathrm{BH}}} &= \langle M_{\mathrm{BH}} \rangle \sqrt{ \exp(\sigma^2) - 1 }. \label{eq:lognormal_s}
\end{align}

The priors are taken over $\mu$ and $\sigma$ and they are assumed to be $P(\mu) = {\cal N}(5,2)$ and $P(\sigma) ={\cal C}auchy(0,1)$.

\subsubsection{Power-law}\label{sec:parametrization_powerlaw}

Power-law distributions often arise in astrophysical processes and can naturally describe the abundance of lower-mass BHs versus the rarity of more massive ones. In this case the distribution is given by:
\begin{align}
    \nonumber
    &P(M_{\mathrm{BH}} | \boldsymbol{\theta}) = P(M_{\mathrm{BH}} | m_{\min}, m_{\max}, \alpha) \\ &= \frac{1 + \alpha}{m_{\max}^{1+\alpha} - m_{\min}^{1+\alpha}} M_{\mathrm{BH}}^{\alpha}, \quad m_{\min} \leq M_{\mathrm{BH}} \leq m_{\max},
\end{align}
and zero otherwise, where $\alpha$ is the power-law index, $m_{\min}$ the minimum mass, and the $m_{\max}$ maximum mass. 

We assumed the priors $P(m_{\min}) = {\cal U}(0.001, 10)$, $P(m_{\max}) = {\cal N}(15, 8)$ with the constraint $m_{\max} > m_{\min}$, and $P(\alpha) = {\cal U}(-10, 10)$ for all samples.

\subsubsection{Decaying Exponential}\label{sec:parametrization_expo}
Finally, based on the work of \citet{Fryer&Kalogera2001}, on the relation between progenitor and remnant masses, \citet{Ozel2010} and \citet{Farr2011} considered a decaying exponential distribution with a minimum BH mass, expected to be consistent with the maximum NS mass. This probability distribution is given by: 
\begin{align}\label{eq:decayingexponential}
    \nonumber
    &P(M_{\mathrm{BH}} | \boldsymbol{\theta}) = P(M_{\mathrm{BH}} | m_{\min}, m_0) \\
    &= \frac{e^{m_{\min}/m_0}}{m_0} \times
    \begin{cases}
        \exp(-M_{\mathrm{BH}}/m_0), & M_{\mathrm{BH}} > m_{\min}, \\
        0, & M_{\mathrm{BH}} < m_{\min}.
    \end{cases}
\end{align}
The parameters sampled in this case are the lower mass limit $m_{\min}$ and a characteristic mass scale $m_0$, for which we assumed the following priors $P(m_{min}) = {\cal U}(0.001, 10)$ and $P(m_0) = {\cal U}(0.1, 30)$.

\section{Results} \label{sec:results}
The posterior distributions were sampled using Hamiltonian Monte Carlo with the No-U-Turn Sampler (HMC/NUTS), implemented via \texttt{PyStan} \citep{pystan}, a Python interface for the Stan probabilistic programming language. We ran 4 independent chains with $5\times10^3$ warm-up and $5\times10^3$ post-warm-up iterations each, for a total of $2\times10^4$ post-warm-up draws. Convergence was assessed through the potential scale reduction factor $\hat{R}$, bulk and tail effective sample sizes (ESS), and the number of divergent transitions. Results were analyzed through the \texttt{ArviZ} package \citep{Arviz}.

We treat the sample in the following ways. The \textit{Combined Sample} includes all 42 BHs found in our research. We also explore the effects on the posterior distributions of excluding the BHs for which the dynamical parameters are either unavailable or poorly constrained, in the called \textit{Refined Sample}. The \textit{LMXB sample} (Section \ref{subsec:lowmass_sample}) comprises only the 28 Galactic low-mass X-ray binaries. In all cases we fit the same five population models described in Section \ref{sec:bayes}. Although not explicitly stated, all parameters in the following subsections referring to masses and their uncertainties are given in units of solar mass ($M_\odot$). Model selection results are discussed in Section \ref{subsec:model_comparison}. The analysis of the lower mass gap in face of the preferred models is presented in Section \ref{sec:lower_mass_gap}.

\subsection{Combined sample}
\label{subsec:comb_sample}

Table \ref{tab:fullsample_results} gives the summary of the marginalized posterior distributions of each parameter derived from our Bayesian analysis for the Combined sample, providing the $5\%, ~15\%,~50\%$ (median), $85\%$ and $95\%$ quantiles. We discuss the results of each model in the following subsections and compare them with previous analyses from \citet{Farr2011}. The plots for all marginalized distributions are presented in Appendix \ref{apx:results_marginalization}.

\begin{table}[ht!]
    \centering
    \caption{Summary of marginalized posterior distribution of all parameters of each model discussed in Section \ref{sec:bayes}, for the Combined (full) sample of BHs listed at Table \ref{tab:list_BHs}. For each parameter we show the $5\%$, $15\%$, $50\%$, $85\%$ and $95\%$ quantiles.}
    \label{tab:fullsample_results}
    \setlength\tabcolsep{3pt}
    \begin{tabular}{ c c c c c c c }
        \hline
        Model & Parameter & $5\%$ & $15\%$ & $\boldsymbol{50\%}$ & $85\%$ & $95\%$ \\
        \hline \hline
        \multirow{2}{4em}{Gaussian}& $\mu$ & $7.99$ & $8.52$ & $\boldsymbol{9.41}$ & $10.31$ & $10.85$ \\
         & $\sigma$ & $4.34$ & $4.65$ & $\boldsymbol{5.26}$ & $5.98$ & $6.48$ \\
        \hline
        \multirow{5}{4em}{Two Gaussian} & $r$ & $0.69$ & $0.75$ & $\boldsymbol{0.84}$ & $0.91$ & $0.94$ \\
                     & $\mu_1$ & $7.01$ & $7.31$ & $\boldsymbol{7.83}$ & $8.42$ & $8.78$ \\
                     & $\sigma_1$ & $1.48$ & $1.73$ & $\boldsymbol{2.26}$ & $2.88$ & $3.26$ \\
                     & $\mu_2$ & $12.50$ & $15.04$ & $\boldsymbol{20.25}$ & $26.62$ & $30.55$\\
                     & $\sigma_2$ & $5.60$ & $7.34$ & $\boldsymbol{10.21}$ & $13.88$ & $16.37$ \\
        \hline
        \multirow{2}{4em}{Log Normal} & $\langle  M_{BH} \rangle$ & $8.08$ & $8.49$ & $\boldsymbol{9.25}$ & $10.14$ & $10.79$\\
                   & $\sigma_{M_{BH}}$ & $3.39$ & $3.78$ & $\boldsymbol{4.57}$ & $5.71$ & $6.60$ \\
        \hline
        \multirow{3}{4em}{Power Law} & $\alpha$ & $-3.29$ & $-2.97$ & $\boldsymbol{-2.47}$ & $-2.03$ & $-1.79$ \\
                 & $m_{min}$ & $4.03$ & $4.27$ & $\boldsymbol{4.68}$ & $5.05$ & $5.26$ \\
                 & $m_{max}$ & $31.99$ & $32.68$ & $\boldsymbol{34.25}$ & $37.10$ & $39.50$ \\
        \hline
        \multirow{2}{4em}{Decaying Exponential} & $m_{0}$ & $3.86$ & $4.32$ & $\boldsymbol{5.24}$ & $6.41$ & $7.22$ \\
                          & $m_{min}$ & $3.38$ & $3.71$ & $\boldsymbol{4.22}$ & $4.65$ & $4.87$ \\
        \hline
    \end{tabular}
\end{table}

\subsubsection{Gaussian}
The parameters sampled in this model are $\mu$ and $\sigma$, whose $90\%$ credible intervals (C.I.) are $7.99 \leq \mu \leq 10.85$ and $4.34 \leq \sigma \leq 6.48$, respectively. The marginalized distributions can be visualized in Figure \ref{fig:Gaussian}. In comparison with \citet{Farr2011}, the distribution of $\mu$ covers basically the same intervals, while the $\sigma$ distribution is displaced to higher values compared to the previous work ($2.24 \leq \sigma \leq 4.68$). This results in a posterior mass distribution much broader than before, given the inclusion of both light and massive BHs in the sample, such as the LMG candidates and Gaia BH3, with $M_{BH} \sim N(32.7, 0.8)$. 

\subsubsection{Two Gaussian}
The model of two Gaussian components is marked by the presence of two well-defined and separated peaks, with the first one being dominant and representing a fraction of $r \sim 0.84$ ($ 0.69 \leq r \leq 0.94$, $90\%$ C.I.) of the probability density function. The first Gaussian ($\mu_1, ~\sigma_1$) has a much narrower distribution than the second component ($\mu_2, ~\sigma_2$), as can be visualized in Figure \ref{fig:TwoGaussian}. The $90\%$ C.I. of the remaining parameters are $7.01 \leq \mu_1 \leq 8.78$, $1.48 \leq \sigma_1 \leq 3.26$, $12.50 \leq \mu_2 \leq 30.55$ and $5.60 \leq \sigma_2 \leq 16.37$, also showing a good agreement with the results in \citet{Farr2011}, except for a much broader distribution for the second component, to accommodate masses in the range of Gaia BH3.

\subsubsection{Log Normal}
The marginalization of the Log Normal model is reported in terms of $\langle M_{BH}\rangle$ and $\sigma_{BH}$, given by Equations \ref{eq:lognormal_m} and \ref{eq:lognormal_s}. The $90\%$ probability of the expected mass ($8.08 \leq \langle M_{BH} \rangle \leq 10.79$) shows a slight difference from results in \citet{Farr2011}, being narrow in our analysis, compared to the previous one. This distribution is naturally asymmetric and can accommodate higher masses without the necessity of very large standard deviations as seen for the Gaussian model. The plots of the marginalized distributions are given in Figure \ref{fig:LogNormal}.

\subsubsection{Power-Law}
The sampled parameters of the power law model are $\alpha$, $m_{min}$ and $m_{max}$, whose $90\%$ C.I. are $-3.29 \leq \alpha \leq -1.79$, $4.03 \leq m_{min} \leq 5.26$ and $31.99 \leq m_{max} \leq 39.50$. In comparison to \citet{Farr2011} the $\alpha$ parameter has a much narrower distribution, with a median at $-2.47$ instead of $-3.23$ , a minimum mass restricted to masses below $35.26~M_\odot$ with $90\%$ confidence (instead of $6.46$ in their work), and a maximum mass above $32~M_\odot$ (instead of $19.11$). These changes on the behavior of the posteriors are also consequences of the inclusion of mass-gap and high mass BHs, together with more accurate mass estimates, revealing a smoother decay up to high mass values. The marginalization we obtained can be visualized in Fig. \ref{fig:PowerLaw}.

\subsubsection{Decaying Exponential}
Finally, in the decaying exponential model, while the cutoff mass, $m_{min}$, has now a broader distribution, the scale mass, $m_{0}$, distributions is narrow when compared with results from \citet{Farr2011}. The median cutoff mass now $4.22$, smaller than the $5.33$ from the previous work, although $m_{min}$ can reach values $\sim 3.86~M_\odot$ with $90\%$ confidence. We find the $90\%$ C.I of the parameters to be $3.86 \leq m_{min} \leq 4.87$ and $3.86 \leq m_0 \leq 7.22$. The marginalized distributions are shown in Fig. \ref{fig:ExpDecay}.

\subsubsection{Refined sample}\label{subsec:refined_sample}
We also define a \textit{Refined} sample that excludes systems with unavailable or poorly constrained dynamical parameters, in order to assess the robustness of the inferred population-level results. In this scenario we deleted MAXI J0637-430, MAXI J1813-095, MAXI J1535-571, XTE J1746-322, GRS 1009-45 and Swift J1727.8-1613 from our sample and explored the impact this change provides in our results in comparison with the Combined sample, finding it to be significantly small in terms of the marginalized posterior distributions of each parameter, as shown in Table \ref{tab:refined_results}. The biggest impact comes in the model comparison, as we discuss in Section \ref{subsec:model_comparison}, solving the divergence found for the Combined case.

\begin{table}[ht!]
    \centering
    \caption{Summary of marginalized posterior distribution of all parameters of each model discussed in Section \ref{sec:bayes}, when considering the Refined sample. For each parameter we show the $5\%$, $15\%$, $50\%$, $85\%$ and $95\%$ quantiles.}
    \label{tab:refined_results}
    \setlength\tabcolsep{3pt}
    \begin{tabular}{ c c c c c c c }
        \hline
        Model & Parameter & $5\%$ & $15\%$ & $\boldsymbol{50\%}$ & $85\%$ & $95\%$ \\
        \hline \hline
        \multirow{2}{4em}{Gaussian}& $\mu$ & $8.02$ & $8.61$ & $\boldsymbol{9.61}$ & $10.65$ & $11.27$ \\
         & $\sigma$ & $4.50$ & $4.84$ & $\boldsymbol{5.52}$ & $6.33$ & $6.89$ \\
        \hline
        \multirow{5}{4em}{Two Gaussian} & $r$ & $0.65$ & $0.72$ & $\boldsymbol{0.82}$ & $0.90$ & $0.93$ \\
                     & $\mu_1$ & $7.01$ & $7.34$ & $\boldsymbol{7.87}$ & $8.48$ & $8.87$ \\
                     & $\sigma_1$ & $1.3$ & $1.57$ & $\boldsymbol{2.18}$ & $2.90$ & $3.36$ \\
                     & $\mu_2$ & $11.82$ & $14.60$ & $\boldsymbol{19.84}$ & $26.01$ & $30.09$ \\
                     & $\sigma_2$ & $5.71$ & $7.40$ & $\boldsymbol{10.33}$ & $13.89$ & $16.30$ \\
        \hline
        \multirow{2}{4em}{Log Normal} & $\langle  M_{BH} \rangle$ & $8.19$ & $8.66$ & $\boldsymbol{9.53}$ & $10.60$ & $11.36$ \\
                   & $\sigma_{M_{BH}}$ & $3.56$ & $3.98$ & $\boldsymbol{4.95}$ & $6.37$ & $7.56$ \\
        \hline
        \multirow{3}{4em}{Power Law} & $\alpha$ & $-3.32$ & $-2.98$ & $\boldsymbol{-2.49}$ & $-2.05$ & $-1.80$ \\
                 & $m_{min}$ & $3.99$ &  $4.25$ & $\boldsymbol{4.65}$ & $5.02$ & $5.23$ \\
                 & $m_{max}$ & $31.98$ & $32.67$ & $\boldsymbol{34.22}$ & $36.94$ & $39.25$ \\
        \hline
        \multirow{2}{4em}{Decaying Exponential} & $m_{0}$ & $3.70$ & $4.20$ & $\boldsymbol{5.24}$ & $6.55$ & $7.48$ \\
                          & $m_{min}$ & $3.41$ & $3.83$ & $\boldsymbol{4.50}$ & $5.09$ & $5.38$ \\
        \hline
    \end{tabular}
\end{table}
 
\subsection{LMXB sample}\label{subsec:lowmass_sample}
In the following, we present and discuss the results of our analysis to the sample of LMXBs only. Table \ref{tab:lmxb_results} shows the $5\%, ~15\%,~50\%, ~85\%$ and $95\%$ quantiles of each parameter. In the following subsections, we discuss the results for each model.

\begin{table}[ht!]
    \centering
    \caption{Summary of marginalized posterior distribution of all parameters of each model discussed in Section \ref{sec:bayes}, when considering a sample only with the LMXBs. For each parameter we show the $5\%$, $15\%$, $50\%$, $85\%$ and $95\%$ quantiles.}
    \label{tab:lmxb_results}
    \setlength\tabcolsep{3pt}
    \begin{tabular}{ c c c c c c c }
        \hline
        Model & Parameter & $5\%$ & $15\%$ & $\boldsymbol{50\%}$ & $85\%$ & $95\%$ \\
        \hline \hline
        \multirow{2}{4em}{Gaussian}& $\mu$ & 7.09 & 7.33 & $\boldsymbol{7.74}$ & 8.17 & 8.45 \\
         & $\sigma$ & 1.07 & 1.23 & $\boldsymbol{1.56}$ & 1.96 & 2.26 \\
        \hline
        \multirow{5}{4em}{Two Gaussian} & $r$ & $0.07$ & $0.14$ & $\boldsymbol{0.34}$ & $0.64$ & $0.82$ \\
                     & $\mu_1$ & $4.41$ & $5.10$ & $\boldsymbol{6.18}$ & $6.98$ & $7.39$ \\
                     & $\sigma_1$ & $0.61$ & $0.98$ & $\boldsymbol{1.57}$ & $2.40$ & $2.99$ \\
                     & $\mu_2$ & $7.31$ & $7.62$ & $\boldsymbol{8.20}$ & $9.01$ & $9.74$ \\
                     & $\sigma_2$ & $0.99$ & $1.72$ & $\boldsymbol{3.22}$ & $7.38$ & $10.64$ \\
        \hline
        \multirow{2}{4em}{Log Normal} & $\langle  M_{BH} \rangle$ & $7.07$ & $7.32$ & $\boldsymbol{7.75}$ & $8.24$ & $8.56$ \\
                   & $\sigma_{M_{BH}}$ & $1.15$ & $1.33$ & $\boldsymbol{1.70}$ & $2.20$ & $2.58$ \\
        \hline
        \multirow{3}{4em}{Power Law} & $\alpha$ & $0.55$ & $1.58$ & $\boldsymbol{3.37}$ & $5.43$ & $6.89$ \\
                 & $m_{min}$ & $0.83$ &  $1.39$ & $\boldsymbol{3.24}$ & $4.41$ & $4.83$ \\
                 & $m_{max}$ & $8.61$ & $8.89$ & $\boldsymbol{9.39}$ & $10.01$ & $10.49$ \\
        \hline
        \multirow{2}{4em}{Decaying Exponential} & $m_{0}$ & $1.68$ & $1.99$ & $\boldsymbol{2.69}$ & $3.59$ & $4.26$ \\
                          & $m_{min}$ & $4.32$ & $4.64$ & $\boldsymbol{5.12}$ & $5.56$ & $5.80
                          $ \\
        \hline
    \end{tabular}
\end{table}

\subsubsection{Gaussian}
When we restrict the analysis to LMXBs only, the most affected parameter in the Gaussian model is the standard deviation, which, as expected, is much more tightly constrained, $1.07 \leq \sigma \leq 2.26$. The mean distribution ($\mu$) is shifted to the left in comparison with the Combined sample, with $7.09 \leq \mu \leq 8.45$ at $90\%$ confidence, since we removed objects with higher masses from the sample. The marginalized distributions can be seen in Figure \ref{fig:Gaussian_LMXB}.

\subsubsection{Two Gaussian}
In comparison to the results in Sec.~\ref{subsec:comb_sample}, the peak of the second component is significantly displaced to lower mass values, with a median now at $8.20$ (instead of $20.25$ for the Combined sample). As we can see in the middle panel of Fig.~\ref{fig:TwoGaussian_LMXB}, the two Gaussian components overlap, so that the mixture effectively reproduces the single-Gaussian fit at the cost of three additional parameters. This behaviour is reflected in the model comparison of Sec.~\ref{subsec:model_comparison}, where the two-Gaussian is essentially tied with the single Gaussian under WAIC and LOO but is penalised by AIC and BIC.

\subsubsection{Log Normal}
In agreement with previous results, the distributions of $\langle M_{BH} \rangle$ and $\sigma_{BH}$ are also displaced to lower values when we consider only LMXBs, besides being very similar to the results for the Gaussian parametrization of LMXBs. The median mass is $7.75$ for this case and its marginalized distribution can be visualized in Figure \ref{fig:LogNormal_LMXB}.

\subsubsection{Power-Law}
The results of the power-law parametrization are also given in Table \ref{tab:lmxb_results}, and the marginalized distributions are shown in Figure \ref{fig:PowerLaw_LMXB}. However, in this work we conclude that this model cannot be trusted when fitted to the LMXB sample. The diagnostics reveal serious problems: $1\,383$ divergent transitions out of $20\,000$ post-warm-up draws ($\sim 7\%$), a recovered slope $\alpha = 3.50(1.92)$ — a \emph{positive} exponent in direct contradiction with the $\alpha \simeq -2.5$ recovered from the broader samples and with the analytical maximum-likelihood solution for the observed $5$--$10~M_\odot$ distribution. This behavior seems to be structural, rather than numerical. The power-law entry is therefore excluded from the LMXB model comparison in Section \ref{subsec:model_comparison}, as reporting its information-criterion values would be both statistically invalid (non-converged chains) and scientifically misleading.

\subsubsection{Decaying Exponential}
When we compare the mass scale $m_{0}$ for the exponential case at Table \ref{tab:lmxb_results} with $m_{0}$ in Table \ref{tab:fullsample_results}, we can clearly notice a shift towards smaller values, with a median $m_{0}$ at $2.69~M_\odot$. The cutoff mass, $m_{min}$, on the contrary, is displaced to the right, with a $90\%$ C.I. $4.32 \leq m_0 \leq 5.80$. The marginalized distributions of both parameters are shown in Fig. \ref{fig:ExpDecay_LMXB}.

\subsection{Model selection}\label{subsec:model_comparison}
Model selection methods are a key procedure to distinguish the most appropriate model among a set of candidates by assessing their relative strengths. They allow us to choose a model that strikes a balance between simplicity and flexibility, avoiding overfitting of the data, at the same time that it provides the best predictions for future observations. A variety of model selection methods exist, each of them based on different assumptions and evaluation criteria. Statisticians recommend applying multiple methods to enhance the rigor and reliability of the model selection, helping to mitigate the limitations of individual methods. If multiple methods consistently point to the same result, it increases the confidence in the selection.

We implement four complementary criteria, all reported on the deviance scale ($-2\log p$). The \textit{Widely Applicable Information Criterion} (WAIC) and the \textit{Leave-One-Out cross-validation} (LOO), both fully Bayesian, as well as the classical \textit{Akaike Information Criterion} (AIC) and \textit{Bayesian Information Criterion} (BIC). The last two methods are respectively described by:
\begin{equation}
    AIC = 2  k - 2 ~ln (\hat{L}),
\end{equation}
\begin{equation}
    BIC =  k ~ ln(N)- 2 ~ln (\hat{L}),
\end{equation}
where $k$ is the number of free parameters in the model, $N$ is the number of data points, and $\hat{L}$ is the maximum likelihood. Lower values of all four criterias indicates a preferred model. 

We assess the relative weight of each model using the Akaike-style normalized weights \citep{PORTET2020111}:
\begin{equation}
    w_i = \frac{\exp(-\tfrac{1}{2}\Delta_i)}{\sum_j\exp(-\tfrac{1}{2}\Delta_j)}, \qquad \Delta_i = \mathrm{IC}_i - \min_j \mathrm{IC}_j.
\end{equation}

Table \ref{tab:model_comparison} shows the results for the Combined and LMXB samples. As discussed in Section \ref{subsec:lowmass_sample}, the power law is excluded from the LMXB comparison on both convergence and scientific grounds.

\begin{table*}[ht!]
    \centering
    \caption{Information-criterion comparison for the five population models, for both Combined and LMXB samples. For each criterion we report its value and the corresponding normalized Akaike-style weight. The single power law was excluded from the LMXB comparison because it shows a bad behavior on that sample.}
    \label{tab:model_comparison}
    \setlength\tabcolsep{5pt}
    \begin{tabular}{| c | c l | c l | c l | c l |}
        \hline
        \multirow{2}{4em}{Model} & \multicolumn{2}{c|}{WAIC} & \multicolumn{2}{c|}{LOO} & \multicolumn{2}{c|}{AIC} & \multicolumn{2}{c|}{BIC} \\
                                 & WAIC & $w_{WAIC}$ & LOO & $w_{LOO}$ & AIC & $w_{AIC}$ & BIC & $w_{BIC}$ \\
        \hline \hline
        \multicolumn{9}{|c|}{Combined sample} \\
        \hline
        Gaussian & 384.49  & $4.5\times10^{-11}$ & 415.57  & $\ll 10^{-10}$ & 347.31 & $9.7\times10^{-10}$ & 350.79  & $2.2\times10^{-9}$ \\
        Two Gaussian & 350.05  & $1.5\times10^{-3}$  & 384.17  & $1.3\times10^{-4}$  & 319.03 & $1.3\times10^{-3}$  & 327.72 & $2.2\times10^{-4}$ \\
        Log Normal & 353.78  & $2.3\times10^{-4}$  & 384.87 & $9.1\times10^{-5}$ & 320.61  & $6.0\times10^{-4}$  & 324.09  & $1.4\times10^{-3}$ \\
        Power Law  & $\boldsymbol{337.06}$ & $\boldsymbol{0.962}$  & $\boldsymbol{366.26}$ & $\boldsymbol{0.998}$ & $\boldsymbol{305.81}$ & $\boldsymbol{0.974}$  & $\boldsymbol{311.02}$ & $\boldsymbol{0.942}$ \\
        Exp.\ Decay  & 343.59 & $0.037$ & 378.45 & $2.3\times10^{-3}$ & 313.17  & $0.025$ & 316.65 & $0.057$ \\
        \hline \hline
        \multicolumn{9}{|c|}{LMXB sample} \\
        \hline
        Gaussian & $\boldsymbol{185.71}$ & $\boldsymbol{0.518}$  & $\boldsymbol{208.07}$ & $\boldsymbol{0.475}$ & $\boldsymbol{164.37}$ & $\boldsymbol{0.843}$ & $\boldsymbol{167.03}$ & $\boldsymbol{0.861}$ \\
        Two Gaussian & 186.18 & $0.410$ & 208.35 & $0.413$ & 171.46 & $0.024$ & 178.12 & $3.4\times10^{-3}$ \\
        Log Normal & 189.65 & $0.072$ & 210.98 & $0.111$ & 168.11 & $0.130$ & 170.77 & $0.133$ \\
        Power Law  & -- & -- & --  & --  & -- & -- & --  & -- \\
        Exp.\ Decay & 201.36 & $2.1\times10^{-4}$  & 221.39 & $6.1\times10^{-4}$ & 175.90  & $2.6\times10^{-3}$ & 178.57   & $2.7\times10^{-3}$ \\
        \hline
    \end{tabular}
\end{table*}

For the Combined sample, all four information criteria unanimously indicate the power law as the best model, with $w \gtrsim 0.94$ under WAIC, AIC and BIC and $w \simeq 1$ under LOO. The truncated exponential decay is the only competitor, but with $\Delta\mathrm{WAIC}=6.5$,  $\Delta\mathrm{LOO}=12.2$, $\Delta\mathrm{AIC}=7.4$, $\Delta\mathrm{BIC}=5.6$, indicating a strong preference for the power law in most of the cases.  Together, the two models account for more than $99\%$ of the total WAIC weight in this sample, leaving the Gaussian, two-Gaussian and log-normal at negligible levels. The margin between both models widens further when the six least-constrained transients are removed in the Refined sample (Sec.~\ref{subsec:comb_sample}). For this case, the power law carries $w \gtrsim 0.99$ under every criterion, with $\Delta\mathrm{IC} \geq 9$ against the exponential decay and $\Delta\mathrm{IC} \geq 16$ against any other alternative (see Table \ref{tab:model_comparison_refined} in Appendix \ref{apx:refinedsample}). The hyper-parameters of the power-law fit are stable between the Combined and Refined runs, both indicating $\alpha\sim -2.5$ to $-2.8$, $m_{\min}\sim 4.7$--$5.1~M_\odot$ and $m_{\max}\sim 34$--$35~M_\odot$. Compared with \citet{Farr2011}, who found the decaying exponential to be the best model for their Combined sample, we now find the power law to be preferred, and attribute this to the wider dynamic range introduced by the presence of a new high-mass object, Gaia~BH3, which populate the high-mass tail that an exponential decay cannot accommodate.

In the LMXB case, the restricted mass range changes the picture entirely. The Gaussian stands as the best model, but the picture is more nuanced. Under the predictive criteria the Gaussian and two-Gaussian are essentially tied ($w_\mathrm{WAIC}=0.52$ vs $0.41$; $w_\mathrm{LOO}=0.48$ vs $0.41$, with $\Delta < 0.5$ in both cases). This means that the LMXB data do not have the resolution to discriminate between a single peak at $\sim 7.8~M_\odot$ and a two-component description in which one component overlaps with the other, making the models statistically indistinguishable. The log-normal comes in third place, and the truncated exponential decay is strongly disfavored ($\Delta\mathrm{WAIC} \simeq 16$), because an exponential tail requires a broad distribution that the LMXB data simply do not populate. However, both AIC and BIC break the WAIC/LOO near-tie in favor of the single Gaussian via the parameter-count penalty (k=2 vs k=5 for the two-Gaussian), driving it to first position. The single power law cannot be reliably fitted to this sample (Sec.~\ref{subsec:lowmass_sample}) and is therefore excluded from the comparison.

\begin{figure*}[htpb]
    \centering
    \begin{minipage}[t]{0.45\linewidth}
        \centering
        \includegraphics[width=\columnwidth]{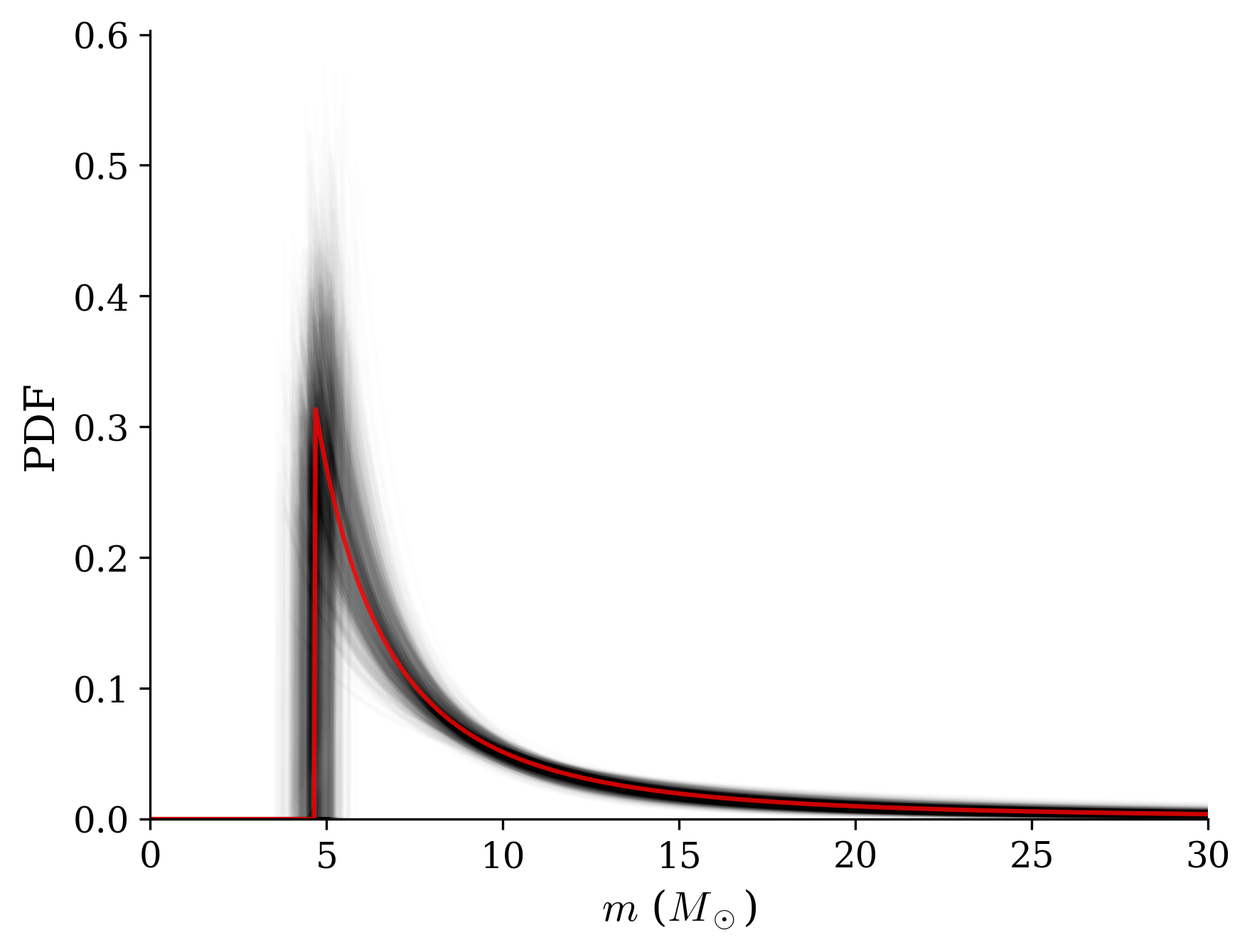}
    \end{minipage}
    ~
    \begin{minipage}[t]{0.45\linewidth}
        \centering
        \includegraphics[width=\columnwidth]{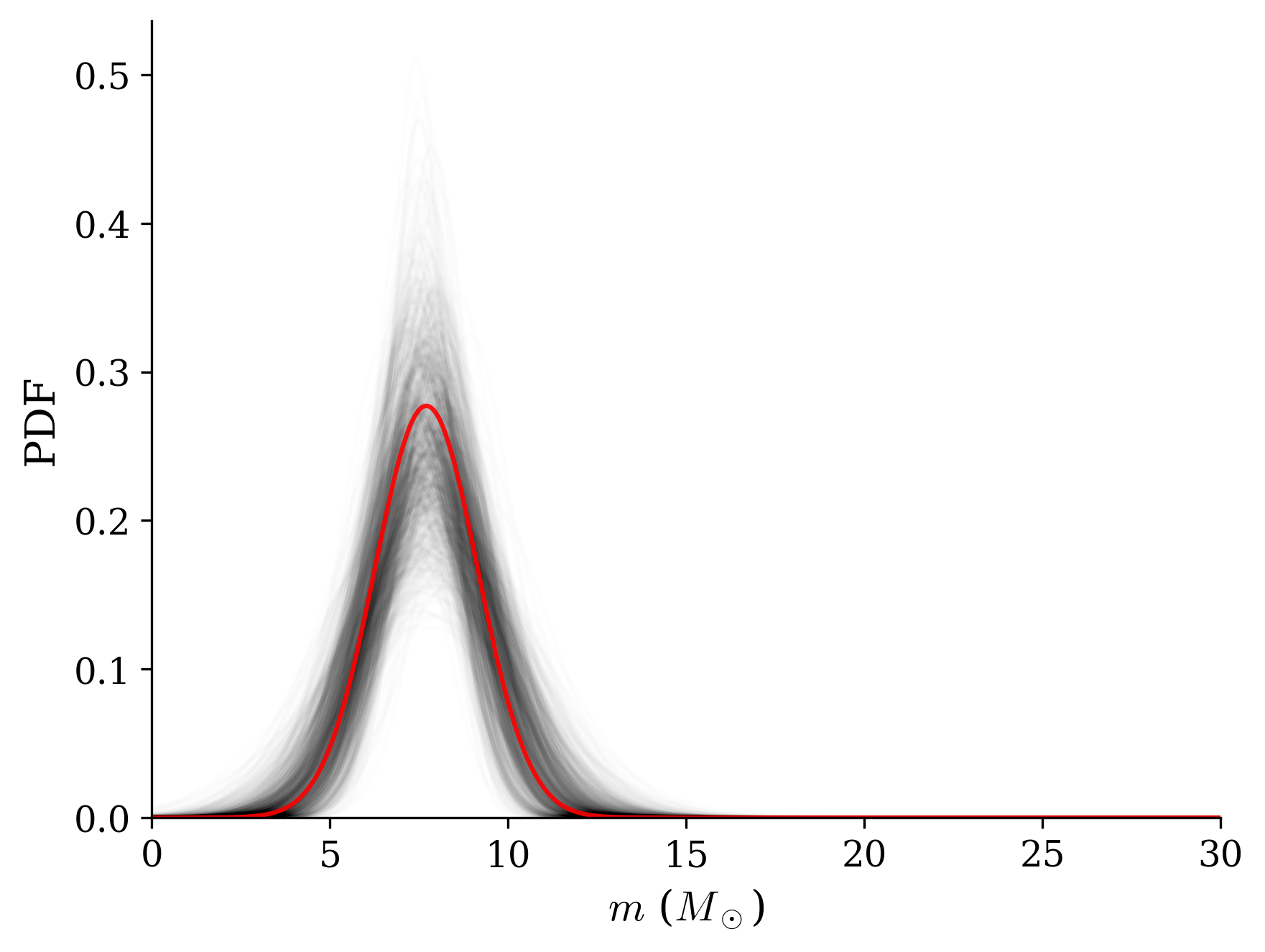}
    \end{minipage}
    \caption{1000 posterior samples (grey lines) drawn from the best-model posterior for the Combined sample (power law, left) and the LMXB-only sample (Gaussian, right). The red curve shows the MAP distribution in each case.}
    \label{fig:Posterior_distributions}
\end{figure*}

Given the best models for both cases, we proceed to obtain draws from the posterior distribution. Fig \ref{fig:Posterior_distributions} shows, on the left panel, 1000 draws (grey curves) from the power law posterior distribution obtained for the Combined sample, giving us a visual representation of the uncertainties of $\alpha$, $m_{min}$ and $m_{max}$. The solid red curve represents the maximum posterior probability (MAP), which corresponds to the mode of the posterior distribution. The right panel shows an equivalent representation for the Gaussian distribution fitted to the LMXB sample.

\section{The Lower Mass Gap}\label{sec:lower_mass_gap}
The first analysis of the mass distribution of galactic BHs was made by \citet{1998ApJ...499..367B} for a sample of 7 LMXBs, and highlighted a significant gap between the masses of these objects and the maximum mass limit of NSs, in the range $2 - 5 ~M_\odot$. The existence of the LMG was later reinforced by the works of \citet{Ozel2010} and \citet{Farr2011}, and potential explanations for its emergence were searched for in, e.g., the timescale of convection growth in supernovae \citep{FryerCompactRemnantMass2012,FryerEffectSupernovaConvection2022,OlejakRoleSupernovaConvection2022}. Since these works, the gap served as a guide to distinguish the nature of compact object based on their masses.

However, as the sensitivity was improved for a variety of instruments, systematics on observational measurements were reduced, and GW detectors started taking data, the number of candidate objects in the gap started to grow. That is, for example, the case of 2MASS J0521+435, with $3.30^{+1.40}_{-0.35}~\mathrm{M}_{\odot}$ \citep[although its BH nature has been questioned; see][]{vandenheuvelCommentNoninteractingLowmass2020,thompsonResponseCommentNoninteracting2020} and G3425, with $3.58^{+0.8}_{-0.47}\,\mathrm{M}_\odot$. 

Candidates have also been found in the extragalactic population through GWs, including the secondary in GW190814, with mass $2.50-2.67\,\mathrm{M}_\odot$ \citep[$90\%$,][]{gw190814}, and the primary in GW230519, with mass $2.5-4.5\,\mathrm{M}_\odot$ \citep[$90\%$][]{gw230529}, well in the middle of the LMG\footnote{Other GW events have components with marginal probability of falling in the LMG, or might have resulted in a LMG object, see the sample in \citet{2022ApJ...941..130D}.}. There are further candidates in the Galaxy, either heavy NSs or COs with an ambiguous nature, such as the $2.09-2.71~\mathrm{M}_\odot$ ($95\%$) companion to the MSP PSR J0514-4002E, and the black widow PSR J0952-0607, with $2.35 \pm 0.17 ~ \mathrm{M}_\odot$ \citep{RomaniFastestHeaviest2022}. These results put into question the absence of compact objects in the LMG, from which \citet{2022ApJ...941..130D} concluded that a depleted gap (with non-negligible density, sparsely populated) is certainly preferred instead of an absolute gap (zero density), and that although it might be an actual feature of some sub-classes, it is not an overall property of the compact object population.

We re-examine the existence of the LMG in the light of the updated hierarchical analysis by following a procedure analogous to that of \citet{Farr2011}: for each sample we generate 1000 draws from the posterior of the preferred model, compute the $1\%$ quantile of each draw, and build the resulting $M_{1\%}$ distribution. For the Combined sample we use the power-law posterior, which is the overwhelmingly preferred model (Section \ref{subsec:model_comparison}). For the LMXB sample, we use the Gaussian posterior, since the power law is excluded from that comparison on convergence grounds. The $M_{1\%}$ quantile is less sensitive to statistical fluctuations and individual objects poorly constrained masses, thus providing a more robust assessment of the lower tail of the inferred population.

\begin{figure*}[htpb]
    \centering
    \begin{minipage}[t]{0.45\linewidth}
        \centering
        \includegraphics[width=\columnwidth]{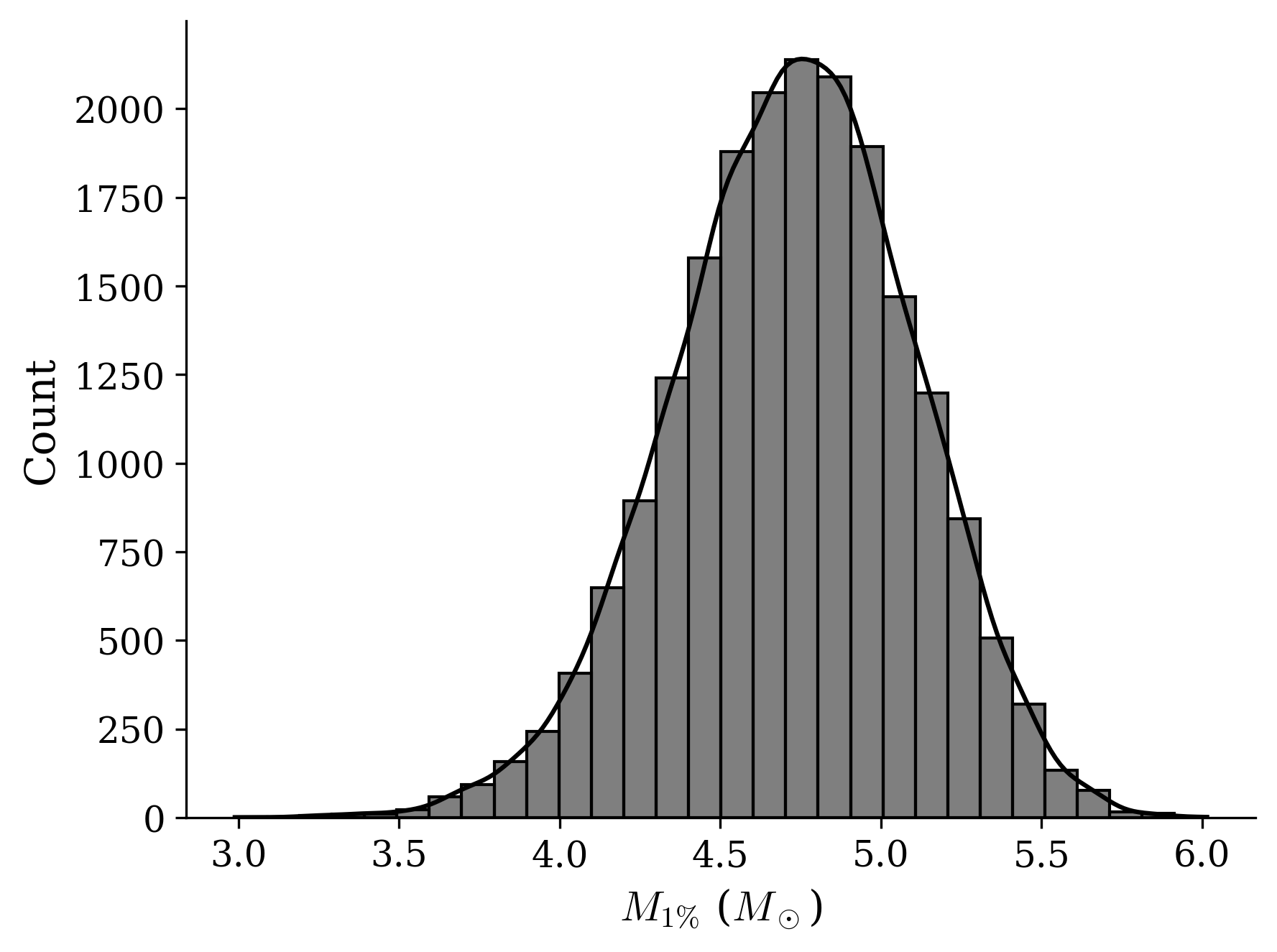}
    \end{minipage}
    ~
    \begin{minipage}[t]{0.45\linewidth}
        \centering
        \includegraphics[width=\columnwidth]{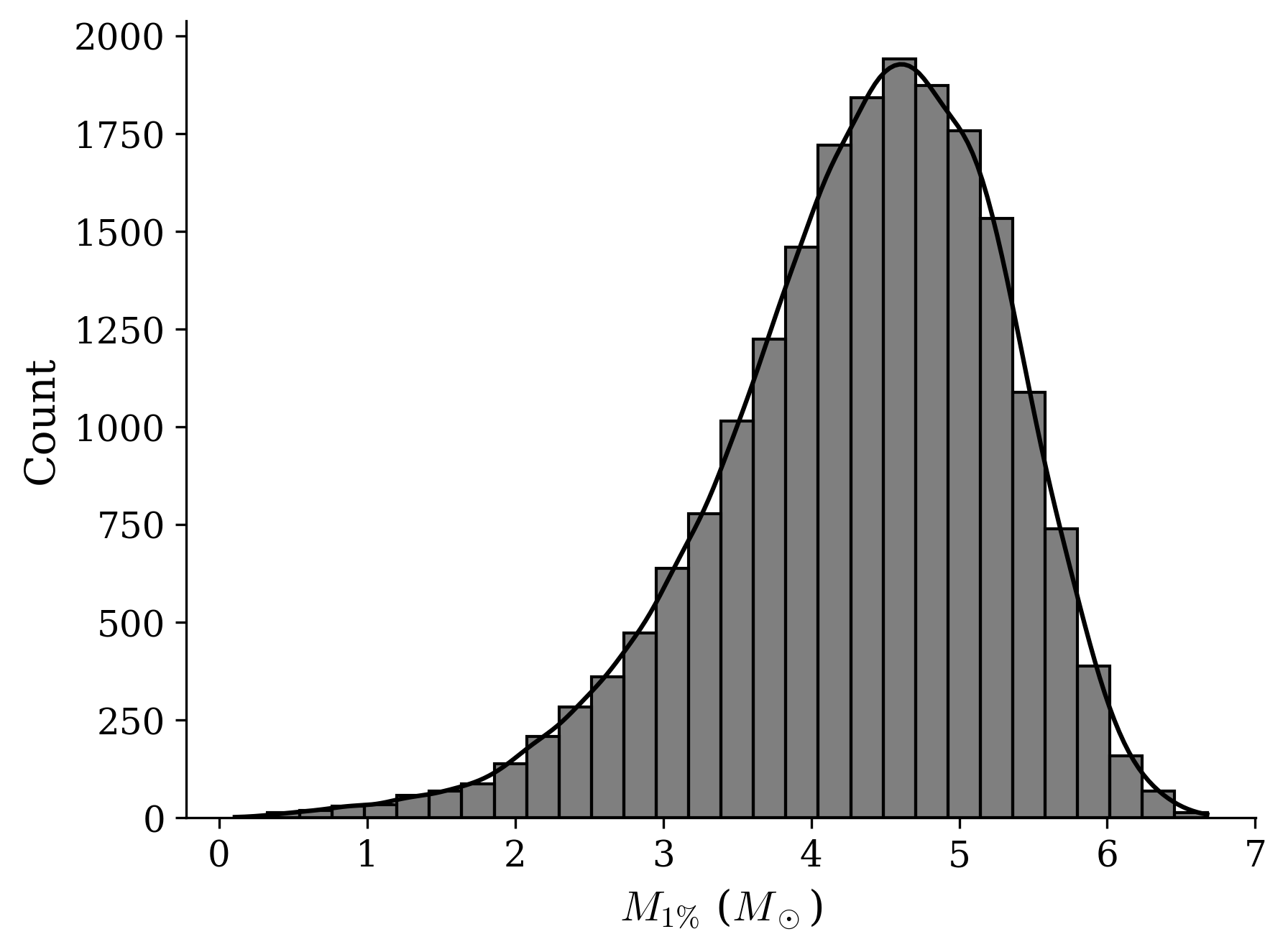}
    \end{minipage}
    \caption{Distribution of the $1\%$ quantile ($M_{1\%}$) built from 1000 posterior draws of the preferred model for each sample, shown in Figure \ref{fig:Posterior_distributions}. Left: power law posterior for the Combined sample. Right: Gaussian posterior for the LMXB sample.}
    \label{fig:Minimum_masses}
\end{figure*}

The $M_{1\%}$ distributions are shown in Figure \ref{fig:Minimum_masses}. For the Combined sample we find $4.06 \leq M_{1\%} \leq 5.29$ at $90\%$ C.I., with the mean value $M_{1\%} \simeq 4.7 \pm 0.37 ~M_\odot$. This represents a discrete \textit{downward} shift relative to \citet{Farr2011}, who found $M_{1\%} > 4.5\,\mathrm{M}_\odot$ at $90\%$ confidence for a much narrower sample, against the $M_{1\%}>5\,\mathrm{M}_\odot$ found by \citet{Ozel2010}. The hierarchical model places the lower edge of the BH posterior distribution around $4.7$$M_\odot$ even when the putative LMG candidates (G3425, 2MASS J0521+435) are included, because their large individual uncertainties prevent them from pulling the population cut-off below $\sim4~M_\odot$. However, when considering LMXB's only, the Gaussian posterior distribution provides $2.56 \leq M_{1\%} \leq 5.68$, in contrast with the $M_{1\%}>4.3\,\mathrm{M}_\odot$ ($90\%$) found by \citet{Ozel2010}. Together with the maximum NS mass constrained to $ 2.0 \leq M_{max} \leq 2.2\,\mathrm{M}_\odot$ ($1\sigma$) \citep{2018MNRAS.478.1377A}, and increasing support for a value as high as $2.56 \pm 0.37 \,\mathrm{M}_\odot$ \citep{2023Univ...10....3R}, our results on the minimum mass distribution of BH's further argue against the existence of an empty LMG.

Although the results between Combined and LMXB samples may appear inconsistent at first glance, the difference should be interpreted with caution. In first place, the $M_{1\%}$ distribution depends on the assumed parametric family. Since the Combined sample is better described by a power-law, it naturally accommodates a sharper cutoff in comparison to a left-tail of a Gaussian distribution, the one preferred for the LMXB sample. The numerical values reported here should be interpreted as model-dependent characterizations of the present data rather than model-independent determinations of a physical cutoff. In addition, the majority of systems with masses compatible with the LMG belong to the LMXB, whereas the systems incorporated in the Combined sample predominantly populate the higher-mass end of the distribution. Since we model the Combined sample using a single parametric population, the addition of more massive systems naturally shifts the inferred distribution toward larger characteristic masses.

In this context, the LMXB provides the strongest observational leverage what can be seen as an absolute minimum of the galactic BH mass distribution. The fact that its preferred model allows $M_{1\%}$ values below $\sim 3~M_\odot$ indicates that the current observations do not support assigning zero probability to BHs within the traditional lower mass gap ($<~5M_\odot$), and challenges the nature discrimination of systems within the old-LMG. An increasing number of dynamically confirmed systems, particularly in the gap interval, together with more flexible population models capable of accounting for multiple formation channels, will be required to precisely determine the structure of the lower end of the galactic BH mass distribution.

The possibility that the LMG emerges directly from CCSNe is further unlikely because of the increasing evidence that low-mass BHs are present in the GW sample. This had already been suggested by the $\sim2.6~\mathrm{M}_\odot$ object in GW190814 \citep{gw190814,ZevinExploringLowerMassGap2020}, and was reinforced by GW230529 \citep{gw230529}, with the inclusion of which \citet{RayHidingOutLowEnd2025} have recently found no support for a LMG in the GW BH mass distribution, with $M_{1\%}=3.13^{+0.18}_{-0.04}~\mathrm{M}_\odot$. Population synthesis studies have similarly found that GW observations favor a SN engine that at least partially fills the LMG \citep{SiegelInvestigatingLowerMassGap2023,XingMassGapBlacKholes2025}.

\section{Formation channels of BHs} \label{sec:formation_channels}

\subsection{BH formation at the turn of the century}

The above compilation of candidate BHs and further analysis attempting to reveal the underlying mass distribution naturally leads to the question of their formation, i.e., which are the mechanism(s) that Nature uses to form these extraordinary stellar corpses. It is well-known that decades of research have consolidated the idea that the formation of a BH needs that an amount of matter ``enters'' its Schwarszchild radius (forgetting for the moment the spin), and forms an event horizon. In practice, the simplest scenario that accomplishes this is the collapse of a massive star, in which gravity becomes extreme in the central regions and follows the evolution first explored by \citet{Oppenheimer1939}. The evolution of numerical simulations and availability of more observational data have now prompted a revision of this simple picture. The goal is to understand in which classes of systems conditions will be sufficient to form actual BHs, and also in which fashion does the formation proceed. In addition, alternative events ending with a BH have been suggested by both theory and observations: extragalactic BH-BH mergers observed by means of GWs \citep{Abbott2019} confirmed that BH masses may change because of these events, although the actual merger rate is very small ($\sim 1 \mathrm{Gyr}^{-1}$) and will not contribute to the bulk of the population in our galaxy. Also, hypothetical scenarios in which a NS is ``pushed'' over its maximum mass limit have been invoked. We shall briefly address these new visions of BH formation below.

Let us start by recalling the widespread scenario considered at the turn of the 20th century. It was believed at the time that, according to the available theory and numerical results, massive stars, born with $M \geq 8 \, M_{\odot}$, ended their lives first by developing a C-O-Ne core which can reach a definite ``Chandrasekhar mass'' and eventually implode to produce the so-called ``electron-capture supernovae'' \citep{Nomoto}, with the production of a light $M \sim 1.25 \, M_{\odot}$ NS and a weak explosion. Only recently an example of this class has been identified and remains under discussion \citep{ECSNobserved}. However, no BHs were thought to be produced in this range of isolated progenitor mass.

Above about $9 \ M_{\odot}$, conditions are sufficient to start carbon burning in full and reach the formation of an iron core. Ultimately, the instability of the central core due to electron capture and photodisintegration caused the implosion later reverted and was believed to be re-energized by neutrino captures \citep[e.g.,][]{JankaConditionsShockRevival2001}. The work of \citet{Heger} found that for progenitors with masses above $25 \, M_{\odot}$, the production of BHs was allowed due to the dynamics of the explosion itself, in which a fallback of a fraction of the ejected matter added to the central object. Moreover, around $40 \, M_{\odot}$ the direct formation of a BH was still expected. This scenario squared with the idea that there must be a threshold (in fact, two of them if the specific dynamics were considered) separating the formation of BHs from the lighter NSs. However, the latest developments on this subject seem to point to a far more complex and qualitatively different outcome, to be outlined in the following.

\subsection{Channels for BH formation in core-collapse}

Hints of a complex, non-deterministic BH formation without a mass threshold are currently present in many simulations \citep[e.g.,][]{SukhboldCoreCollapseSupernovae2016,Ertl}, not only from isolated but also from binary supernovae calculations, which bring a set of challenges not yet solved. This is at odds with the orthodox pre-21st century view sketched above. An attempt to determine the systematics of the different formation channels emerging from simulations has been presented by \citet{Adam}. The group identified four different outcomes, namely

\begin{itemize}
    \item \textbf{Channel 1:} Strong asymmetric supernova explosions (exemplified by two of their solar metallicity models, with $19.56 \, M_{\odot}$ and $40 \, M_{\odot}$), both preceded and followed by progenitors forming neutron stars. The lowest mass yielded a BH inside the LMG;
    \item \textbf{Channel 2:} Weak supernova explosion and fallback BH, corresponding to their model of $23 \, M_{\odot}$ followed for long times and producing also a low-mass BH;
    \item \textbf{Channel 3:} Pair Pulsational Instability producing a $\sim 37 \, M_{\odot}$ BH from a $100 \, M_{\odot}$ progenitor, featuring a short formation time of a few tenths of second;
    \item \textbf{Channel 4:} Complete direct collapse, without explosion, found for their models of $12.25 \, M_{\odot}$ and $14 \, M_{\odot}$, which produced BHs with a slightly lower mass than the progenitors, which are also close in mass to the exploding-neutron-star-producing ones.  
\end{itemize}

With all caution, and taking into account the reasons found for this {\it a priori} unorthodox behavior \citep{Adam}, the channels point to a stochastic and not deterministic outcome, as suggested by previous works. A recent revision by \citet{Janka} concludes that the progenitors producing BH cannot be predicted with certainty, and a deeper analysis of the mass accretion rate following bounce, the equation of state and the efficiency of the neutrino transfer are needed, since differences between results of groups using different codes remain.

Given that a large, if not dominant, fraction of supernovae occur in binary systems, the consideration of the binary environments is also crucial to construct a complete picture of stellar BH remnants. Up to a few years ago most of the binary effects were just simulated through the artificial loss of the hydrogen envelope of the progenitor star. This rough procedure is progressively being improved by several groups, \citep{Patton, Stan1}, with the aim of determining the BH initial mass function. Intermittency is also present in many calculations \citep{Ertl}, adding support to the stochastic behavior extended to the binary case. 

On the observational side, one of the most intriguing possibilities suggested by calculations, but amenable now to direct testing, is the direct collapse of light progenitors (instead of the heavy ones expected at the time of the work of \citeauthor{Heger}). With the large surveys underway and planned in the near future, evidence for the plain disappearance of massive stars \citep{Villa} reported a few years ago have features that could correspond to the direct collapse predicted by simulations. However, no case could be confirmed yet and the search continues, since this important channel cannot be dismissed. It is also worth pointing out that there is no definitely accepted signature of the event itself that can be identified in large transient surveys \citep{Zwicky}, although some events were tentatively identified as such. It is worthwhile to remark that 
this was the channel favored by \citet{MirabelRodrigues} long time ago, and in fact \citet{Felix} has stressed that there is no evidence of the formation of stellar black holes in supernova explosions yet, in spite of the 
consensus from the numerical simulations results.

Another important feature of CSSNe are the net momentum kicks imparted onto the nascent CO by the asymmetric explosion. It has traditionally been expected that BHs should receive significantly lower kicks than NSs at birth, or in some cases no kick at all, due to the "implosive" nature of their formation: in order for a proto-NS not to collapse into a BH, significant amounts of matter must be ejected. BHs, in contrast, have no upper limit on how much matter can be accreted, allowing for the accretion of all or of a significant fraction of their progenitor's mass, be it in the form of a direct collapse or of fallback matter \citep[as in, e.g., the commonly adopted ][remnant mass prescription]{FryerCompactRemnantMass2012}. That does not exclude, however, the formation of BHs with significant explosions and imparted kicks, as has been found to be the case in some simulations of the collapse of a high-compactness progenitor in \citep{BurrowsTheoryNSBHSpins2024,Adam}. It is conceivable that such a mass-kick correlation could explain the dearth of low-mass BHs in LMXBs (where accretion puts an upper limit on separation), as has been suggested before \citep{MandelBinaryPopulationSynthesis2021,Adam}.

\subsection{Additional BH formation channels} 

In addition to the core-collapse channels listed above, the galactic population of BHs may harbor a subset of objects formed by alternative processes, which are as uncertain as the collapsing ones. And while the rate of formation of the latter is certainly high, possibly up to a few per millennium, the former frequencies are much more difficult to assess. 

The first notorious channel, confirmed in gravitational wave data, is the merging NS-NS, with an initially  predicted rate much smaller than the collapse-produced BHs \citep{Predict}. However, the direct empirical estimates \citep{AbbottGWTC3Pop2023} are much higher than the former number, and the question of the real rate and the events that contribute to this channel remains under discussion \citep{Bel,MandelRatesCompactObject2022,BOSSA}. NS-NS mergers have, for example, been suggested as a significant channel of LMG BH formation in dense environments, where dynamical interactions play a significant role \citep{YeGlobularStarClusters2024}, and as a potential progenitor of the low-mass BH in GW230529 \citep{MahapatraPossibleBinaryNeutronStar2025}. Regarding the BH population, the related channels NS-BH and BBH (which dominate the rate) can be considered as events that reprocess the masses of the entering BHs which are already present \citep{Zhu}.

Several types of binaries can contribute to the BH population by pushing a NS above its maximum mass limit. This Accretion-induced Collapse of NSs has been studied sporadically, but the total contribution to the BH population is quite uncertain. As an example, \citet{Chen4} presented a scenario of Ultra-compact BH X-Ray Binaries formation starting from a NS progenitor in a LMXB, in which a small amount of accreted mass would produce ``light'' BHs, close to the maximum NS mass static limit. A similar event, but originated in a spider binary, was discussed by \citet{HorvathSciChi}. A recent work by \citet{Manu} also concludes that light BHs may be formed in some of these systems. It is expected that long accretion times ($\geq \, Gyr$) favor the Accretion-induced Collapse phenomenon in the so-called {\it redback} stage. In both cases further accretion would happen after a long time and would add mass to the BH to populate the LMG region, although the actual number of BHs produced by this channel is uncertain to the point that we can not be sure whether the contribution is negligible or truly relevant.

In summary, we have presented some of the main channels contributing to the galactic population sampled and studied here. These formation scenarios should therefore be seen as qualitative interpretations of the mass-spectrum features inferred in Sections \ref{sec:results} and \ref{sec:lower_mass_gap}. Even with the small number of systems, we hope that a glimpse of the whole population of BHs in the galaxy can stimulate further work in this direction.

\section{Conclusions}\label{sec:conclusions}

We have updated the Galactic BH mass sample from \citet{galaxiesBHs2023deSa} and performed a Bayesian analysis of the distribution. Relative to the sample update, the most significant changes are the removal of LB-1, previously described as hosting a $\sim68~\mathrm{M}_\odot$ BH \citep{Liu2019} and later shown to consist in a stripped star-Be binary \citep{Shenar2020}, the addition of the dormant BH G3425, with $3.58^{+0.8}_{-0.47}~\mathrm{M}_\odot$, to the lower edge of the mass spectrum, and of Gaia BH3 and IC 10 X-1, with $32.7\pm0.82~\mathrm{M}_\odot$ and $28.86^{+14.61}_{-8.11}~\mathrm{M}_\odot$, respectively, to the upper edge. Systems 3A 1524-617, Gaia BH2, MAXI J0637-430, Swift J1727.8-1613, HD 130298, VFTS 243 were included in the sample, while the SS 433 was removed, and the masses of GRO J0422+32, GRO J1655-40, XTE J1859+226, GRS 1716-249, M33 X-7 and OGLE-2011-BLG-0462 revised. The combined sample now consists of 42 BHs, of which 28 are in LMXBs, 7 in detached binaries (dormant BHs), 6 in HMXBs and one is an isolated BH. A living catalog of galactic BHs, \texttt{CatNoir (v 1.0.0)}, is presented.

Following the methodology of \citet{Ozel2010} and \citet{Farr2011}, we performed a fully hierarchical Bayesian analysis of the sample to infer an underlying mass distribution. We tested a Gaussian, two-Gaussian, log-normal, power-law, and decaying exponential parametrization against the Combined sample (all 42 systems), and the LMXB sample (28 systems). Furthermore, we consider a Refined sample ($N \simeq 36$, removing the most poorly constrained transients) to access the effect that unconstrained systems have on final results.  Model selection was performed via four information criteria (WAIC, LOO, AIC, BIC) with relative weights.

Computing the four information criteria jointly, we find that the Combined sample strongly favors a power law with $m_{\min}=4.68^{+0.58}_{-0.65}~\mathrm{M}_\odot$, $m_{\max}= 34.25^{+5.25}_{-2.26}~\mathrm{M}_\odot$ and slope $\alpha=-2.47^{+0.68}_{-0.82}$, followed  by a decaying exponential with $m_{\min}=4.22^{+0.65}_{-0.84}~\mathrm{M}_\odot$ and scale mass $m_0=5.24^{+1.98}_{-1.38}~\mathrm{M}_\odot$ (all uncertainties with $90\%$ C.I.). An even firmer distinction between models will be possible once the sample is increased and/or the mass measurements of BHs are tightly constrained.

For the LMXB case, we disconsidered the power-law parametrization since results lead us to significant divergencies and a bad behavior of latent masses, pushing results to a positive slope. From the remaining parametrizations, the Gaussian stands as best-model with $\mu = 7.74^{+0.71}_{-0.65}$ and $\sigma = 1.56^{+0.7}_{-0.49}$, although from WAIC and LOO metrics it is considered statistically indistringuishable from the two-Gaussian model, but mildly preferred if we consider the AIC and BIC criterias. Again, a break of degeneracy will only be possible once the sample and the data accuracy increases.

To evaluate the significance of the partial filling of the LMG, we compute the $1\%$-quantile for the preferred model of both samples, finding $4.06 < M_{1\%} < 5.29$ ($90\%$ C.I.) for the power-law in the Combined sample, and $2.56 < M_{1\%} <5.68$ ($90\%$ C.I.) for the Gaussian LMXB sample. This is a significant shift downward from previously reported $M_{1\%}>4.3~\mathrm{M}_\odot$ and $M_{1\%}>5~\mathrm{M}_\odot$ \citep{Farr2011}, and adds evidence for the LMG to be at least partially filled \citep{2022ApJ...941..130D,SiegelInvestigatingLowerMassGap2023,XingMassGapBlacKholes2025}, as has been recently supported by CCSNe simulations \citep[e.g.][]{Adam} and by GW data, from which \citet{RayHidingOutLowEnd2025} have found $M_{1\%}=3.13_{-0.04}^{+0.18}~\mathrm{M}_\odot$. 

Finally, one should highlight that the significance of the presence of a $\sim \, 7~\mathrm{M}_\odot$ peak in the Galactic BH mass distribution, and the more recent addition of the $\sim33~\mathrm{M}_\odot$ Gaia BH3, is severely obscured by the incompleteness of the sample, which counts on only 42 of an expected $\gtrsim10^7$ BHs in the Galaxy. Of those 42, only 2 stand to have mass $>20~\mathrm{M}_\odot$ within $90\%$ credibility (in fact, 3 if the new value for Cyg X-1 of \citet{new} is adopted). Therefore, while a comparison with the peaks at $\sim10~\mathrm{M}_\odot$ and $\sim35~\mathrm{M}_\odot$ found in the primary mass distribution of extragalactic BBH mergers \citep{AbbottGWTC3Pop2023} is tempting, no conclusive compatibility test can be performed at this time. This is specially true because, while the selection effects of current GW detectors are very well understood and can be corrected for in the mass distribution with confidence, the same cannot be said for the Galactic population, for which particular features may still be a consequence of observational biases. While it can be expected that some, if not all, BH formation channels should be represented in both the Galactic and extragalactic samples, determining which they are will require both (a) modeling of selection effects for Galactic populations \citep[see, e.g.,][for Gaia detections]{ElbadryGenerativeModelGaia2024,NagarajanRealisticPredictionsGaiaBH2025}, and (b) understanding the origins of particular features in the distribution \citep[through, e.g., population synthesis, as in][]{SiegelInvestigatingLowerMassGap2023,XingMassGapBlacKholes2025}. In any case, an extensive search for galactic BHs, as well as improvements on mass determinations, are necessary to deepen our knowledge on the field.


\begin{acknowledgments}
The authors acknowledge the support of Fundação de Amparo à Pesquisa do Estado de São Paulo (FAPESP) through the Grant 2024/16892-2. L. S. R. acknowledges FAPESP through project 2023/08649-8. J.E.H. thanks the CNPq Federal Agency, Brazil for a Research Scholarship. N. P. acknowledges FAPESP through the grant \#2025/01653-5. L. M. S. acknoledges support by the CNPq, grant number 140794/2021-2, and by the Alexander von Humboldt Foundation. M.G.B.A. acknowledges the São Paulo Research Foundation (FAPESP) through the grant \#2025/25588-8. B. B. M. acknowledges FAPESP through the grant \#2025/01641-7. L.G.B. acknowledges the support of Coordenação de Aperfeiçoamento de Pessoal de Nível Superior (CAPES) through the grant 8887.200829/2025-00. G.R.C.S acknowledges the University of São Paulo (USP) through the grant 694/2025. D.V.R acknowledges the University of São Paulo (USP) through the grant 307/2025. R.M.G.T acknowledges CNPq through the grant 2025/1510. The authors would like to thank the Brazilian Ministry of Science, Technology and Innovation (MCTI) and the Brazilian Space Agency (AEB), which supported the present work under the PO 20VB.0009.
\end{acknowledgments}

%



\software{The Gaussian Skewed fit in this work was made using the {\texttt{LMFIT v1.3.3}} Python library, \citep{LMFIT}. This work made use of \texttt{Mathematica} \citep{Mathematica}  and Python libraries: \texttt{numpy} \citep{harris2020array}, \texttt{scipy} \citep{2020SciPy-NMeth}, \texttt{astropy} \citep{2022ApJ...935..167A} and \texttt{seaborn} \citep{Waskom2021}.
Software citation information aggregated using \texttt{\href{https://www.tomwagg.com/software-citation-station/}{The Software Citation Station}} \citep{software-citation-station-paper,software-citation-station-zenodo}.}



\bibliography{BHs}
\bibliographystyle{aasjournal}

\clearpage
\appendix
\section{Visual representation of BH binary systems}\label{apx:schematic_visu}
\begin{figure}
    \centering
    \caption{Visual representation of the systems analyzed in the combined sample. All systems were scaled according to the distance from the Sun to Mercury.  The color of the companion star in each system is related to its spectral type, this relation is shown in the upper right corner, with the companion stars with unknown spectral type being drawn as white with a black contour. The BHs are represented as a black dot. The systems with accretion are shown with a line in direction to the accretion disk (gray ellipsis), indicating the matter flow to the BH.}
    \label{fig:individual_BHs1}
    \vspace{-1cm}
    \includegraphics[width=1.2\linewidth, page=1, angle=90,trim={0 0 1cm 0},clip]{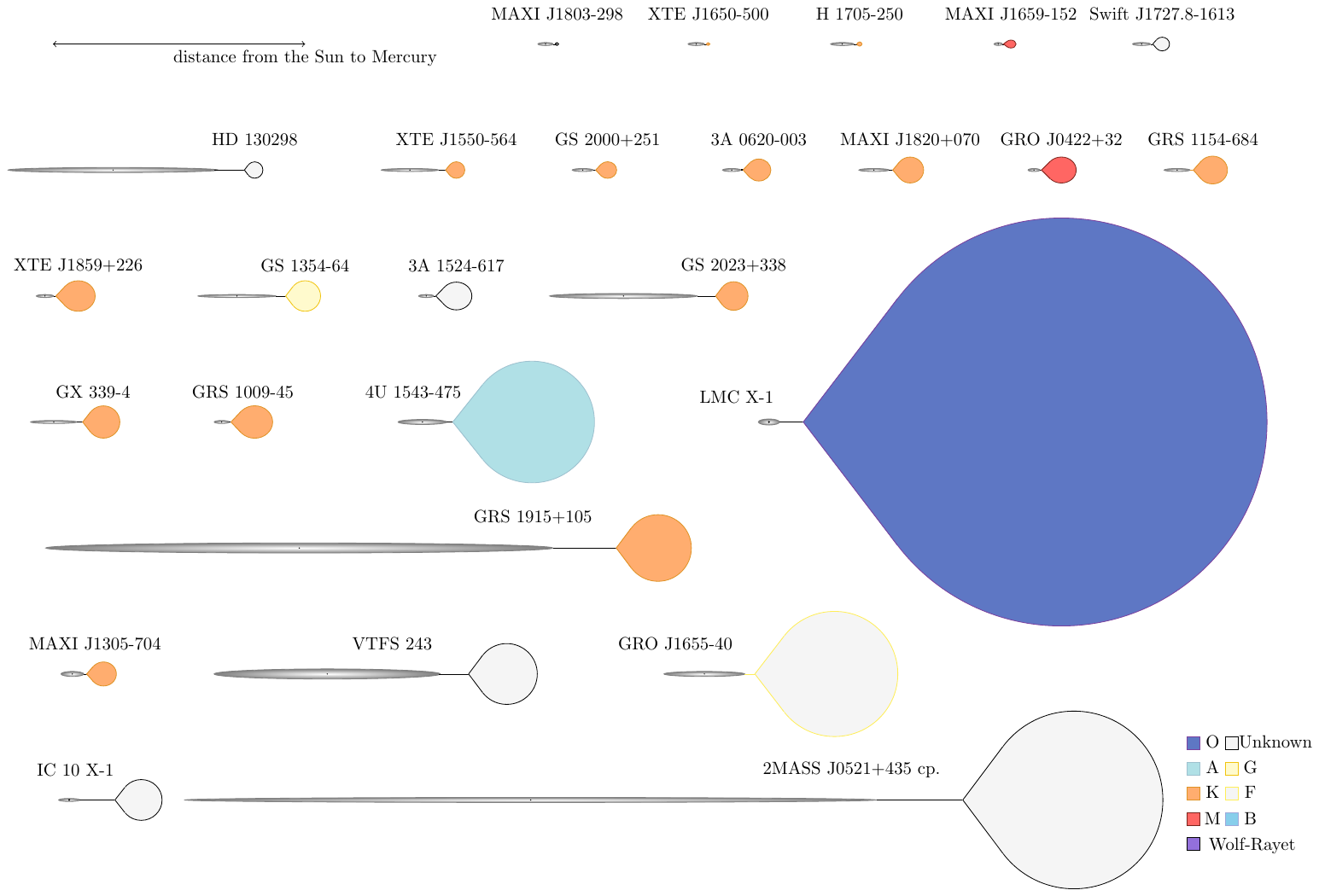}
\end{figure}

\begin{figure}
    \centering
    \caption{Visual representation of the systems analyzed in the combined sample (continued). Note that a different scale for the same distance was employed for the systems below the dashed line to allow a fit in the page.}
    \label{fig:individual_BHs2}
    \includegraphics[width=1.2\linewidth, page=2, angle=90]{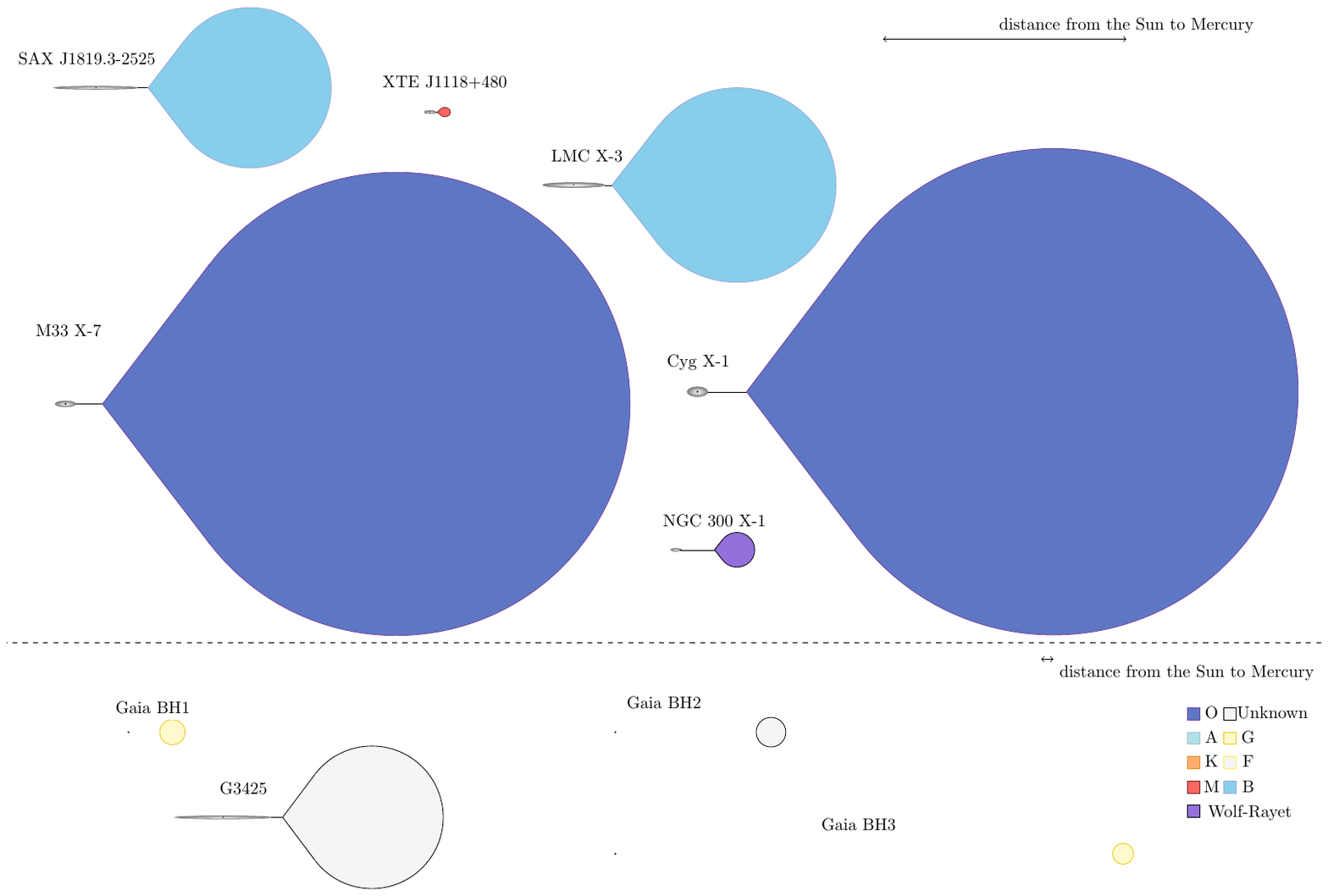}
\end{figure}

For a better visualization of the scale between the galactic binary systems analyzed in this work, we present a pictoric representation shown in Figures \ref{fig:individual_BHs1} and \ref{fig:individual_BHs2}. The BHs are represented by small black dots, centered inside grey ellipses depicting the accretion disk in case of interacting systems. The distance between the BH and the companion star is proportionally given by their orbital period and scaled according to the distance from the Sun to Mercury. The companion stars are colored according to their spectral type as reported in BlackCAT \citet{corralsantana2016}, and their radius follows the approximated relation $1\ R_{\odot}\sim 1\ M_{\odot}$, as suggested by \citet{demory2009}. The Wolf-Rayet stars had their radius given by \citet{langer1989} results. For design purposes, the LMXB, Non-Interacting, and Isolated systems were scaled according to LMC X-3 mass, while the HMXB, according to Cyg X-1 mass.

\section{Mass estimates of individual BH masses}
\label{sec:estimated_masses}

In the Bayesian analysis described troughout this work, the individual masses of BHs were treated as latent variables, and sampled accordingly, in a ``population-informed'' treatment. However, it is also possible to estimate their masses directly from the mass function equation (Eq. \ref{eq:mass_function}), as we present here. Following the procedure in \citet{galaxiesBHs2023deSa} the individual masses of each BH were re-estimated via Monte Carlo (MC) calculations based on their orbital parameters (when available), which were discussed in the previous subsections and presented in Tables \ref{tab:BHs_ecc} and \ref{tab:list_BHs}. The eccentricity is set to zero when not reported. Given the distribution of each parameter ($f, q, i$), random values are generated, resulting in a mass distribution,
which we fit with a skewed-Gaussian, instead of an asymmetric one:

\begin{align}\label{eq:SwdGsn}
    P(m; A, \mu, \sigma, \gamma) &= \frac{A}{\sigma \sqrt{2\pi}} \exp \left[{-\frac{(m - \mu)^2}{2\sigma^2}}\right]\times \\ &\times\left\{ 1 + \operatorname{erf}\left[ \frac{\gamma(m - \mu)}{\sigma \sqrt{2}} \right] \right\},
\end{align}
where $A$ is a normalization factor, $\mu$ is the center location, $\sigma$ is the scale and $\gamma$ the skewness. We will interchangeably refer to the skewed-gaussian as asymmetric gaussian also. 

\begin{figure*}[ht]
    \centering
    \begin{minipage}[t]{0.45\linewidth}\label{fig:HistAndChiSquareWithoutHighMasses}
        \centering
        \includegraphics[width=\columnwidth]{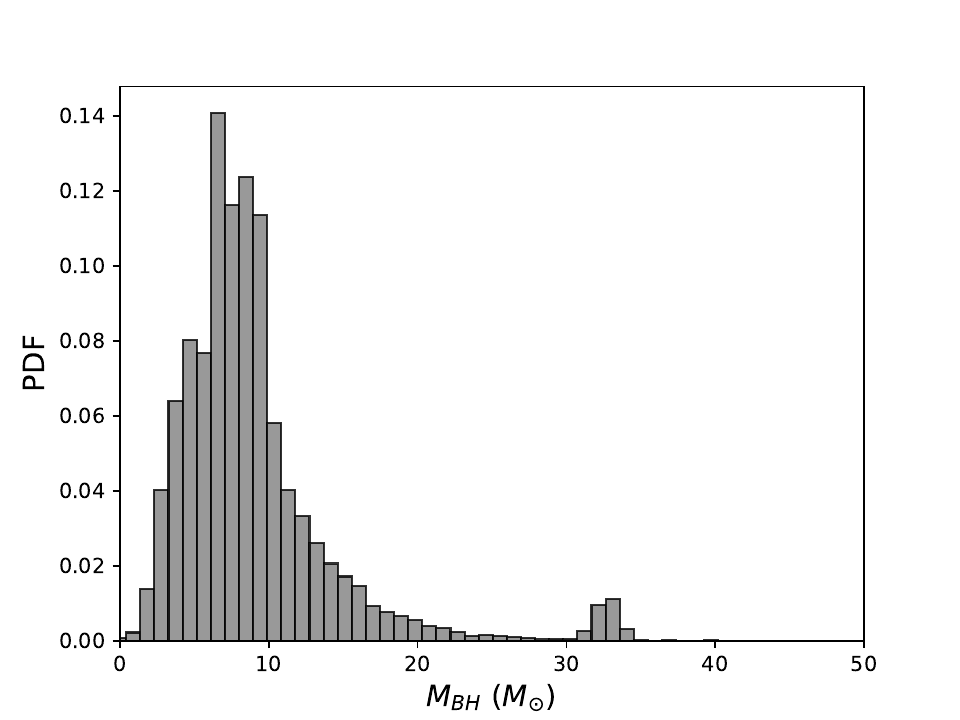}
    \end{minipage}
    ~
    \begin{minipage}[t]{0.45\linewidth}\label{fig:histMC}
        \centering
        \includegraphics[width=\columnwidth]{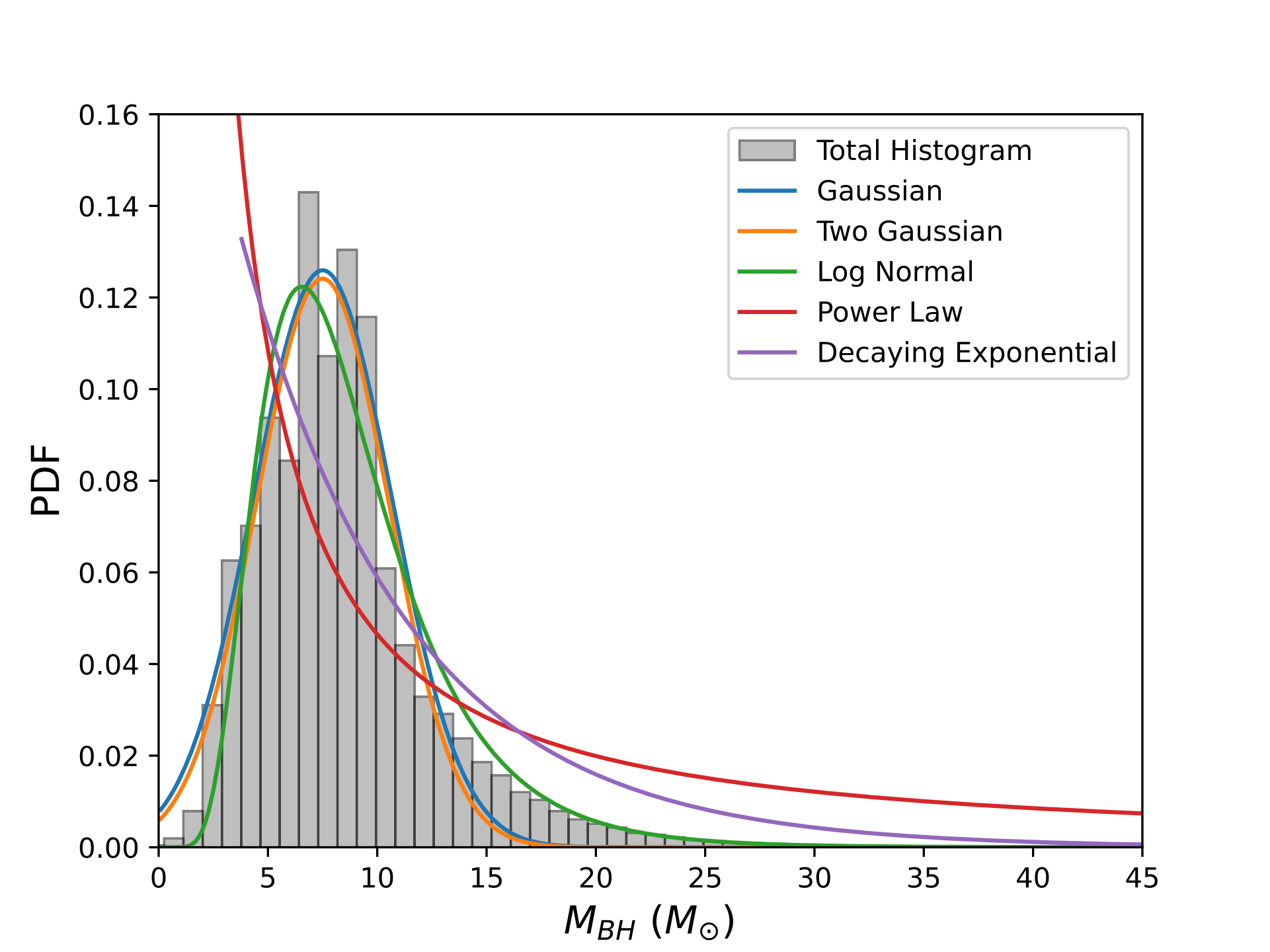}
    \end{minipage}
    \caption{Left: For each individual BH, 50000 masses are sampled from the skewed-Gaussian (given at Eq. \eqref{eq:SwdGsn}), and combined into the histogram plotted above, so uncertainties are taken into account. The parameters values of each skewed-Gaussian are informed at Table \ref{tab:parsskewed}. The histogram is presented with 50 bins. Right: Different models fitted (within $\sim$10000 degrees of freedom) excluding systems with higher masses. The histogram is presented in 50 bins. In this histogram, the high mass objects were deleted. The objects IC 10 X-1 and Gaia BH3 were excluded from the sample.}
    \label{fig:MC_histograms}
\end{figure*}

\begin{figure}[ht]
    \centering
    \includegraphics[width=\linewidth]{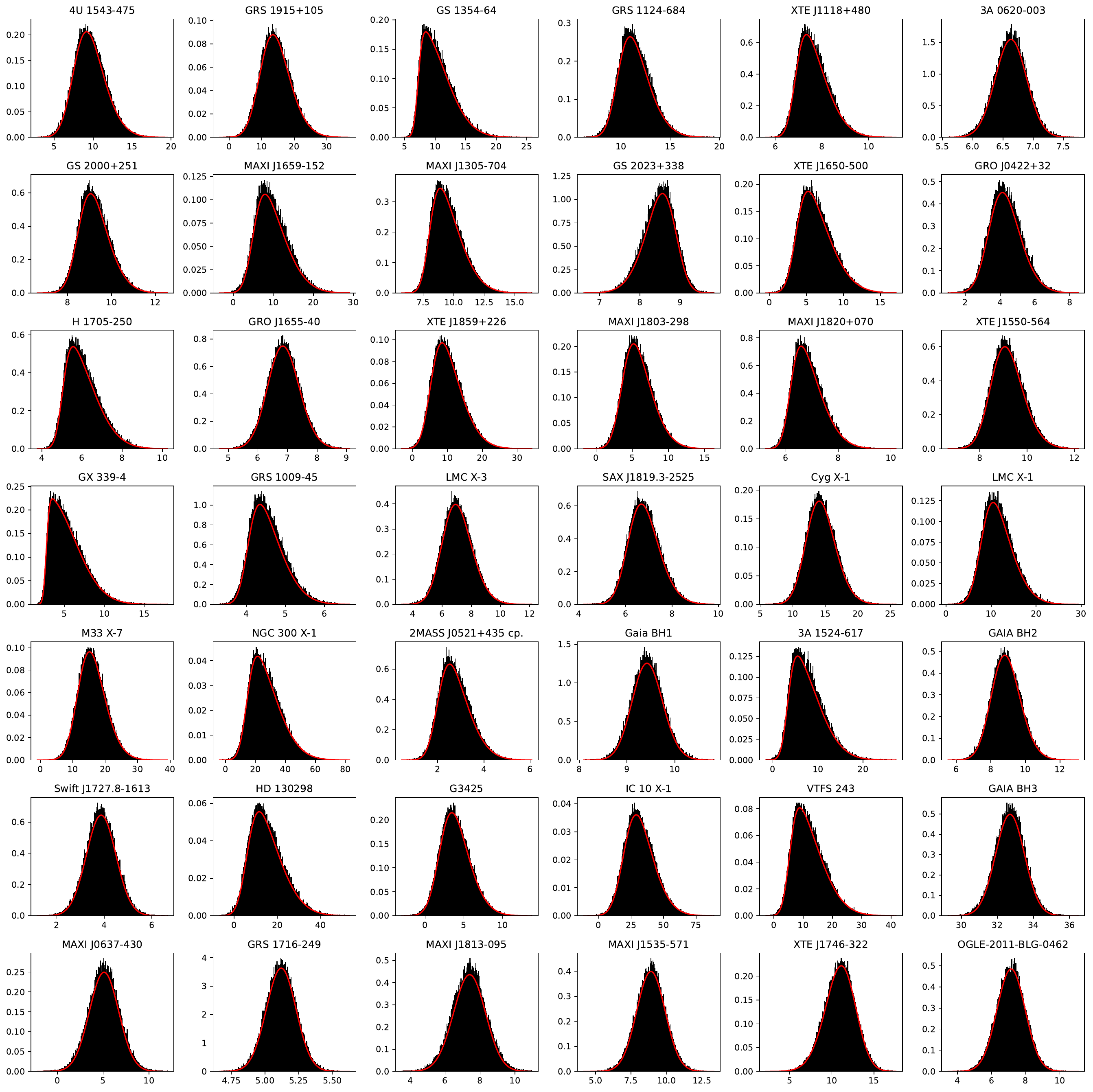}
    \caption{Resulting fits using a Skewed Gaussian as described in eq. \eqref{eq:SwdGsn}. The histogram bins are represented by the black bars. A total of 1000 bins were considered, indicating the probable BH mass, while its height shows its Probability Density Function (PDF). The red line shows the resulting fit.}
    \label{fig:appAllSwdBHs}
\end{figure}

In the last column of Table \ref{tab:list_BHs} we list only the mode and $68\%$ interval of the resulting fit, performed with the Python library \texttt{LMFIT} \citep[v. 1.3.3,][]{LMFIT}, and extracted from the distributions shown in Figure \ref{fig:appAllSwdBHs}. The full list of resulting parameters can be found in Table \ref{tab:parsskewed}. To visualize the population distribution resulting from this approach, taking uncertainties into account, we drew 50000 sampled masses from each individual mass distribution, through a Monte Carlo sampling, and combined them into the histogram shown in the left panel of Figure \ref{fig:MC_histograms}. 

The presence of high-mass objects suggests the possible presence of a second peak in the distribution, or a much smoother decay, although more data is necessary for a firm conclusion. The extraction of the ``true'' underlying distribution becomes mode involved. Because most of the points (40 out of 42) belong to the region in which the original peak (now shifted) stands, we have performed an analysis excluding the higher masses systems of the Combined Samplem, IC 10 X-1 and Gaia BH3. The resulting histogram and fitted curves are presented in the right panel of Figure \ref{fig:MC_histograms}. When fitting the models presented in eqs. \eqref{eq:gaussian}--\eqref{eq:decayingexponential}, and performing a simple analysis using just the $\chi^2$ results, we find that the log-normal distribution with parameters $\sigma_{\text{log-normal}}=0.45$ and $\mu_{\text{log-normal}}=2.08$ is better than any other expression to fit the distribution, featuring a low $\chi^2$ value, while the Bayesian analysis of the combined sample shows a preference to a power-law parametrization, as we show in Section \ref{subsec:model_comparison}. This indicates that the high-mass bins force a distortion of the former log-normal preferred distribution, which becomes much worse in the bulk of data points. Of course, more complex fits can be attempted provided more objects are discovered in the high-mass region.

\begin{table}  
\centering
\caption{Resulting parameters from the fit using a Skewed-Gaussian defined in eq. \eqref{eq:SwdGsn}. $R^2$ refers to the coefficient of determination. $\chi^2$ was evaluated comparing the data with the given fit model. The systems are divided between LMXBs, HMXB, Non-interacting (or dormant BHs) and Isolated.}
\label{tab:parsskewed}
\begin{tabular}{l l ccccccc}
\hline
Name & Type & $A$ & $\mu$ & 
$\sigma$& $\gamma$ & $\chi^{2}$ & $R^{2}$ \\
\hline
4U 1543--475 & LMXB & 1.00540116 & 7.61513403 & 2.87162584 & 1.94917292 & 0.02705231 & 0.99364362 \\
GRS 1915+105 & LMXB & 1.00490618 & 9.91746777 & 6.56634113 & 1.78687921 & 0.00396054 & 0.99463822 \\
GS 1354--64 & LMXB & 1.00422431 & 7.17865453 & 4.19014254 & 7.06492009 & 0.02578601 & 0.99333365 \\
GRS 1124--684 & LMXB & 1.00612552 & 9.71556955 & 2.38164218 & 2.42607955 & 0.11429774 & 0.98957174 \\
XTE J1118+480 & LMXB & 1.0058643 & 6.86360434 & 1.04129816 & 3.35920148 & 0.41701984 & 0.99238305 \\
3A 0620--003 & LMXB & 0.99977605 & 6.63263174 & 0.25806643 & -0.00181737 & 2.52727494 & 0.9920093 \\
GS 2000+251 & LMXB & 1.00522807 & 8.51308772 & 1.02222596 & 2.15245921 & 0.43102117 & 0.99135763 \\
MAXI J1659--152 & LMXB & 1.00647992 & 5.01156277 & 6.32807025 & 3.17981492 & 0.02825312 & 0.98696059 \\
MAXI J1305--704 & LMXB & 1.00740783 & 8.01436139 & 1.99105089 & 3.53379152 & 0.11030439 & 0.99245074 \\
GS 2023+338 & LMXB & 1.00492782 & 8.85966774 & 0.55652174 & -1.94887849 & 2.61386139 & 0.98813191 \\
XTE J1650--500 & LMXB & 1.00628567 & 3.58606998 & 3.63103972 & 3.37397355 & 0.02832496 & 0.99227726 \\
GRO J0422+32 & LMXB & 1.0061899 & 3.44886731 & 1.27739373 & 1.76814831 & 0.41188687 & 0.98885374 \\
H 1705--250 & LMXB & 1.0058057 & 5.03084444 & 1.32704155 & 4.46459841 & 0.40908355 & 0.99064591 \\
GRO J1655--40 & LMXB & 1.00078514 & 6.51315133 & 0.6535822 & 1.01424005 & 0.38575866 & 0.99361997 \\
XTE J1859+226 & LMXB & 1.00662968 & 5.24171708 & 6.7424974 & 2.86597084 & 0.00449884 & 0.99451716 \\
MAXI J1803--298 & LMXB & 1.00542531 & 3.63212695 & 3.07051976 & 2.44523611 & 0.02932482 & 0.99332944 \\
MAXI J1820+070 & LMXB & 1.00613935 & 6.17490531 & 0.92078501 & 3.44413778 & 0.47362261 & 0.99244494 \\
XTE J1550--564 & LMXB & 1.00397909 & 8.55126049 & 0.93984679 & 1.66070217 & 0.38194696 & 0.99208862 \\
GX 339--4 & LMXB & 1.00394358 & 2.71213805 & 3.47028264 & 11.0681863 & 0.02610708 & 0.99443039 \\
GRS 1009--45 & LMXB & 1.00570177 & 4.04691239 & 0.66882723 & 3.24124892 & 2.68458896 & 0.98661986 \\
LMC X-3 & LMXB & 1.00345708 & 6.18077896 & 1.36633333 & 1.46896658 & 0.10003477 & 0.99408081 \\
SAX J1819.3--2525 & LMXB & 1.00395486 & 6.17163524 & 0.9469017 & 1.79737892 & 0.35147352 & 0.99299076 \\
3A 1524--617 & LMXB & 1.00614835 & 3.30407473 & 5.74633786 & 4.77563625 & 0.02852373 & 0.98912854 \\
MAXI J0637--430 & LMXB & 0.9990865 & 5.09100861 & 1.59584664 & 0.0034024 & 0.10278745 & 0.99038938 \\
GRS 1716--249 & LMXB & 1.00075604 & 5.15311769 & 0.11496905 & -0.39812854 & 10.7517887 & 0.99305399 \\
MAXI J1813--095 & LMXB & 0.99924477 & 7.77376638 & 0.99442142 & -0.56133217 & 0.38997785 & 0.98888377 \\
MAXI J1535--571 & LMXB & 0.99857859 & 8.90218166 & 0.99909061 & -0.0050846 & 0.09994633 & 0.9942394 \\
XTE J1746--322 & LMXB & 1.00178802 & 12.3394527 & 2.29878986 & -1.18389622 & 0.02568185 & 0.99443401 \\
Cyg X-1 & HMXB & 1.00412277 & 12.4110113 & 3.0251226 & 1.49555138 & 0.02436556 & 0.99373217 \\
LMC X-1 & HMXB & 1.00643756 & 7.8640127 & 5.17338695 & 2.52686077 & 0.02642629 & 0.98975147 \\
M33 X-7 & HMXB & 1.0051808 & 11.9315015 & 6.02315205 & 1.82105382 & 0.00413574 & 0.9951779 \\
NGC 300 X-1 & HMXB & 1.00605914 & 14.3629908 & 16.6279874 & 3.82217842 & 0.00125156 & 0.99285384 \\
Swift J1727.8--1613 & HMXB & 1.00004004 & 3.8660762 & 0.61946128 & -8.843 & 0.43101541 & 0.99215661 \\
IC 10 X-1 & HMXB & 1.00534146 & 20.1048105 & 16.6602056 & 2.05674408 & 0.00101518 & 0.99339367 \\
2MASS J0521+435 cp. & Non-int. & 1.00669271 & 2.0140484 & 1.01702613 & 2.65347429 & 0.49769785 & 0.9906238 \\
Gaia BH1 & Non-int. & 1.00166548 & 9.2354679 & 0.37657206 & 0.87620909 & 2.58166778 & 0.99013402 \\
Gaia BH2 & Non-int. & 1.00291592 & 8.21818812 & 1.08975324 & 1.30834686 & 0.09871184 & 0.99525444 \\
Gaia BH3 & Non-int. & 1.00015372 & 32.6965689 & 0.80100022 & 0.00384962 & 0.3801105 & 0.99118432 \\
HD 130298 & Non-int. & 1.00648838 & 6.04544044 & 12.1380098 & 3.25598999 & 0.00386963 & 0.99122099 \\
G 3425 & Non-int. & 1.00603776 & 1.97839628 & 2.78414886 & 2.02666913 & 0.02495614 & 0.99455157 \\
VFTS 243 & Non-int. & 1.00714176 & 5.33233775 & 8.82056643 & 4.40084675 & 0.00483124 & 0.99281443 \\
OGLE-2011-BLG-0462 & Isolated & 0.99985171 & 7.1433586 & 0.82220765 & 0.00319976 & 0.10297398 & 0.99516076 \\
\hline
\end{tabular}
\end{table}

\section{Marginalized posterior distributions}\label{apx:results_marginalization}
\subsection{Combined sample}\label{apx:dist_full}
\begin{figure*}[ht]
    \centering
    \begin{minipage}[t]{0.45\linewidth}
        \centering
        \includegraphics[width=\columnwidth]{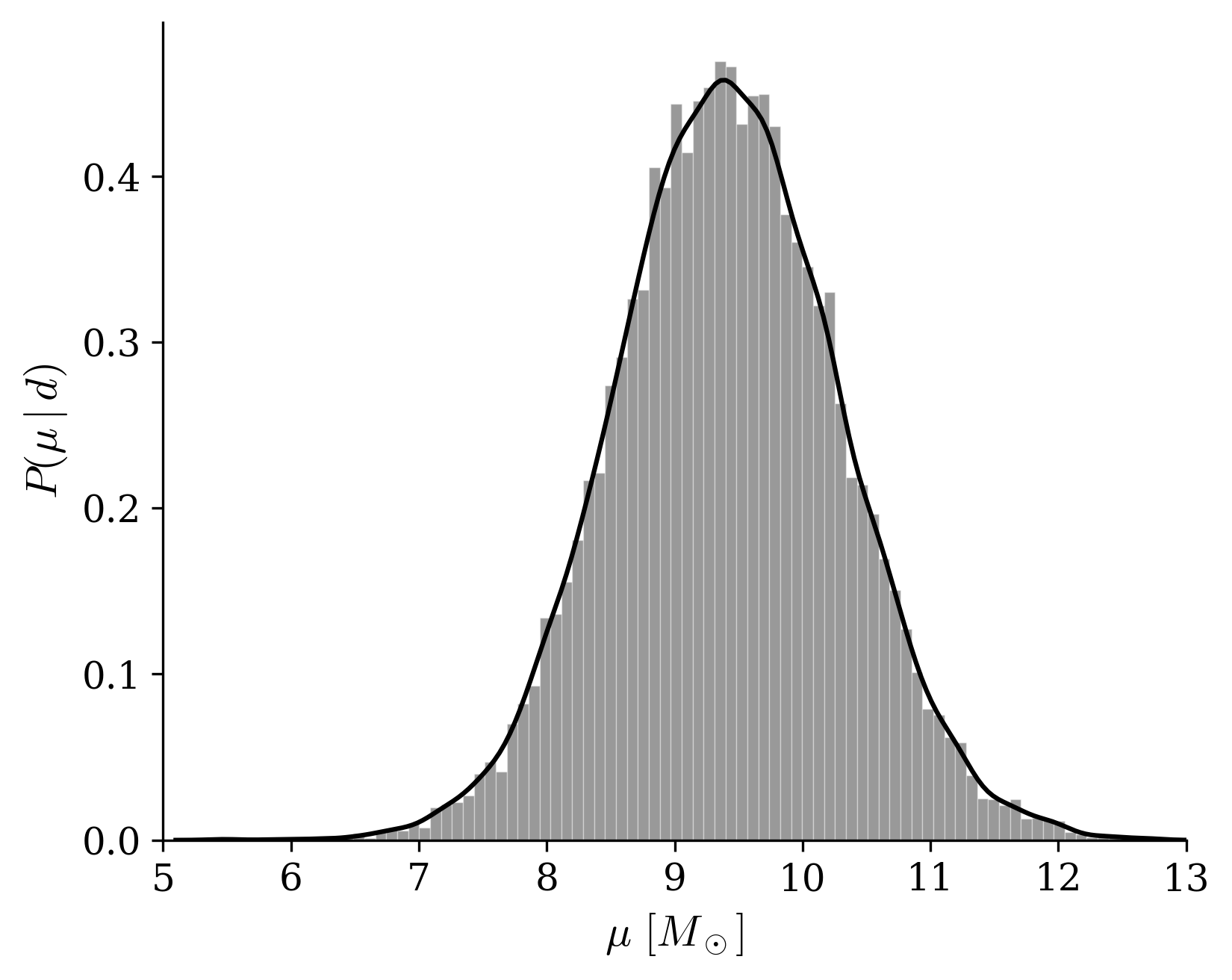}
    \end{minipage}
    ~
    \begin{minipage}[t]{0.45\linewidth}
        \centering
        \includegraphics[width=\columnwidth]{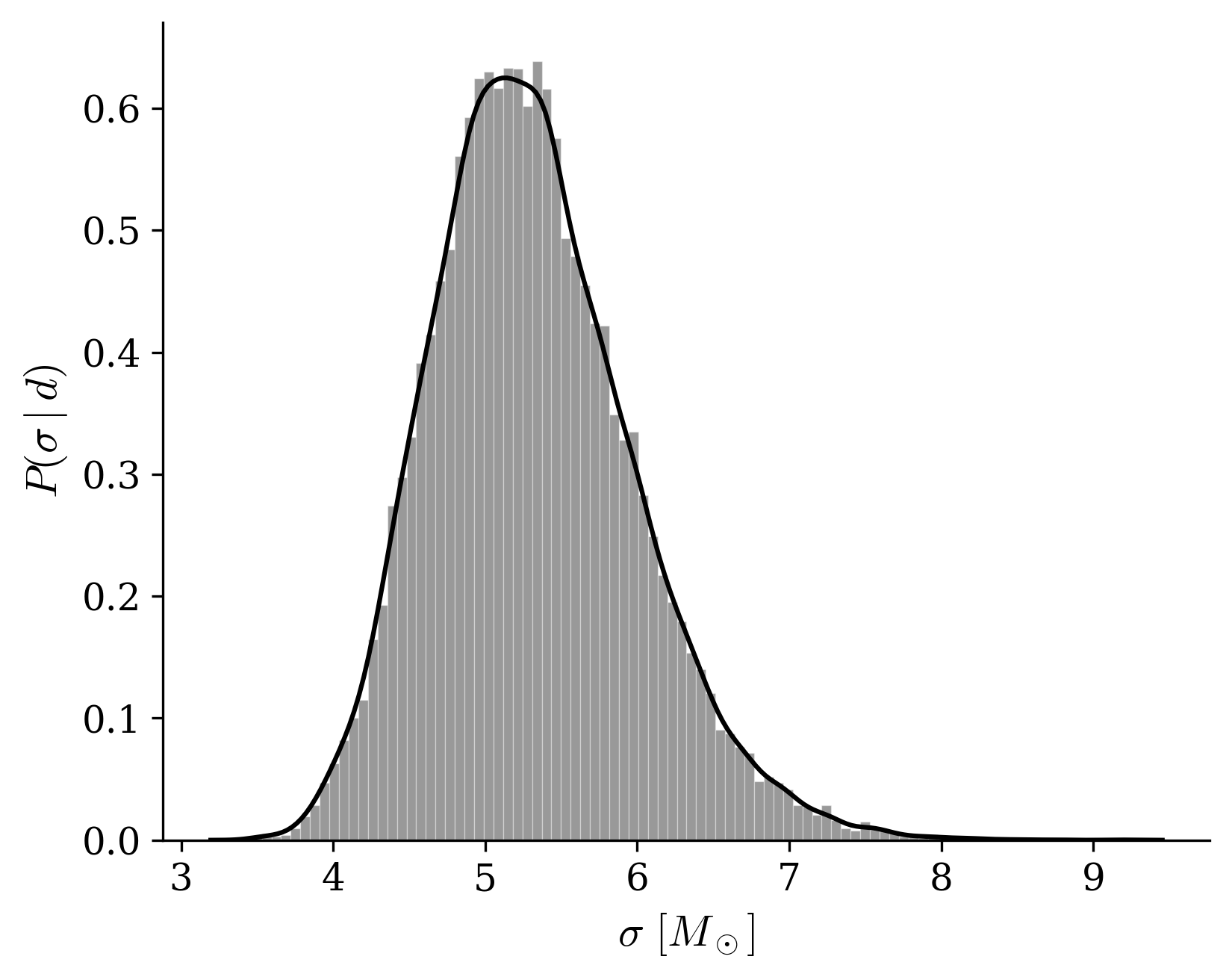}
    \end{minipage}
    \caption{Marginalized distributions of $ \mu $ (left graph) and $ \sigma $ (right graph) for the Gaussian distribution for the Combined Sample.}
    \label{fig:Gaussian}
\end{figure*}

\begin{figure*}[ht]
    \centering
    \begin{minipage}[t]{0.3\linewidth}
        \centering
        \includegraphics[width=\linewidth]{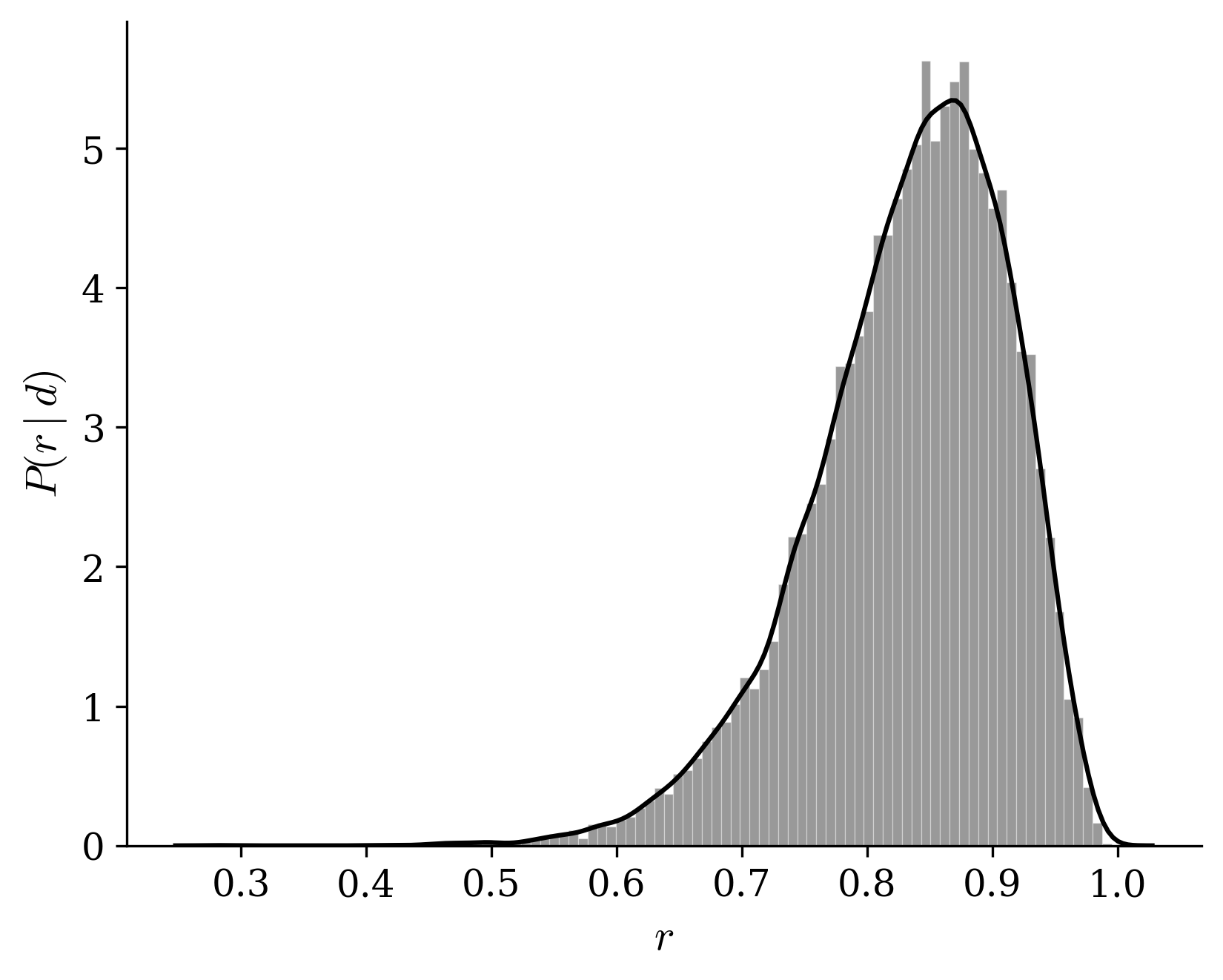}
    \end{minipage}
    ~
    \begin{minipage}[t]{0.3\linewidth}
        \centering
        \includegraphics[width=\linewidth]{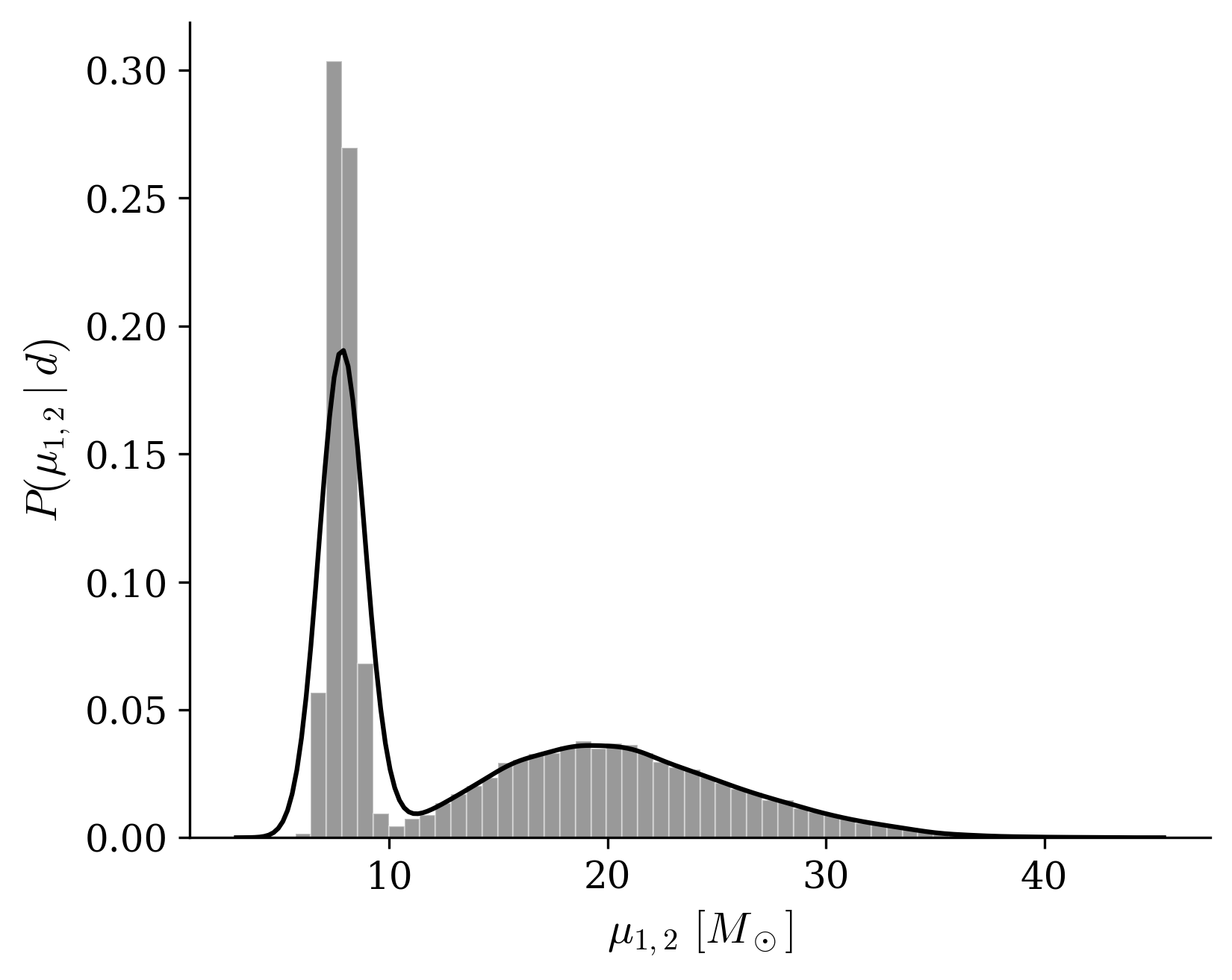}
    \end{minipage}
    ~
    \begin{minipage}[t]{0.3\linewidth}
        \centering
        \includegraphics[width=\linewidth]{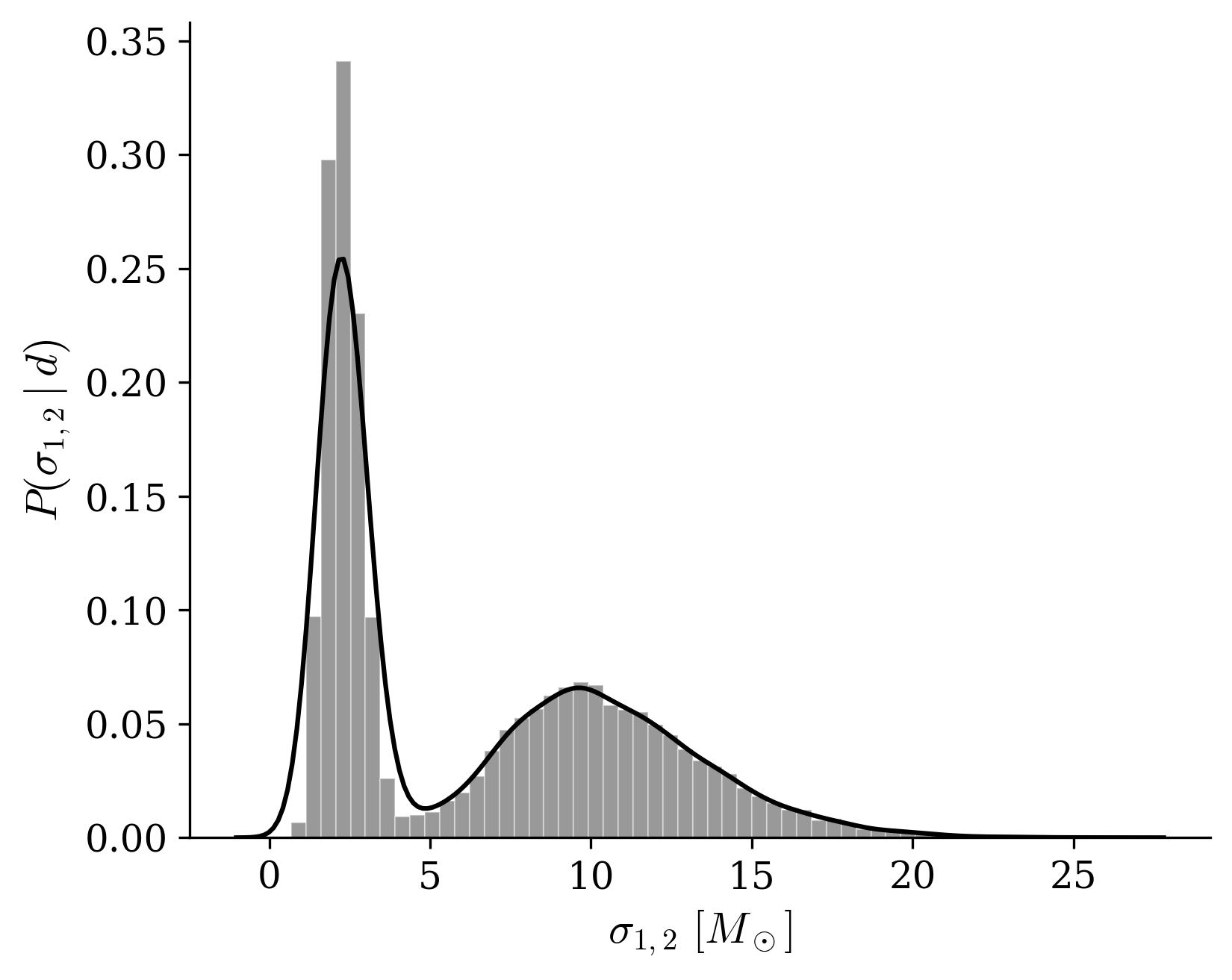}
    \end{minipage}
    \caption{Marginalized distributions of $r$ (left graph), $\mu_{1,2}$ (middle graph, in which the lighter graph indicated the $\mu_1$ distribution while the darker one, $\mu_2$) and $\sigma_{1,2}$ (right graph, in which the lighter graph indicated the $\sigma_1$ distribution while the darker one, $\sigma_2$) for the Two Gaussian distribution for the Combined Sample.}
    \label{fig:TwoGaussian}
\end{figure*}

\begin{figure*}
    \centering
    \begin{minipage}[t]{0.45\linewidth}
        \centering
        \includegraphics[width=\columnwidth]{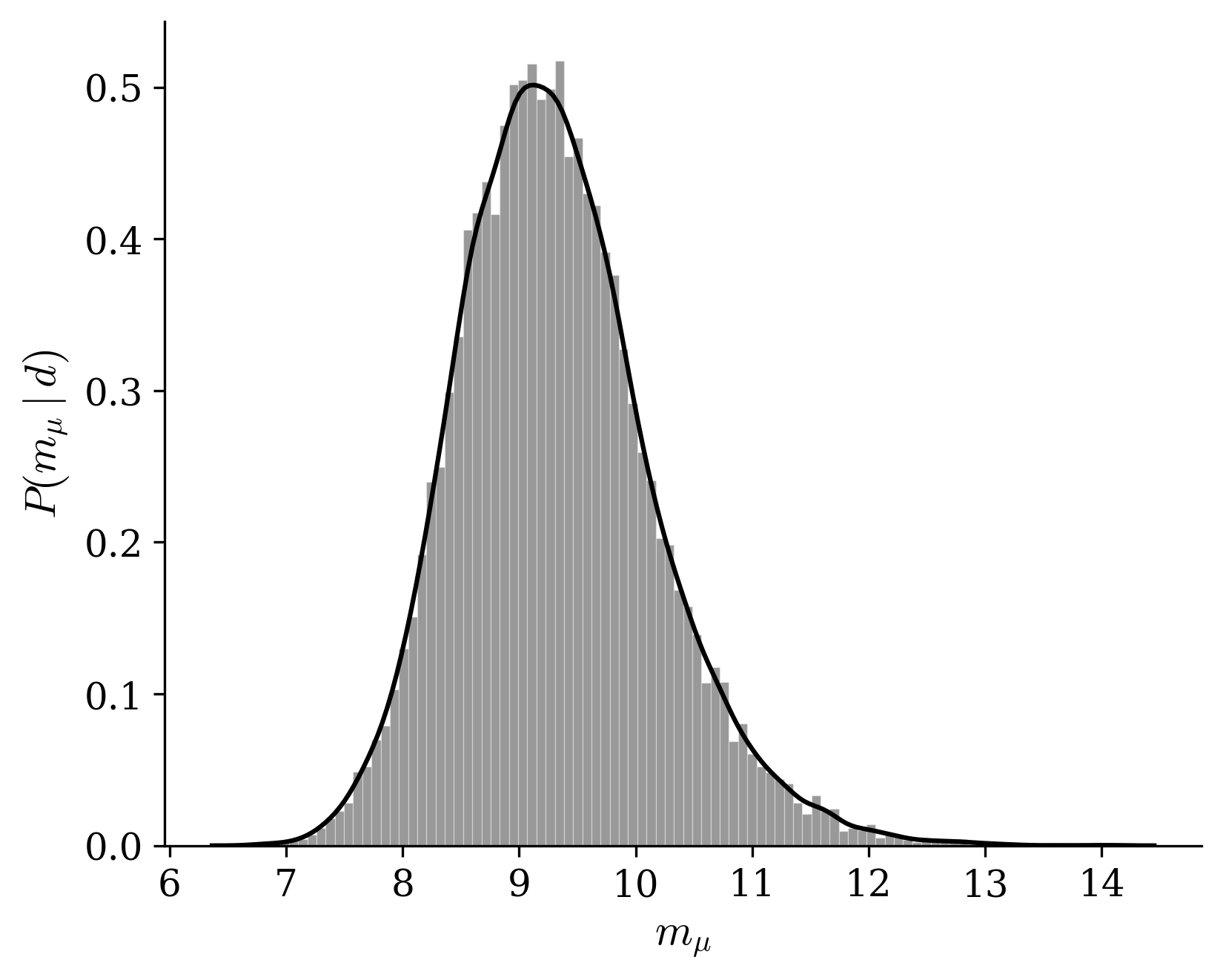}
    \end{minipage}
    ~
    \begin{minipage}[t]{0.45\linewidth}
        \centering
        \includegraphics[width=\columnwidth]{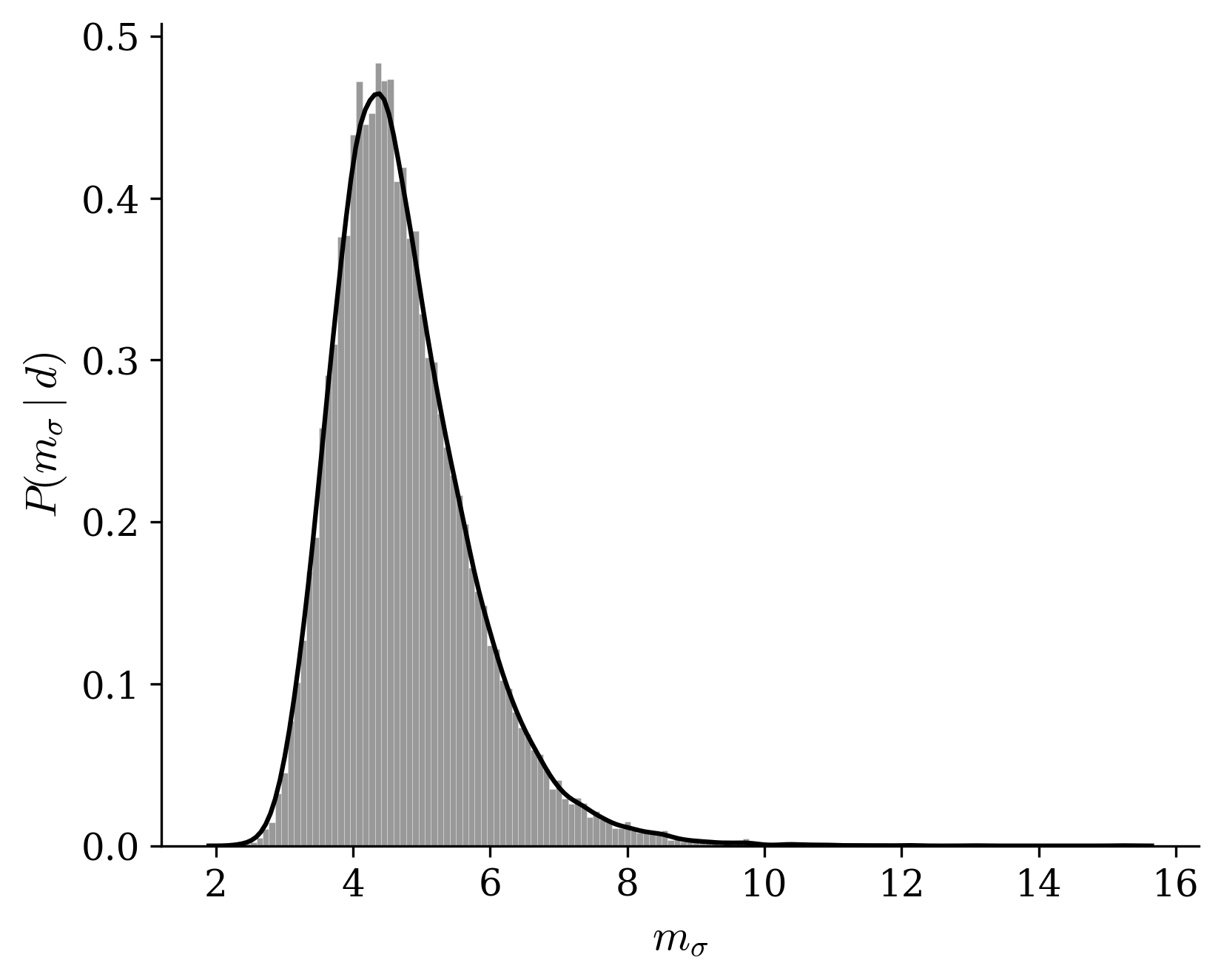}
    \end{minipage}
    \caption{Marginalized distributions of the average BH mass $\langle M_{BH}\rangle$ (left graph) and the uncertainty of the BH mass $\sigma_{M_{BH}}$ (right graph) for the Log Normal distribution for the Combined Sample.}
    \label{fig:LogNormal}
\end{figure*}

\begin{figure*}[ht]
    \centering
    \begin{minipage}[t]{0.3\linewidth}
        \centering
        \includegraphics[width=\linewidth]{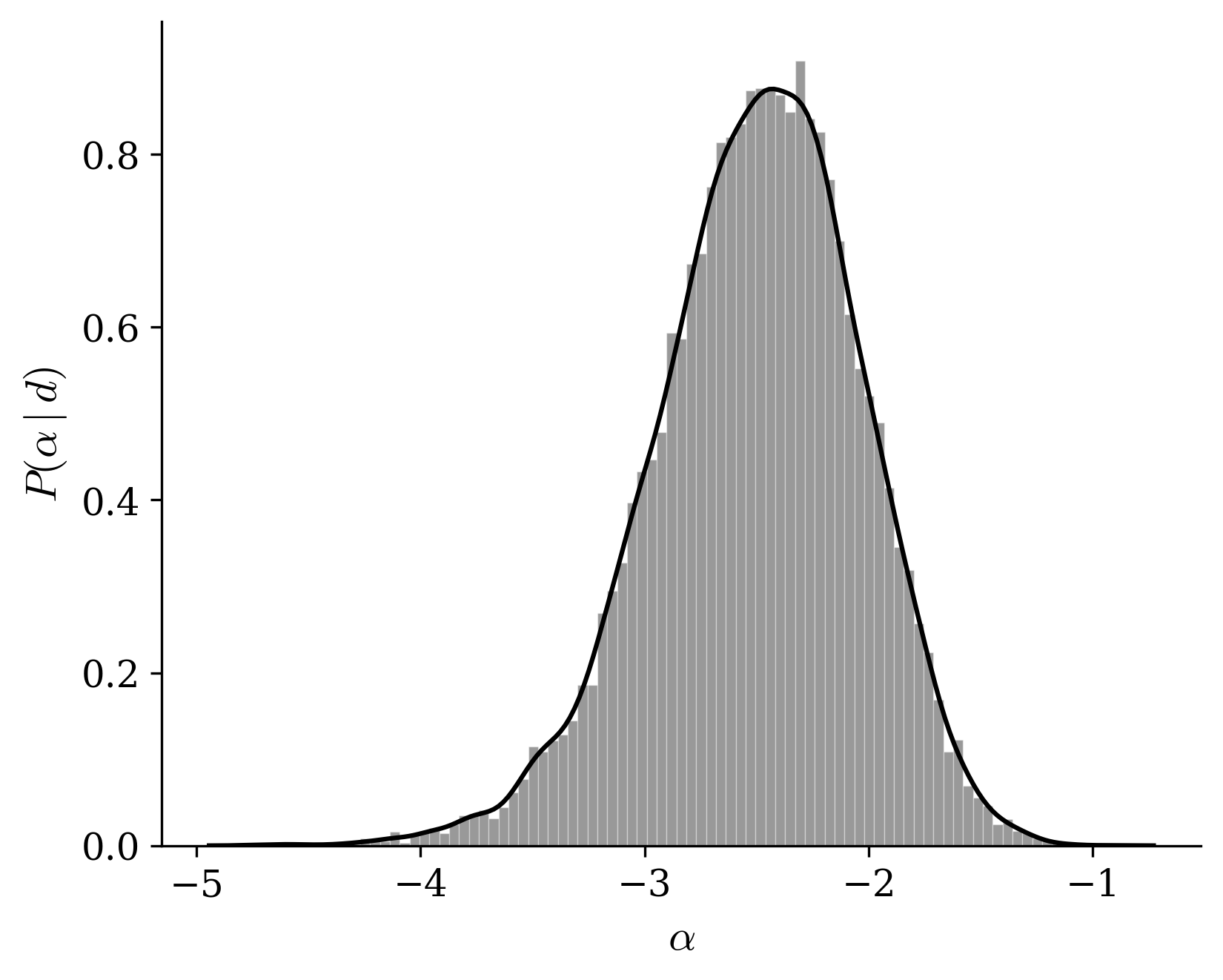}
    \end{minipage}
    ~
    \begin{minipage}[t]{0.3\linewidth}
        \centering
        \includegraphics[width=\linewidth]{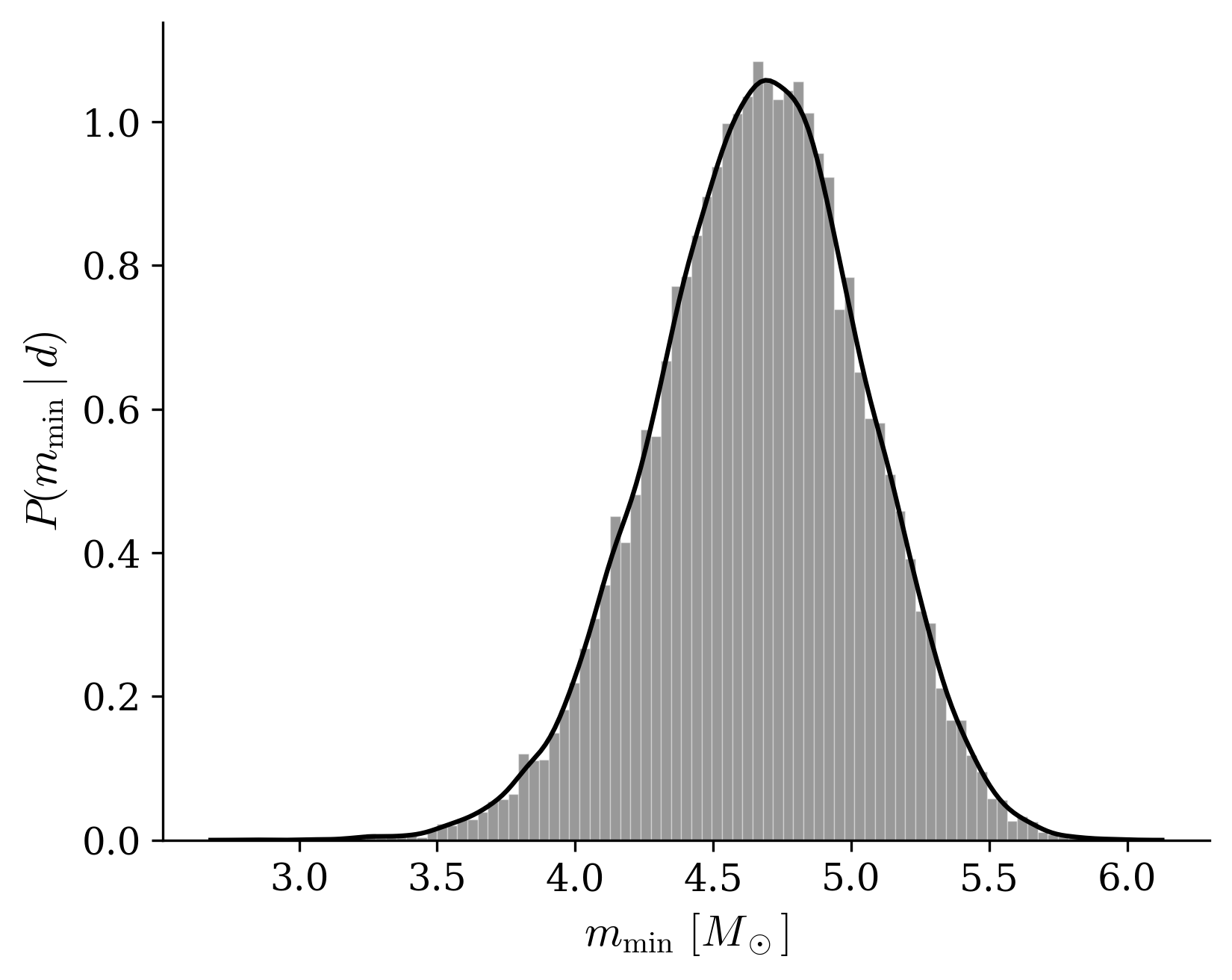}
    \end{minipage}
    ~
    \begin{minipage}[t]{0.3\linewidth}
        \centering
        \includegraphics[width=\linewidth]{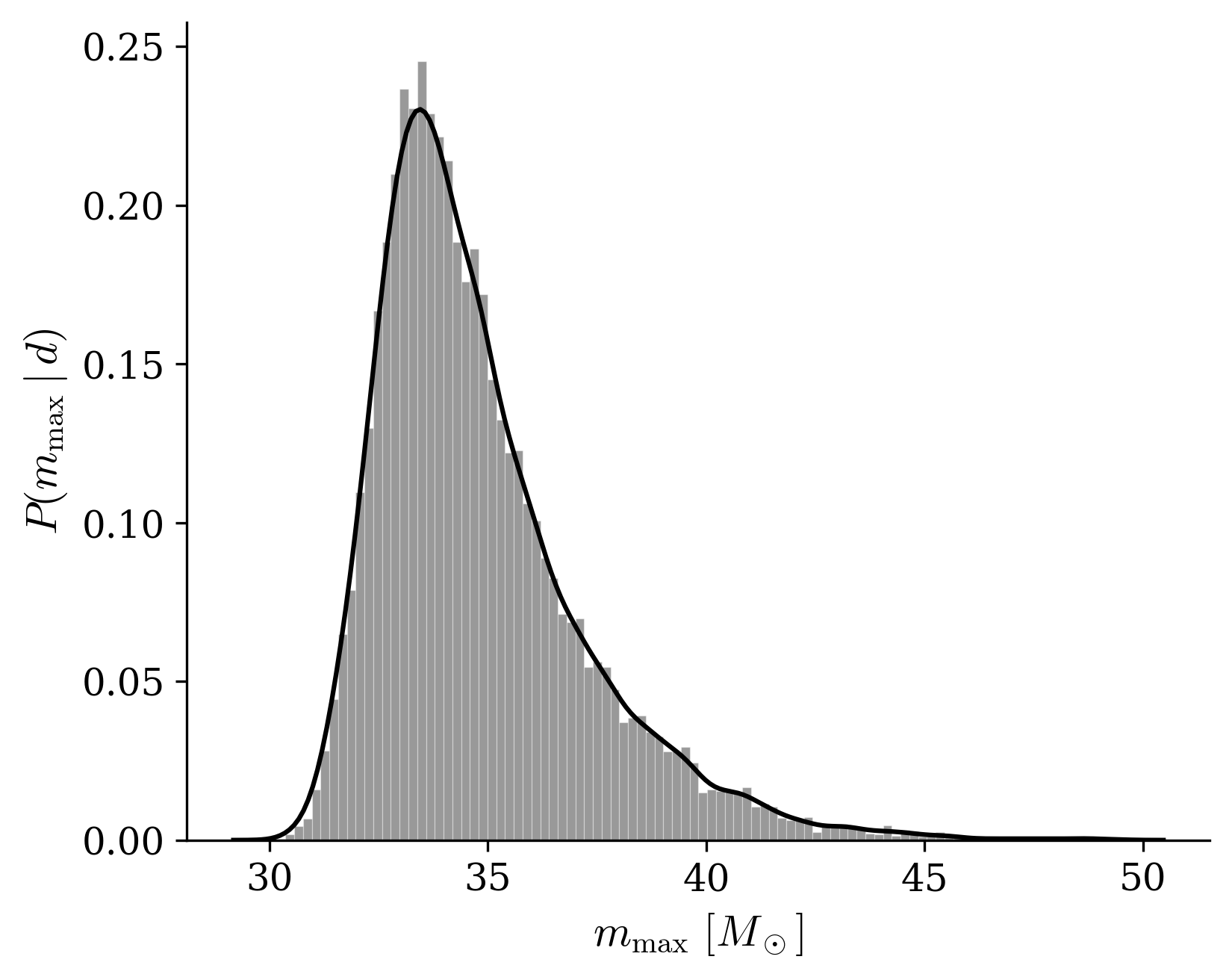}
    \end{minipage}
    \caption{Marginalized distributions of $\alpha$ (left graph), $m_{\min}$ (middle graph) and $m_{\max}$ (right graph) for the Power Law distribution for the Combined Sample.}
    \label{fig:PowerLaw}
\end{figure*}

\begin{figure*}[ht]
    \centering
    \begin{minipage}[t]{0.45\linewidth}
\centering
\includegraphics[width=\columnwidth]{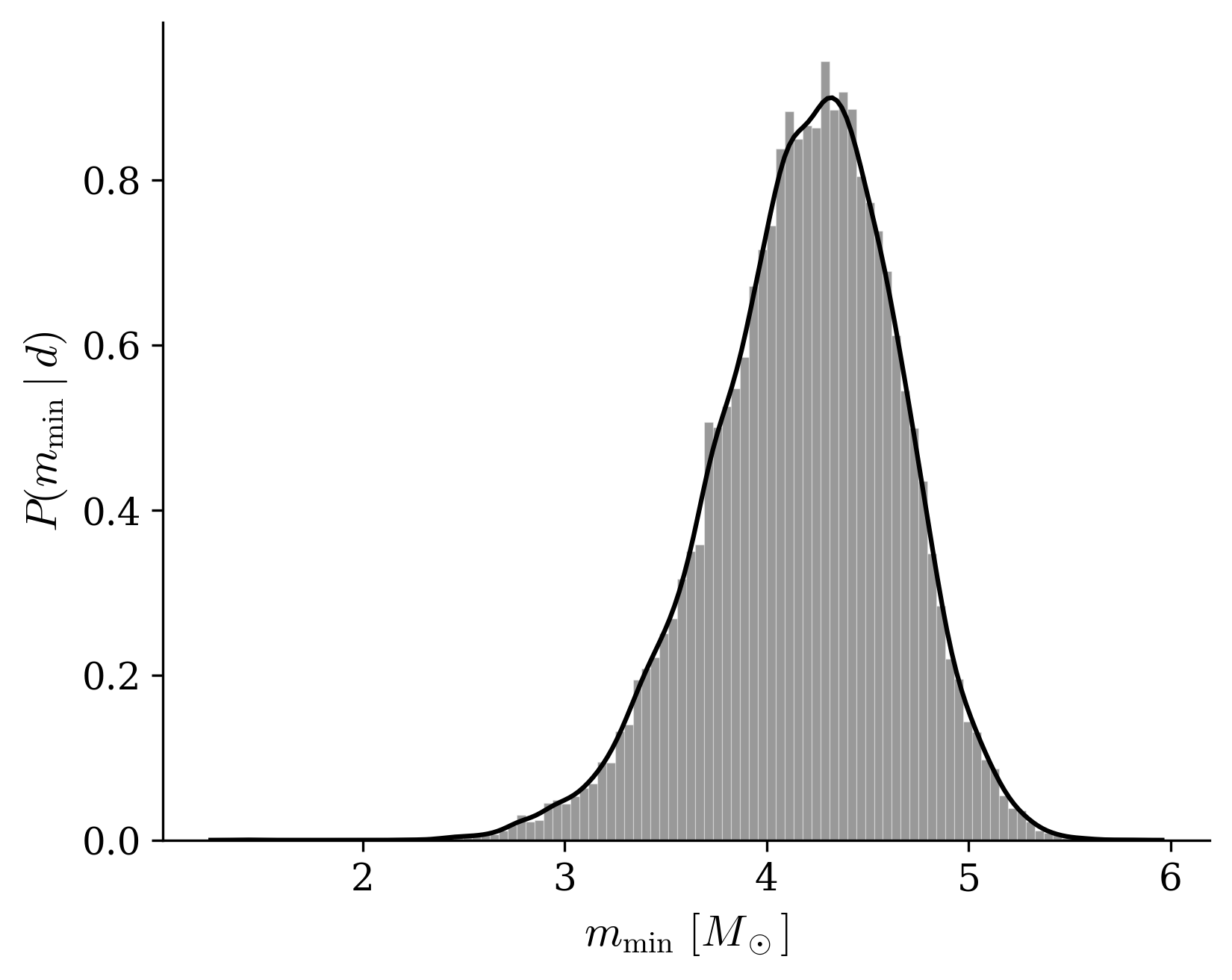}
    \end{minipage}
    ~
    \begin{minipage}[t]{0.45\linewidth}
        \centering
        \includegraphics[width=\columnwidth]{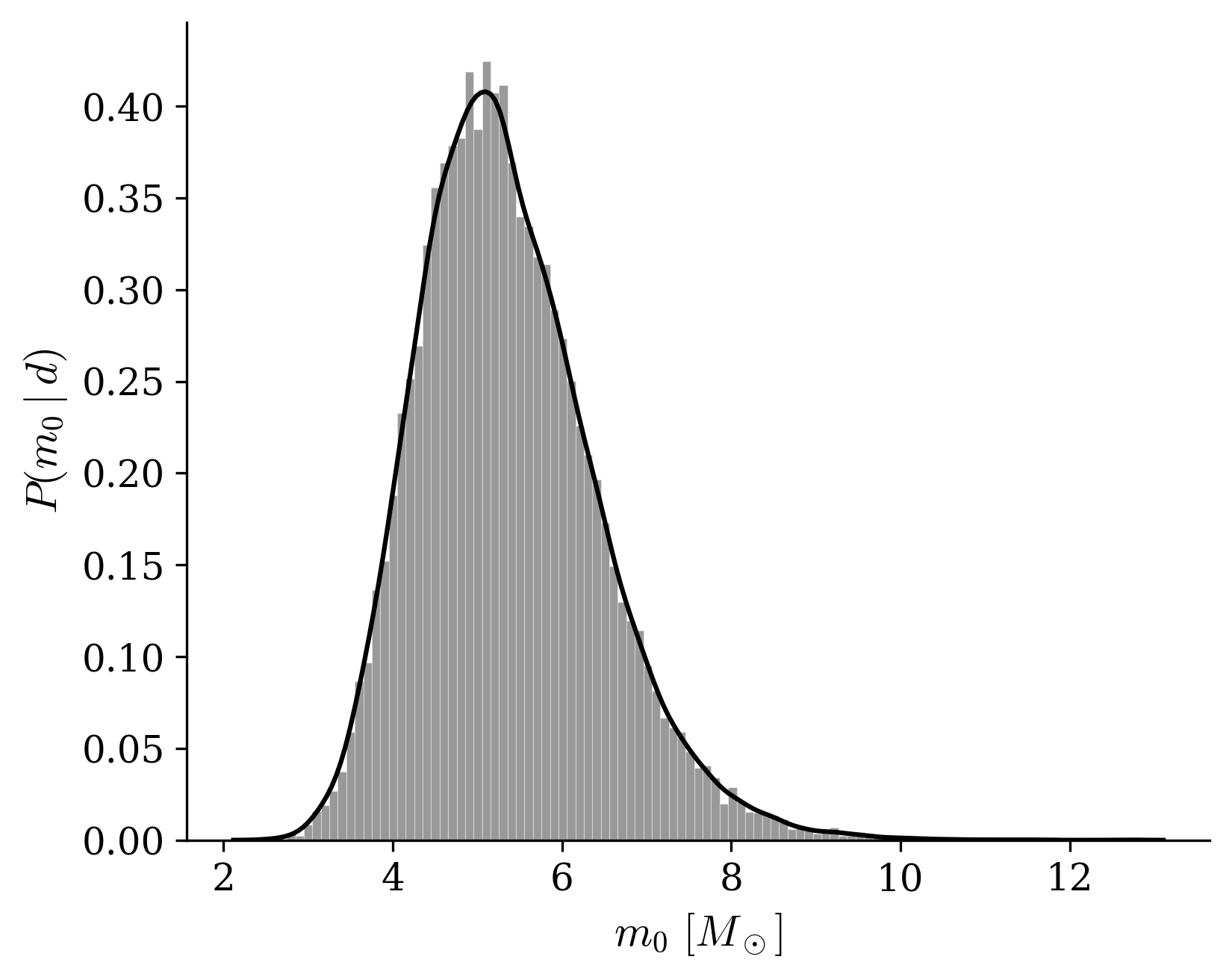}
    \end{minipage}
    \caption{Marginalized distributions of $m_{\min}$ (left graph) and $m_0$ (right graph) for the Decaying Exponential distribution for the Combined Sample.}
    \label{fig:ExpDecay}
\end{figure*}

Figures \ref{fig:Gaussian} -- \ref{fig:ExpDecay} shows the marginalized posterior distributions in the Combined sample framework, for each of the parameters in all five models considered in our work, and summarized in Table \ref{tab:fullsample_results}. 

\newpage
\subsection{LMXB sample}\label{apx:dist_lmxb}
\begin{figure*}[ht]
    \centering
    \begin{minipage}[t]{0.45\linewidth}
        \centering
        \includegraphics[width=\linewidth]{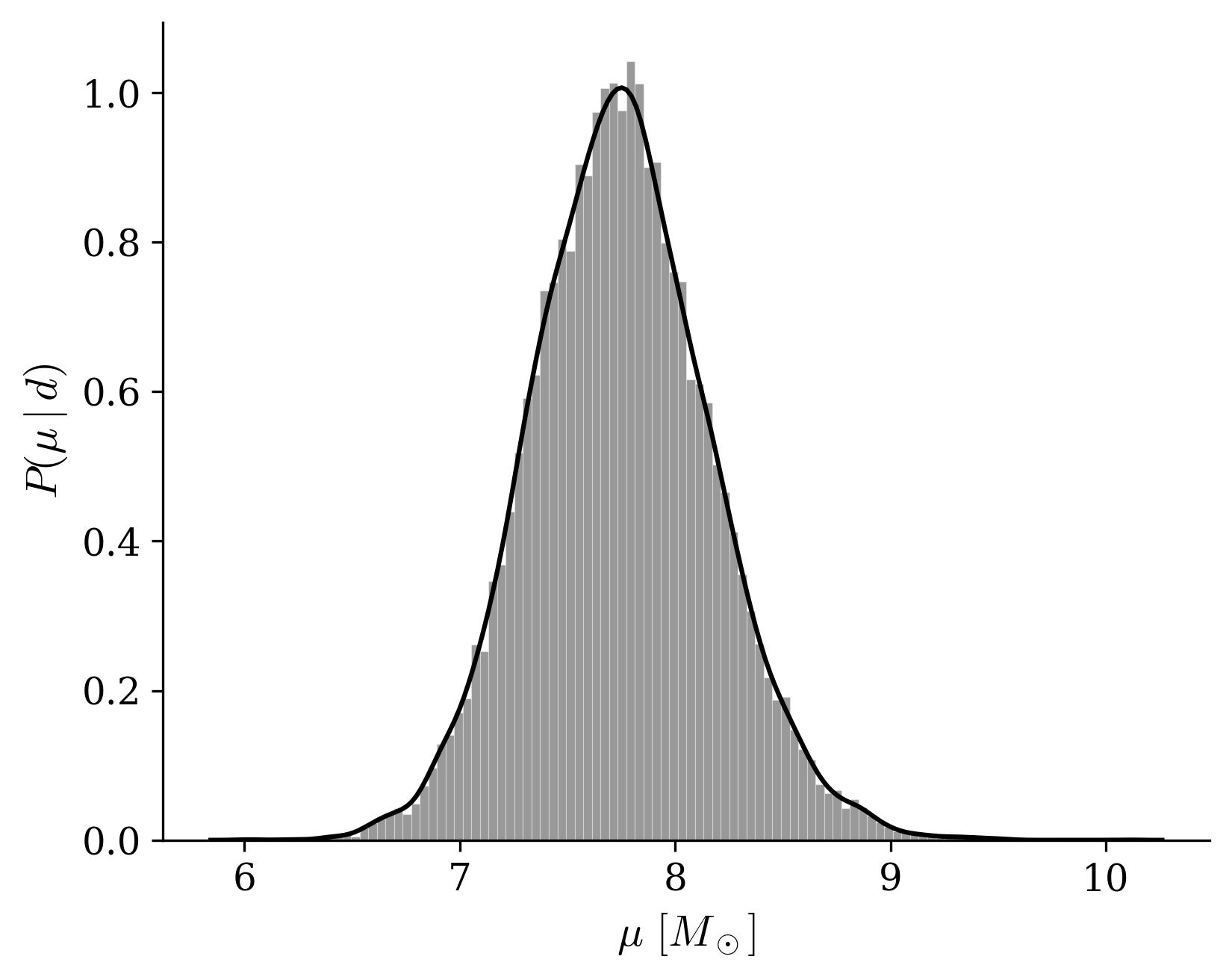}
    \end{minipage}
    ~
    \begin{minipage}[t]{0.45\linewidth}
        \centering
        \includegraphics[width=\linewidth]{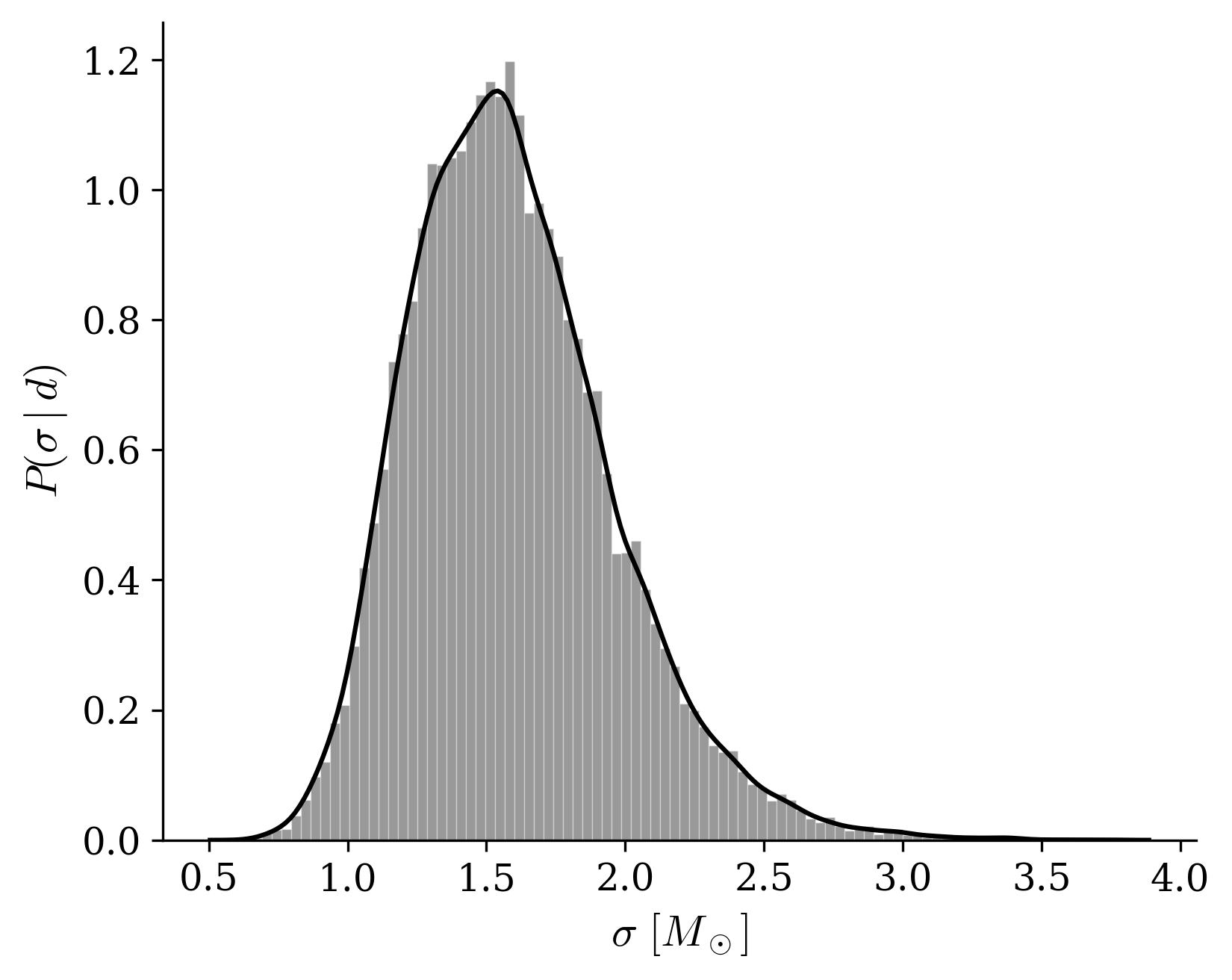}
    \end{minipage}
    \caption{Marginalized distributions of $\mu$ (left graph) and $\sigma$ (right graph) for the Gaussian distribution for the LMXB Sample.}
    \label{fig:Gaussian_LMXB}
\end{figure*}

\begin{figure*}[ht]
    \centering
    \begin{minipage}[t]{0.3\linewidth}
        \centering
        \includegraphics[width=\linewidth]{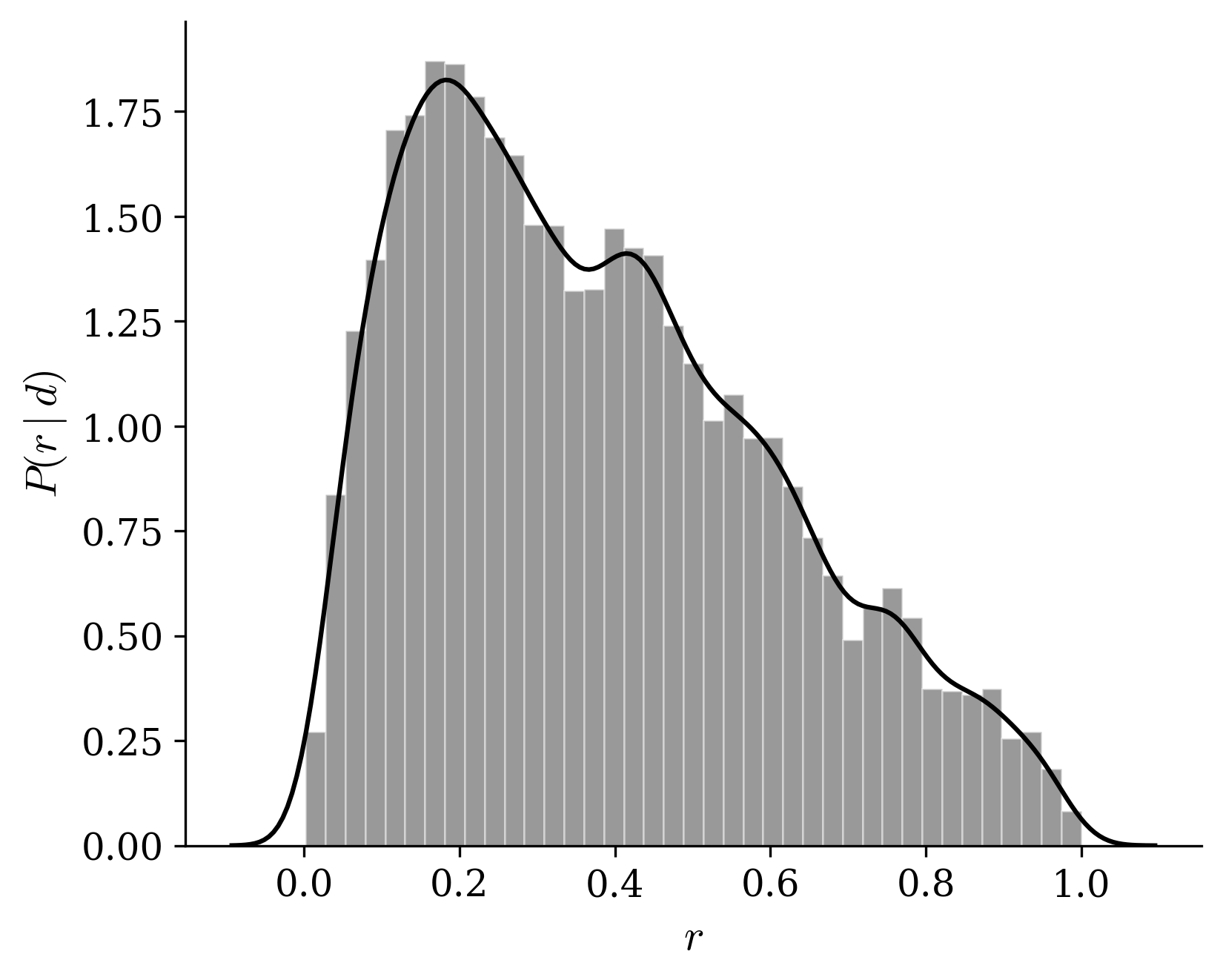}
    \end{minipage}
    ~
    \begin{minipage}[t]{0.3\linewidth}
        \centering
        \includegraphics[width=\linewidth]{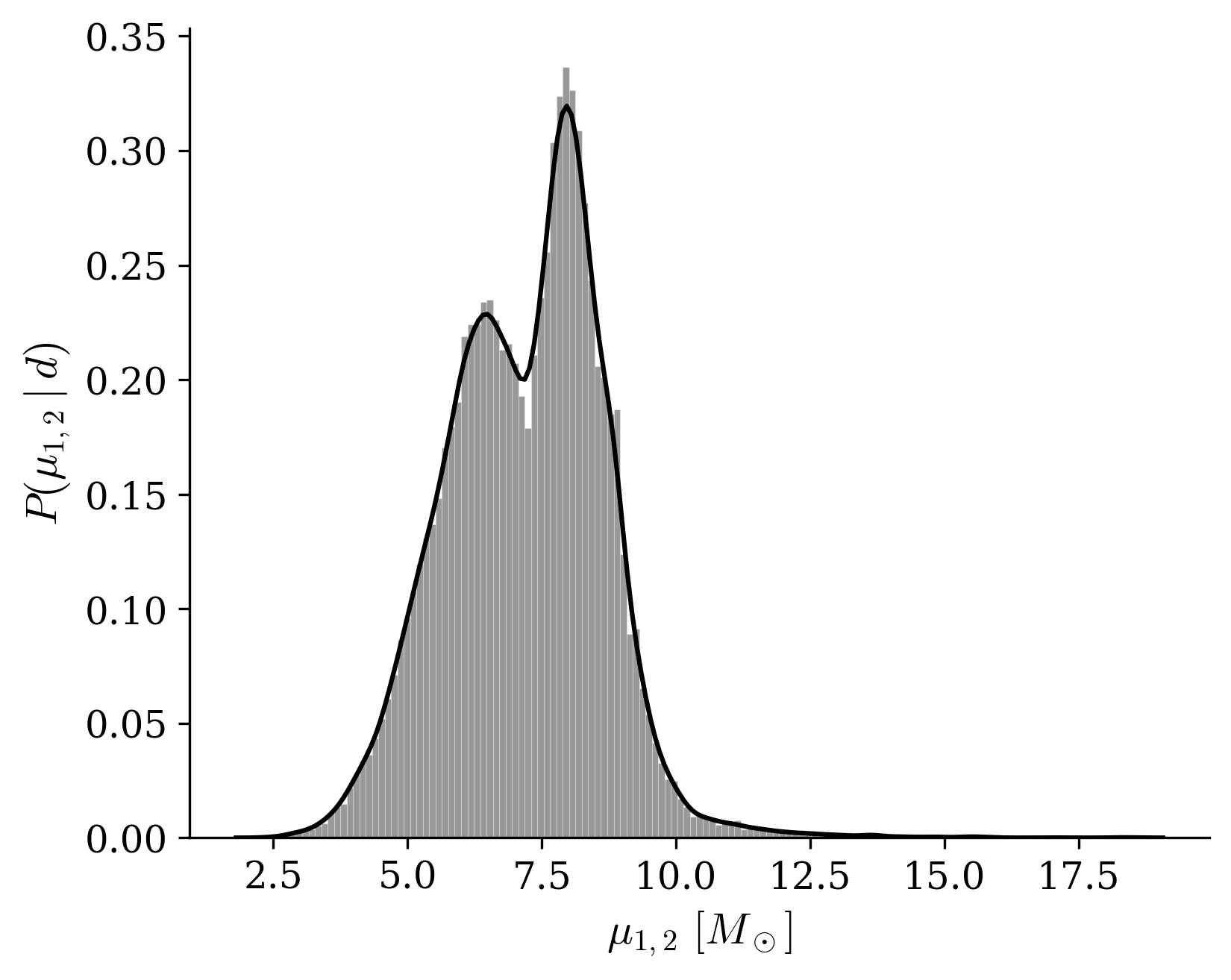}
    \end{minipage}
    ~
    \begin{minipage}[t]{0.3\linewidth}
        \centering
        \includegraphics[width=\linewidth]{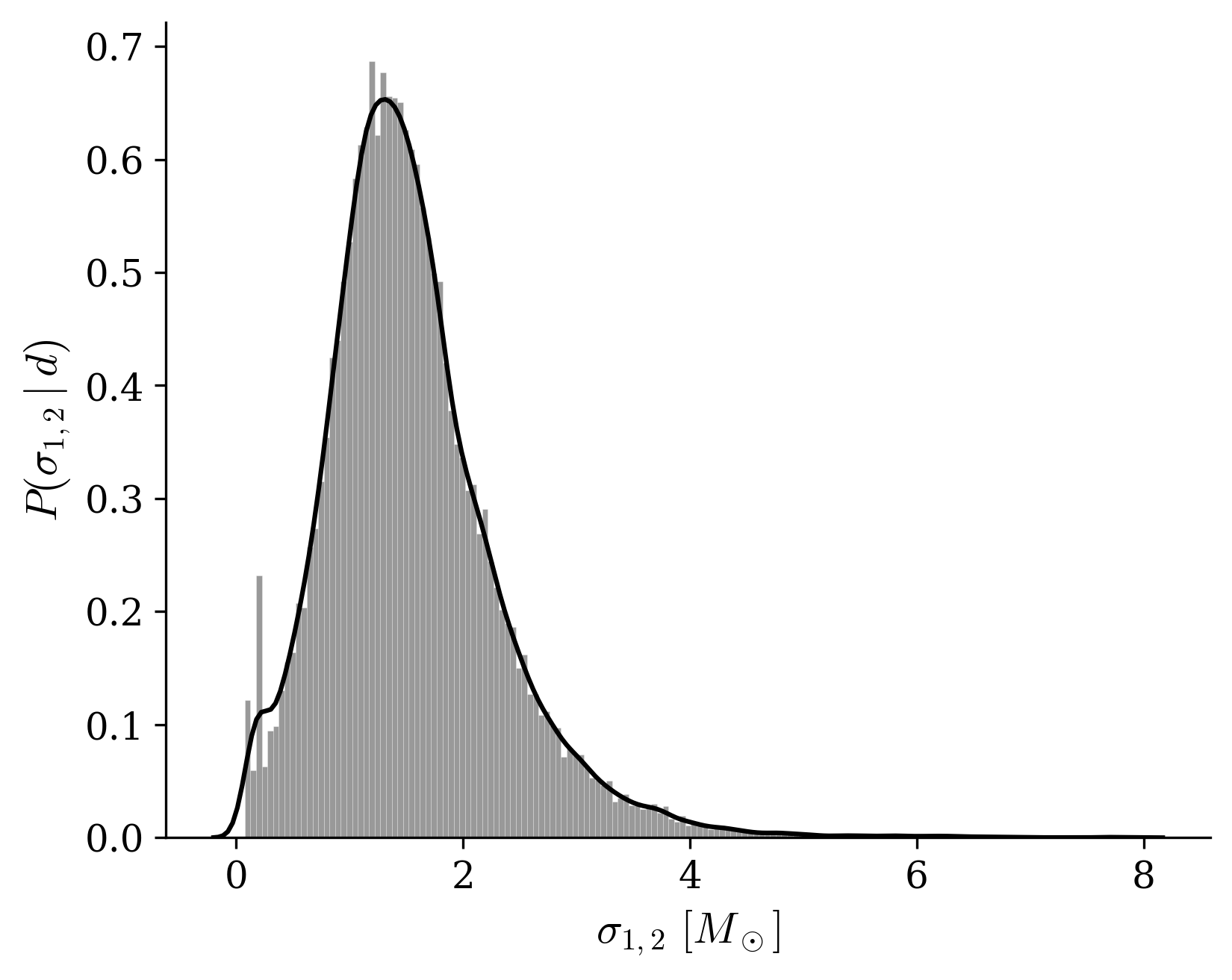}
    \end{minipage}
    \caption{Marginalized distributions of $r$ (left graph), $\mu_{1,2}$ (middle graph, in which the lighter graph indicated the $\mu_1$ distribution while the darker one, $\mu_2$) and $\sigma_{1,2}$ (right graph, in which the lighter graph indicated the $\sigma_1$ distribution while the darker one, $\sigma_2$) for the Two Gaussian distribution for the LMXB Sample.}
    \label{fig:TwoGaussian_LMXB}
\end{figure*}

\begin{figure*}
    \centering
    \begin{minipage}[t]{0.45\linewidth}
        \centering
        \includegraphics[width=\linewidth]{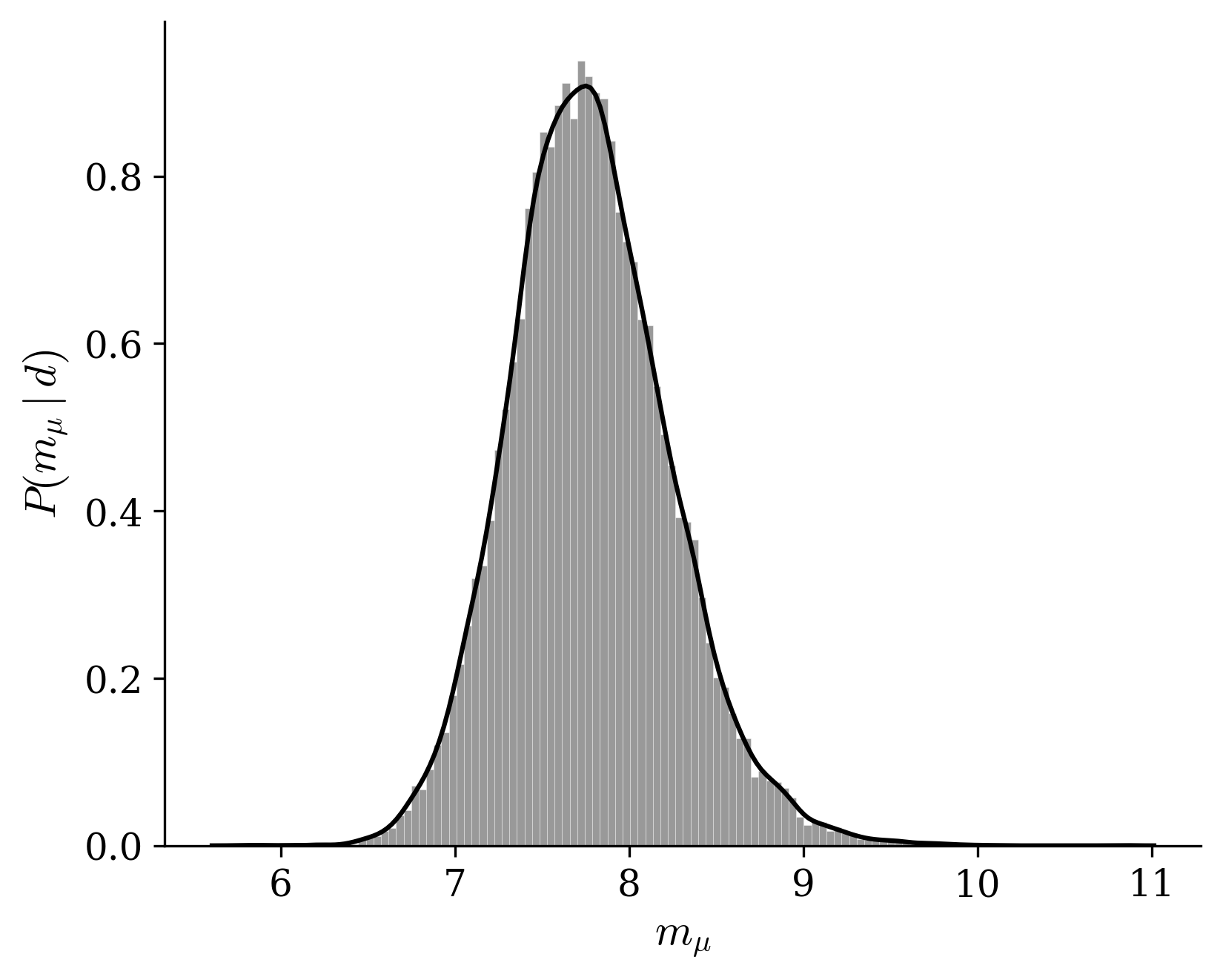}
    \end{minipage}
    ~
    \begin{minipage}[t]{0.45\linewidth}
        \centering
        \includegraphics[width=\linewidth]{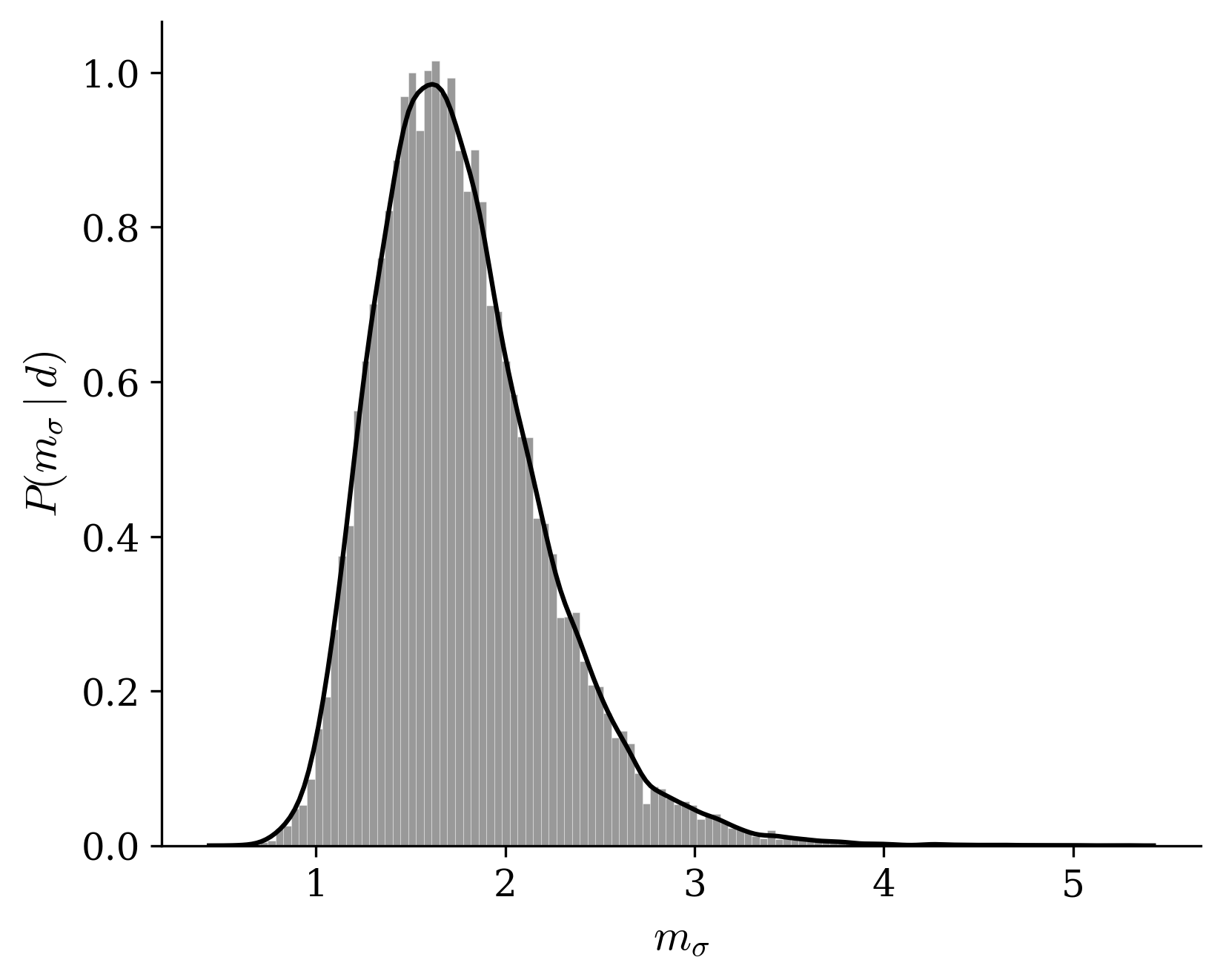}
    \end{minipage}
    \caption{Marginalized distributions of the average BH mass $\langle M_{BH}\rangle$ (left graph) and the uncertainty of the BH mass $\sigma_{M_{BH}}$ (right graph) for the Log Normal distribution for the LMXB Sample.}
    \label{fig:LogNormal_LMXB}
\end{figure*}

\begin{figure*}[ht]
    \centering
    \begin{minipage}[t]{0.3\linewidth}
        \centering
        \includegraphics[width=\linewidth]{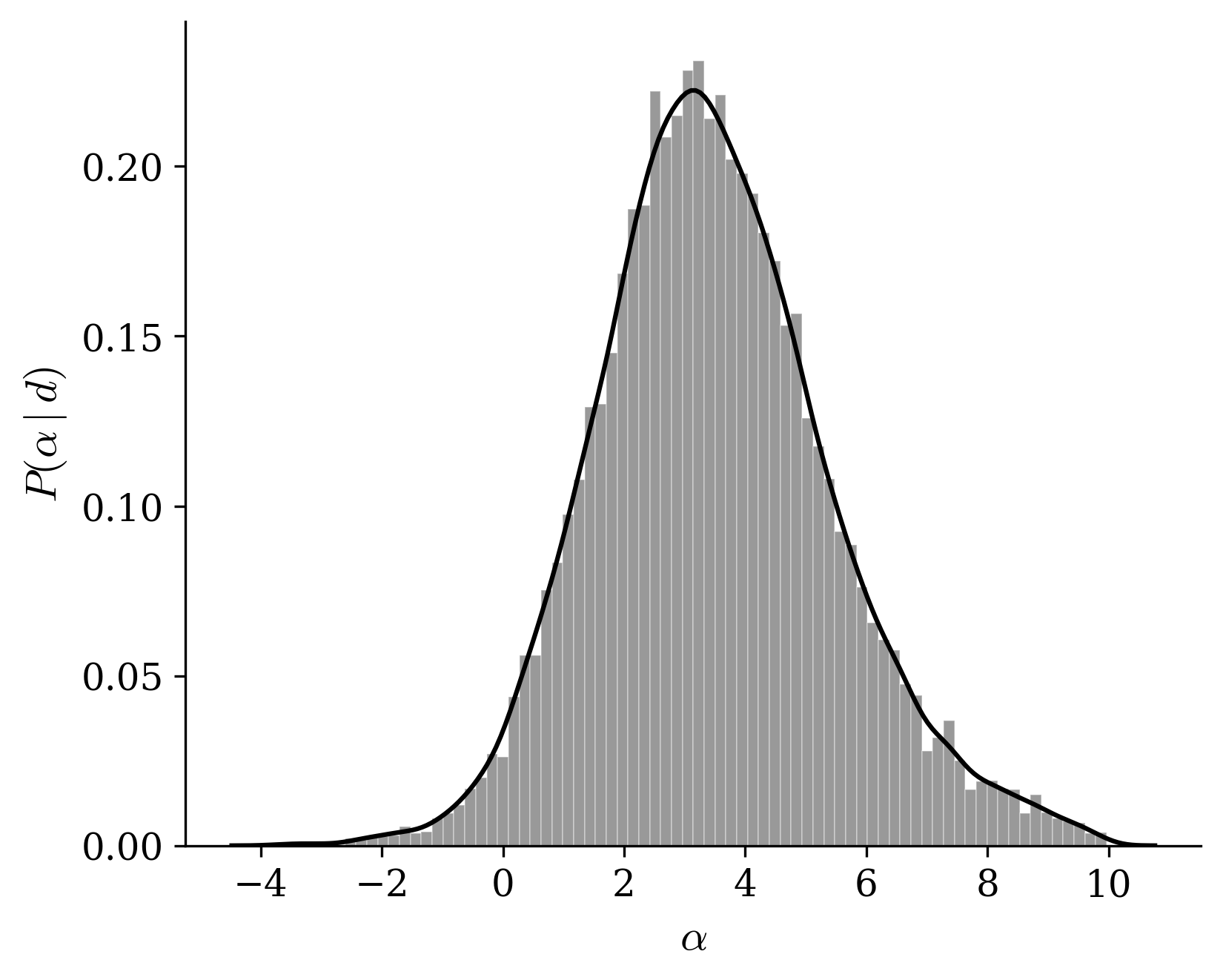}
    \end{minipage}
    ~
    \begin{minipage}[t]{0.3\linewidth}
        \centering
        \includegraphics[width=\linewidth]{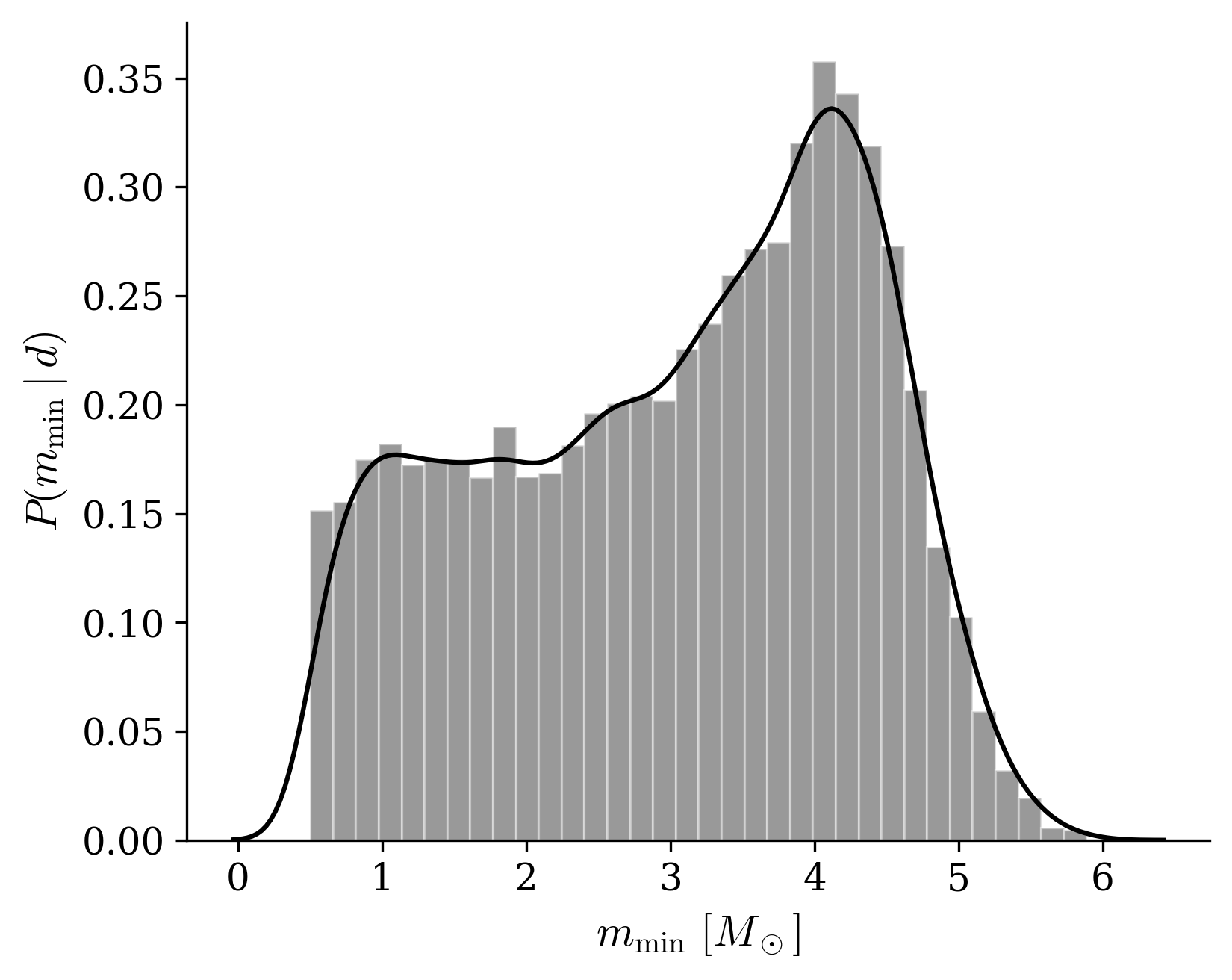}
    \end{minipage}
    ~
    \begin{minipage}[t]{0.3\linewidth}
        \centering
        \includegraphics[width=\linewidth]{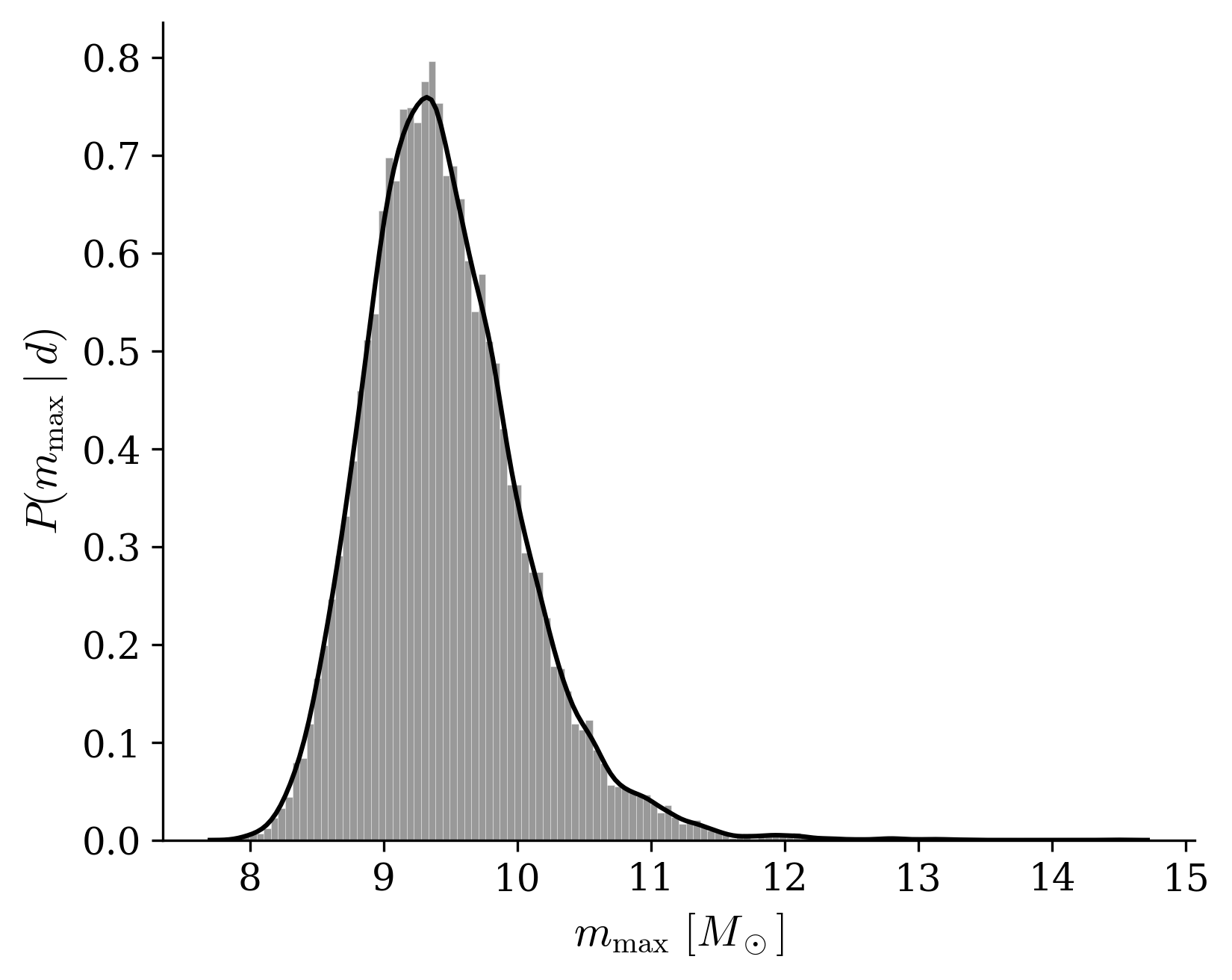}
    \end{minipage}
    \caption{Marginalized distributions of $\alpha$ (left graph), $m_{\min}$ (middle graph) and $m_{\max}$ (right graph) for the Power Law distribution for the LMXB Sample.}
    \label{fig:PowerLaw_LMXB}
\end{figure*}

\begin{figure*}[ht]
    \centering
    \begin{minipage}[t]{0.45\linewidth}
        \centering
        \includegraphics[width=\linewidth]{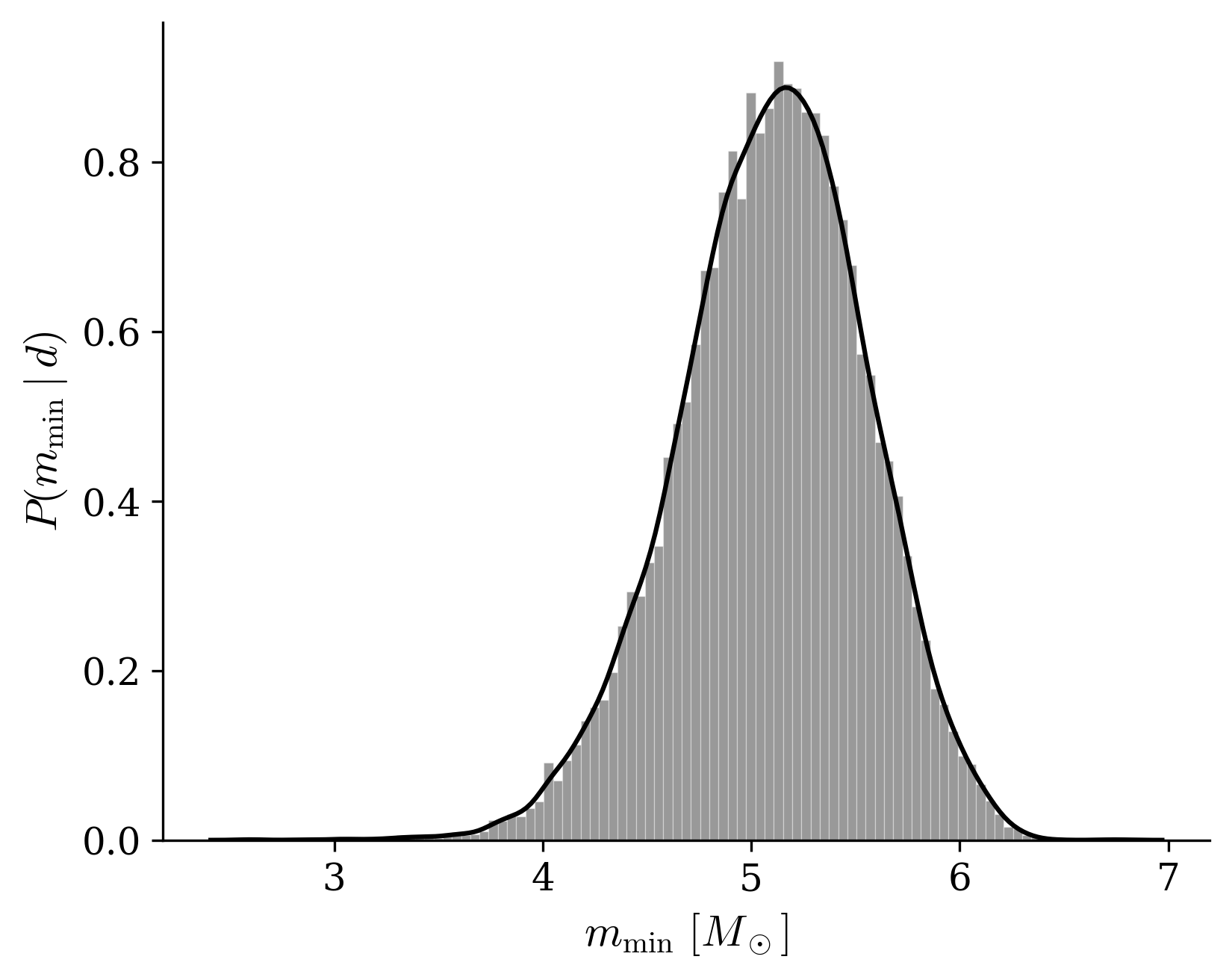}
    \end{minipage}
    ~
    \begin{minipage}[t]{0.45\linewidth}
        \centering
        \includegraphics[width=\linewidth]{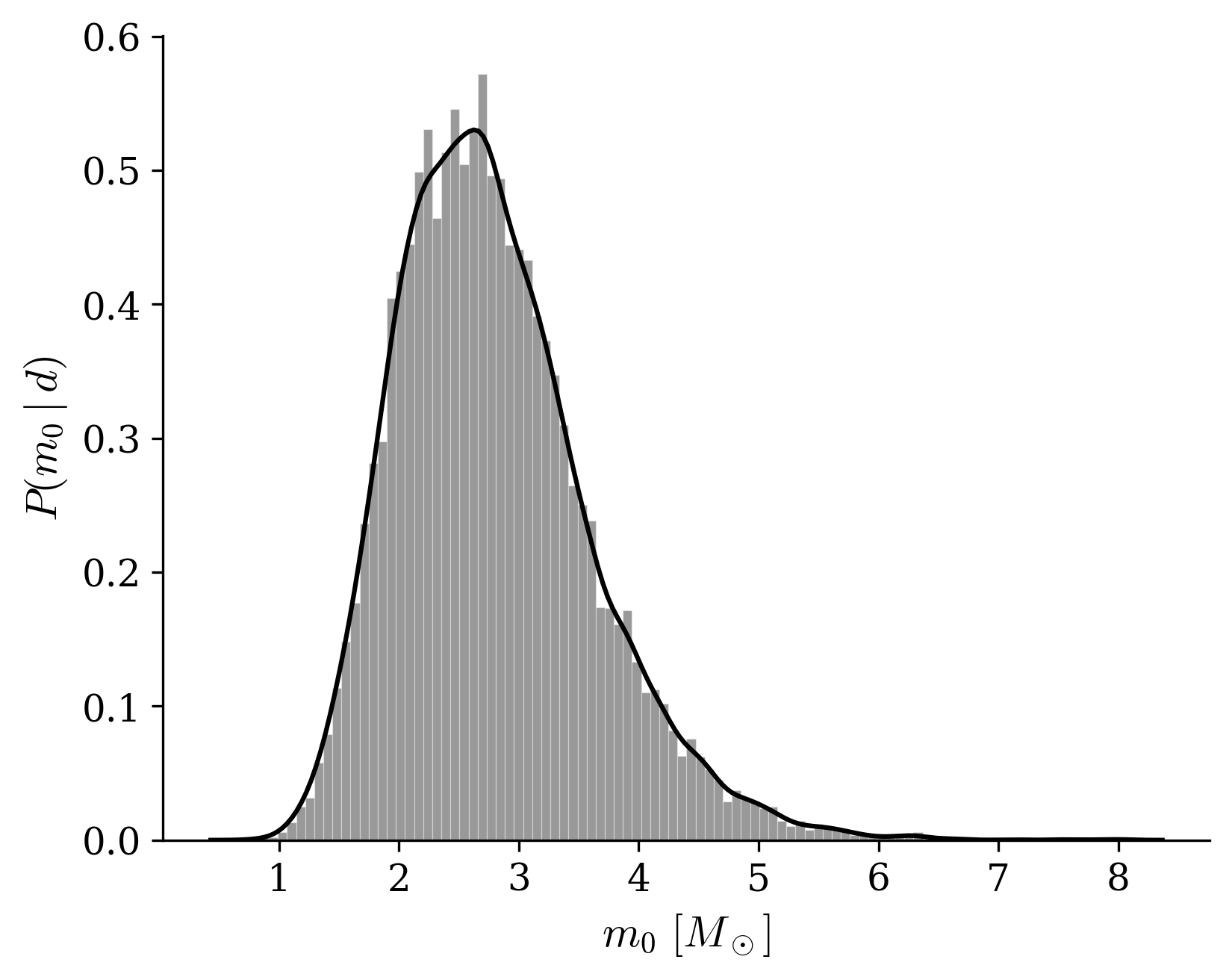}
    \end{minipage}
    \caption{Marginalized distributions of $m_{\min}$ (left graph) and $m_0$ (right graph) for the Decaying Exponential distribution for the LMXB Sample.}
    \label{fig:ExpDecay_LMXB}
\end{figure*}

Figures \ref{fig:Gaussian_LMXB} -- \ref{fig:ExpDecay_LMXB} shows the marginalized posterior distributions we derived for the LMXB sample, for each of the parameters in all five models considered in this work, and summarized in Table \ref{tab:lmxb_results}. 

\subsection{Refined Sample}\label{apx:refinedsample}
As stated in Section \ref{subsec:refined_sample}, in face of questionable or model-dependent measurements of orbital parameters, we test the Bayesian analysis for a sample without a few objects, and inspect the effect this has on final results. The systems MAXI J0637-430, MAXI J1813-095, MAXI J1535-571 and XTE J1746-322 lack information on the orbital parameters, preventing us from re-estimating the individual masses.

For the GRS 1009-45, the authors show both an upper constraint for the inclination angle of $i<80$°, due to the absence of eclipses, and $i\approx 78^\circ$ when assuming a K7-K8 secondary star \citep{filippenko1999}. Given the high dependence the mass determination has on the inclination angle \citep{kreidberg2012}, we decided to test the distribution without the presence of this systems.

In the case of Swift J1727.8-2525, the authors derived a minimum BH mass from the limiting case $q\sim0$ \citep{sanchez2024}. Then, for evaluating the BH mass, we supposed that $q=\mathcal{U}(0,1)$ (see Table \ref{tab:list_BHs}), which can heavily change the BH mass. Therefore we also decided to test the results in the absence of this system.

\begin{figure*}
    \centering
    \begin{minipage}[t]{0.45\linewidth}
        \centering
        \includegraphics[width=\linewidth]{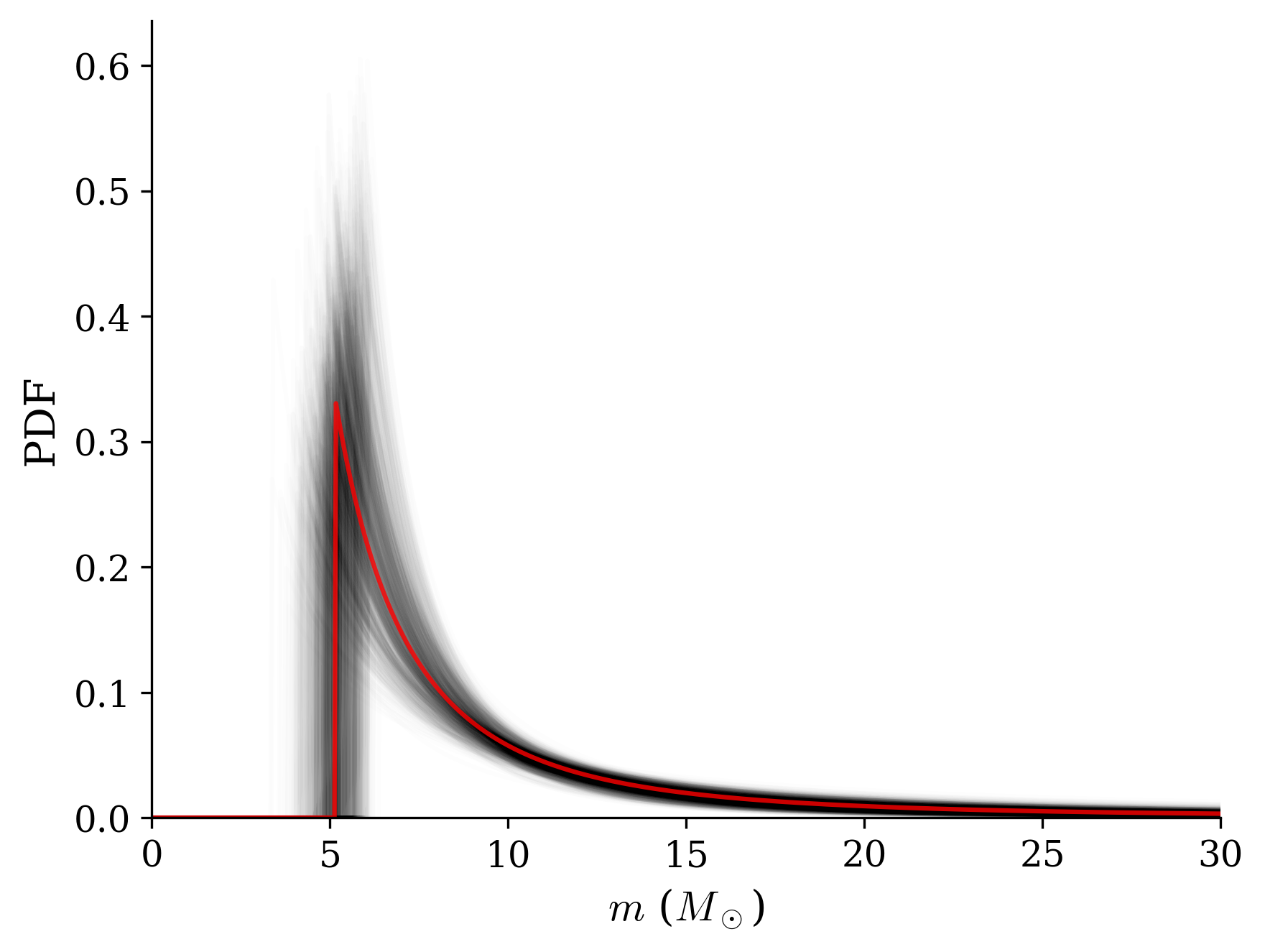}
    \end{minipage}
    ~
    \begin{minipage}[t]{0.45\linewidth}
        \centering
        \includegraphics[width=\linewidth]{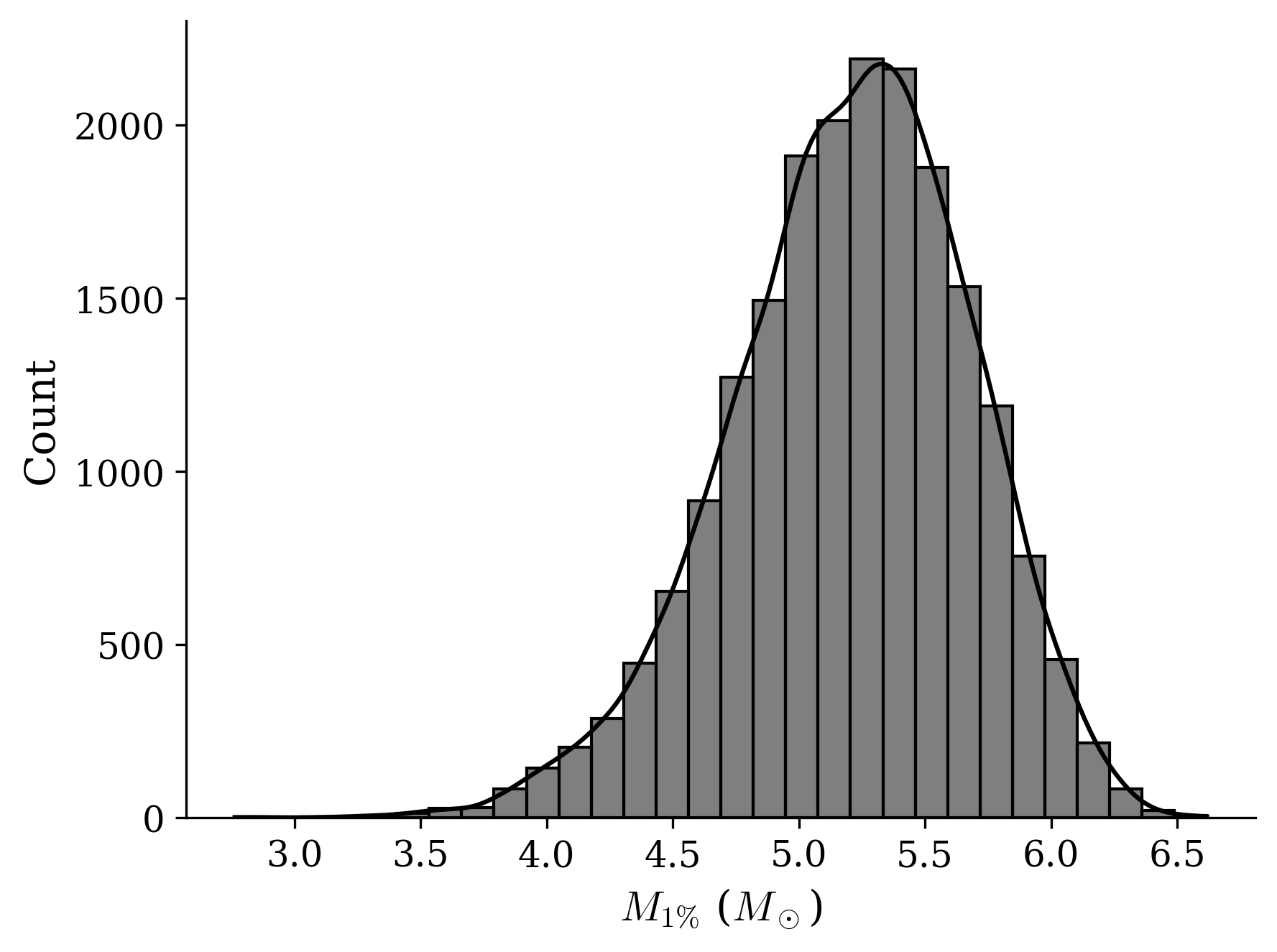}
    \end{minipage}
    \caption{Left: Predictive posterior distribution obtained from the power-law model in the Refined sample case. Grey lines represent 1000 posterior samples. The red curve shows the MAP distribution. Right: Distribution of the $1\%$ quantile built from the predictive posterior distribution. }
    \label{fig:Refined_posterior}
\end{figure*}

Table \ref{tab:model_comparison_refined} summarizes the model comparison we obtained for the Refined sample, confirming that when removing the mentioned systems from the analysis, the power law is strongly favored over other parametrizations. The predictive posterior distribution is shown in the left panel of Figure  \ref{fig:Refined_posterior}, basically indistinguishable from what we found for the combined sample, shown in the left panel of Figure \ref{fig:Posterior_distributions}. The $M_{1\%}$ distribution is defined in the $[4.37, 5.93]$ interval with $90\% C.I.$, as shown in the right panel of Figure \ref{fig:Refined_posterior}

\begin{table*}[ht!]
    \centering
    \caption{Information-criterion comparison for the five population models on the Refined sample. For each criterion we report its value and the corresponding normalised Akaike-style weight.}
    \label{tab:model_comparison_refined}
    \setlength\tabcolsep{5pt}
    \begin{tabular}{| c | c l | c l | c l | c l |}
        \hline
        \multicolumn{9}{|c|}{Refined sample} \\
        \hline \hline
        \multirow{2}{4em}{Model} & \multicolumn{2}{c|}{WAIC} & \multicolumn{2}{c|}{LOO} & \multicolumn{2}{c|}{AIC} & \multicolumn{2}{c|}{BIC} \\
                                 & WAIC & $w_{WAIC}$ & LOO & $w_{LOO}$ & AIC & $w_{AIC}$ & BIC & $w_{BIC}$ \\
        \hline
        Gaussian     & 336.23          & $1.3\times10^{-11}$ & 366.51          & $4.4\times10^{-13}$ & 302.99          & $2.0\times10^{-10}$ & 306.21          & $4.3\times10^{-10}$ \\
        Two Gaussian & 304.05          & $1.3\times10^{-4}$  & 333.25          & $7.3\times10^{-6}$  & 279.02          & $3.2\times10^{-5}$  & 287.07          & $6.2\times10^{-6}$ \\
        Log Normal   & 303.77          & $1.5\times10^{-4}$  & 329.82          & $4.1\times10^{-5}$  & 276.29          & $1.3\times10^{-4}$  & 279.46          & $2.8\times10^{-4}$ \\
        Power Law    & \textbf{286.11} & $\mathbf{0.997}$    & \textbf{309.60} & $\mathbf{0.999}$    & \textbf{258.36} & $\mathbf{0.995}$    & \textbf{263.11} & $\mathbf{0.990}$ \\
        Exp.\ Decay  & 297.77          & $2.9\times10^{-3}$  & 323.50          & $9.6\times10^{-4}$  & 269.17          & $4.5\times10^{-3}$  & 272.39          & $9.6\times10^{-3}$ \\
        \hline
    \end{tabular}
\end{table*}

\end{document}